\documentclass[10pt]{article}

\usepackage{amsmath,amsfonts}
\usepackage{amsbsy}
\usepackage{color,soul}
\usepackage{verbatim}
\usepackage{setspace}
\usepackage{rotating}
\usepackage{array}
\usepackage{graphicx}
\usepackage{pdflscape}
\usepackage{booktabs}
\usepackage{colortbl}
\usepackage{xcolor}
\usepackage{multirow}
\definecolor{lightgray}{gray}{0.9}
\usepackage{xcolor}

\usepackage[bordercolor=white,backgroundcolor=gray!30,linecolor=black,colorinlistoftodos]{todonotes}

\usepackage{enumerate}
\usepackage{url} 

\newcommand{\bz}{{\bf z}}

\newcommand{\bbe}{\mbox{\boldmath{$\beta$}}}
\newcommand{\bmu}{\mbox{\boldmath{$\mu$}}}
\newcommand{\bSig}{\mbox{\boldmath{$\Sigma$}}}
\newcommand{\bgam}{\mbox{\boldmath{$\gamma$}}}

\newcommand{\bzeta}{\mbox{\boldmath{$\zeta$}}}
\newcommand{\bZ}{{\bf Z}}
\newcommand{\ba}{{\bf a}}

\newcommand{\bV}{{\bf V}}
\newcommand{\bW}{{\bf W}}

\newcommand{\bD}{{\bf D}}
\newcommand{\bG}{{\bf G}}
\newcommand{\bH}{{\bf H}}
\newcommand{\bX}{{\bf X}}

\newcommand{\bU}{{\bf U}}

\newcommand{\bg}{{\bf g}}
\newcommand{\Var}{\mbox{Var}}

\newcommand{\Cov}{\mbox{Cov}}

\newcommand{\mc}{\multicolumn}
\newcommand{\real}{\mathbb{R}}

\newcommand{\bth}{{\mbox{\boldmath{$\theta$}}}}
\newcommand{\bph}{{\mbox{\boldmath{$\phi$}}}}

\numberwithin{equation}{section}
\usepackage[]{graphicx}
\chardef\bslash=`\\ 

\def\bc{\begin{center}}
\def\ec{\end{center}}

\title{Estimation in the Cox Survival Regression Model with Covariate Measurement Error and a Changepoint}
\author{Sarit Agami\\
Department of Statistics\\
Hebrew University, Mount Scopus, Jerusalem, Israel\\
email:sarit.agami@mail.huji.ac.il\\
\and
David M. Zucker\\
Department of Statistics\\
Hebrew University, Mount Scopus, Jerusalem, Israel\\
\and
Donna Spiegelman\\
Departments of Epidemiology, Biostatistics, Nutrition and Global Health\\
Harvard T.H. Chan School of Public Health, Boston MA, USA}

\begin{document}
\maketitle

\begin{abstract}
The Cox regression model is a popular model for analyzing the relationship between a covariate vector and a survival endpoint.
The standard Cox model assumes a constant covariate effect across the entire covariate domain. However, in many epidemiological and other applications, the covariate of main interest is subject to a threshold effect: a change in the slope at a certain point within the covariate domain. Often, the covariate of interest is subject to some degree of measurement error. In this paper, we study measurement error correction in the case where the threshold is known.
Several bias correction methods are examined: two versions of regression calibration (RC1 and RC2, the latter of which is new), two methods based on the induced relative risk under a rare event assumption (RR1 and RR2, the latter of which is new), a maximum pseudo-partial likelihood estimator (MPPLE), and simulation-extrapolation (SIMEX). We develop the theory, present simulations comparing the methods, and illustrate their use on data concerning the relationship between chronic air pollution exposure to particulate matter PM$_{10}$ and fatal myocardial infarction (Nurses Health Study (NHS)), and on data concerning the effect of a subject's long-term underlying systolic blood pressure level on the risk of cardiovascular disease death (Framingham Heart Study (FHS)). The simulations indicate that the best methods are RR2 and MPPLE.
\end{abstract}

\maketitle                   






\section{Introduction}
The {Cox} model is a popular model for analyzing the relationship between a covariate vector and a survival endpoint.
The Cox model expresses the hazard function as
\begin{equation}
\lambda(t|\bzeta(t)) = \lambda_0(t) \exp(\bbe^T \bzeta(t)),
\label{CoxA}
\end{equation}
where $\lambda_0(t)$ is a baseline hazard function of unspecified form, $\bzeta(t)$ is the covariate vector
(which can depend on time), and
$\bbe$ is a vector of regression coefficients to be estimated.
Model (\ref{CoxA}) implies the assumption that the covariate effects are constant across the entire covariate domain.
In many epidemiological and other applications, however, there is interest in considering the possibility
that the covariate of primary interest is subject to a threshold effect, that is, a change in the slope at a certain
point within the covariate domain. Our interest in this issue was prompted, as described in Zucker {\it{et al}}.\ (2013),
by some instances of threshold effects observed in the Nurses' Health Study (NHS). For example, a threshold effect was observed in the relationship between air pollution and fatal myocardial infection (MI). Often the covariate
of primary interest is not measured exactly, but rather is subject to some degree of measurement error.
This leads to the problem addressed in this paper, of estimating threshold effects in the presence of covariate measurement error.

The problem of covariate measurement error in standard regression models without threshold effects has
been extensively studied. Fuller (2009) provides a comprehensive treatment of covariate error in classical
linear regression models, and Carroll {\it{et al}}.\ (2006) provides a comprehensive treatment of measurement error
correction in nonlinear models. There is a substantial literature, starting from Prentice (1982), on
measurement error correction methods in the specific context of the Cox model. Zucker (2005) provides
a review of these methods.

Several papers have considered threshold models with covariate error in the context of linear regression models. Gbur and Dahm (1985) considered the case of a segmented linear model with known changepoint and suggested a moment estimator for the regression coefficients. Kukush and Van Huffel (2004) considered the multivariate measurement error model. Additional
correction methods for the linear case have been examined by Storck and Vencovsky (1994) and Grimshaw (1992). K{\"u}chenhoff and Carroll (1997) considered threshold regression in the generalized linear model with unknown changepoint.
They studied the regression calibration (RC) method and the SIMEX method (Carroll {\it{et al}}., 2006, Chapters 4 and 5).
They found that the RC estimator usually has more bias but smaller variance than the SIMEX estimator; this is contrary to the case without a changepoint in which these two procedures behave similarly. Staudenmayer and Spiegelman (2002) examined the direction of the bias in the estimated changepoint in segmented regression with covariate measurement error in main study/validation study designs. G{\"o}ssl and K{\"u}chenhoff (2001) considered threshold logistic regression from a Bayesian perspective. Quintana {\it{et al}}.\ (2005) considered Bayesian identification of the threshold by clustering algorithms. As far as we are aware, threshold models with covariate error in the context of survival data has not been previously studied.

Working a threshold Cox model in the presence of covariate error presents two challenges. The first challenge is dealing with the covariate error.
In the absence of measurement error, if the changepoint is known the analysis reduces to a simple application of standard Cox model methodology.
In the presence of measurement error, the problem becomes considerably more difficult, even when the changepoint is known. Beyond that, there
is the second challenge of estimating the changepoint.
\\
\indent There are some cases where the potential changepoint is known, at least
approximately. One notable example is body mass index, which has been found to be protective against breast cancer risk before the age of menopause, and harmful afterwards (Pitsavos et al. (2005)). Another example is alcohol intake, which exhibits a well established changepoint in relation to cardiovascular disease risk with a protective effect observed up to 15 g/day, and a harmful effect after (Chen et al. (2016)). These are considered as examples for known changepoint since they are estimates of the data from literally millions of participants, in some cases through meta-analyses of all available data up to the publication date.
In this paper, we examine methods for correcting the estimated regression coefficients in the presence of measurement error under the Cox model with a threshold effect, where the changepoint is known.
We consider the case where the covariate is dependent on time, unless mentioned otherwise.
Studying the case of known changepoint
provides useful insight into the effect of measurement error in fitting a threshold Cox model and the performance of various approaches to dealing
with the measurement error. The known changepoint case provides a benchmark for the unknown changepoint case.
In a follow-up paper currently in preparation, we examine estimation of the changepoint in the Cox threshold model with covariate error.

Let $X(t)$ denote the covariate of main interest and $\bZ(t) \in \real^p$ the vector of additional covariates.
The main covariate $X(t)$ is subject to measurement error, while the additional covariates $\bZ(t)$ are error-free.
The measurement error in $X(t)$ is assumed to be additive, that is, we observe $W(t) = X(t) + U$, where $U$ is a random
variable such that $E(U|X(t))=0$. Define $u_+ = \max(u,0)$. The model we consider in this paper is then given by
\begin{equation}
\lambda(t|x(t),\bz(t)) = \lambda_0(t) \exp(\bgam^T \bz(t) + \beta x(t) + \omega (x(t)-\tau)_+),
\label{CP}
\end{equation}
for a known changepoint $\tau$ in the covariate domain. We seek to estimate $\beta$, $\omega$, and $\bgam$.
Note that the special case of zero effect below the changepoint is obtained with $\beta=0$.
With $\tau$ known, the model can be put in the framework of the standard Cox model (\ref{CoxA}) by setting
$\zeta_j(t) = Z_j(t), j = 1, \ldots, p, \zeta_{p+1}(t)=X(t)$, and $\zeta_{p+2}(t) = (X(t)-\tau)_+$.

Thus, in the changepoint setup, the effect of the covariate $X$ involves two regression terms, one involving $X_1=X$ and one involving $X_2 = (X-\tau)_+$. There is a deterministic functional relationship between these two terms. In the measurement error situation, simple substitution leads to $W_1=W$ and $W_2 = (W-\tau)_+$. In Section 4, we present simulations comparing the results of naive analysis and regression calibration in
the changepoint setting just described to those obtained in a contrasting setting that we refer to as the ``2 variables" setting. In the ``2 variables" setting, the random vector $(X_1, X_2, W_1, W_2)$ is generated as a multivariate normal vector with the same covariance structure as for the vector $(X_1, X_2, W_1, W_2)$ in the changepoint setting. Thus, in the ``2 variables" setting, $X_1$ and $X_2$ are correlated, and so are $W_1$ and $W_2$, but there is no deterministic relationship between $X_1$ and $X_2$. If we were working with
a linear model, the limiting values of the regression parameter estimates would be completely determined
by the covariance structure, so that the results for the changepoint setting and those for the ``2 variables" setting would be essentially the same. But the Cox model is a nonlinear model, and the limit
of the Cox partial likelihood score function under a naive analysis involves terms that are nonlinear
in the covariates. Hence the results in the changepoint setting are not necessarily the same as those
as in the ``2 variables" setting, and indeed in the simulations reported in Secion 4 we find that the results for the two settings differ. Thus, the Cox changepoint model with covariate error is not simply a special case of the Cox model with two error-prone covariates.

The purpose of this paper is to develop two new methods for covariate measurement error correction in the Cox model with a threshold effect, and
adapt several existing ones to this context. Section 2 presents the notation and background, and describes the methods examined. Section 3 presents the asymptotic properties of the methods.
Section 4 presents a simulation study comparing the various methods.
Section 5 presents two examples. The first example involves data from the Nurses' Health Study
(NHS) on the relationship between air pollution, expressed in terms of exposure to particulate matter of diameter 10 $\mu g$/$m^3$ or less (PM$_{10}$),
and fatal myocardial infarction. The second example involves data from the Framingham Heart Study (FHS) on the effect of a subject's long-term underlying systolic blood pressure level on the risk of cardiovascular disease death. Section 6 presents a discussion, and Section 7 presents a brief summary.

Often the research context involves an event of main interest and other competing events which may be related to this event but are not themselves of interest. Lau, Cole, and Gange (2009) describe the various approaches to analyzing survival data
in the presence of competing risks, and discuss the advantages and disadvantages of each. They state that the choice of approach should be driven by the scientific question. In many situations, the quantity of interest is the cause-specific hazard, which expresses the probability that an individual experiences the event of interest within a short interval after the present time given that the individual is still under observation. For example, many of the analyses carried out in the Nurses' Health Study, such as Hart et al. (2013), Palacios  et al. (2014), and Hart et al.(2018),
have focused on the cause-specific hazard for a specific health event.
In such situations, the event of interest can be analyzed without any modeling of the competing events.
This is the situation that we address in the present paper.
There are other situations where the various competing events are on an equal footing
and there is interest in examining the cumulative incidence rates of the various events, but this setting is beyond the scope of the present paper.
For simplicity, we carry out our simulations in the setting of a single event, but our examples involve
competing events.

\section{Methods}

\subsection{Setting, Notation, and Background}

We assume we have data on $n$ individuals, with the data on these individuals
independent and identically distributed.
Time is measured relative to a specified zero point, such as birth or time of disease diagnosis.
We allow for left truncation, such as occurs in studies where the time metameter is age (time zero = birth) and
people enter the study at different ages, as in NHS.
We denote by $\tilde{T}_i$ the time at which individual $i$ first comes under observation, by $T_i$ the time at the end of follow-up on individual $i$, and by $\delta_i$ a 0--1 variable indicating whether the event of interest was (1) or was not (0) observed on individual $i$.
The maximum possible observation time is denoted by $t^{*}$.
For a given individual $i$, $(\bZ_i(t),X_i(t))$ denotes the true covariate vector. Again,
$X(t)$ is the covariate of primary interest, which is subject to a possible threshold effect, while $\bZ(t)$ is a vector of additional covariates. The main covariate $X(t)$ is measured with error, while the additional covariates $\bZ(t)$ are assumed error-free.

We define $Y_i(t)=I(\tilde{T}_i < t \leq T_i)$, which is a 0--1 variable indicating whether individual $i$ is (1) or is not (0) under observation at time $t$, and $N_i(t)=\delta_i I(T_i \leq t)$, which is a 0--1 variable  indicating whether individual $i$ has (1) or has not (0) experienced the event of interest before or at time $t$. If individual $i$ experiences the event of interest at time $t$, follow-up is taken to end at time $t$, so that $Y_i(s) = 0$ for all $s>t$. In addition, follow-up can end due to censoring or competing events. We work with a model for the cause-specific hazard (Putter, Fiocco, and Geskus, 2006, Section 3.3.1). Let $\mathcal{F}_t = \sigma(N_i(s), Y_i(s), X_i(s), \bZ_i(s), s \in [0,t], i = 1, \dots, n)$ denote the observed study history up to time $t$. We assume that $X_i$ and $\bZ_i$ are predictable with respect to $\mathcal{F}_t$. Our model says that
$$
\lim_{h \downarrow 0} h^{-1} \Pr(N_i(t+h)-N_i(t-)=1|\mathcal{F}_{t-}) = Y_i(t)\lambda(t|X_i(t),\bZ_i(t))
$$
with the cause-specific hazard $\lambda(t|x(t),\bz(t))$ given by (\ref{CP}).

The observed version of $X(t)$ is denoted by $W(t)$. We assume a classical normal additive measurement error model where
$W_i(t) = X_i(t) + U_i(t)$, where the conditional distribution of $X_i(t)$ given
$\bZ_i(t)=\bz$ is $N(\mu_x(\bz),\sigma_x^2)$ and the $U_i(t)$'s are i.i.d.\
$N(0,\sigma_u^2)$, independent of $\mathcal{F}_{t}$.
We assume that $\mu_x(\bz)$ is of the form $\mu_x(\bz) = \alpha_0 + \alpha_1 \bz$.
We write $\lambda_i(t) = \lambda(t|X_i(t),\bZ_i(t))$ and
$d\tilde{F}\left(t\right)=E\left[Y_{i} \left(t\right)\lambda _{i} \left(t\right)\right]dt$ .
We work under the main study/external reliability design.
We write $\bth = (\bgam^T, \beta, \omega)^T$ and $\sigma_w^2 = \Var(W(t)|\bZ(t)) = \sigma_x^2 + \sigma_u^2$.

In studies with time-dependent covariates, typically the covariates are not measured on a continuous basis, but rather at specific time
points $t_1^*, \ldots, t_K^*$. One popular approach for analyzing data of this type is through a joint survival/longitudinal model
as described by Rizopoulos (2012). However, as discussed in Liao et al.\ (2011), in many cases the joint modeling approach is
hard to apply, in which case a reasonable alternative is to carry forward the most recent measurement of the covariate. Although this carry-forward approach itself induces measurement error,
in epidemiological studies it is commonly the case that
the point exposures are subject to considerable measurement error, while the error induced by carrying
forward the most recent cumulative exposure value is less serious. The methods described in this paper are aimed at this context.
The vector of true covariate values is ${\bf X} = (X(t_1), \ldots, X(t_K))$ and the corresponding vector of measured
covariate values is ${\bf W} = (W(t_1), \ldots, W(t_K))$. When we write $X(t)$ or $W(t)$, we mean the value of $X$ or $W$ at the largest $t_j$ value less than $t$.

The classical normal additive measurement error model is assumed with
$W_i(t_j) = X_i(t_j) + U_i(t_j)$, where the conditional distribution of $X_i(t_i)$ given
$\bZ_i(t_k)= \bz$ is $N(\mu_x( \bz),\sigma_x^2)$ and the $U_i(t_j)$'s are i.i.d.\
across $j$ $N(0,\sigma_u^2)$, independent of the $X(t)$'s and the $\bZ(t)$'s.
The mean $\mu_x(\bz)$ is assumed to be of the form $\mu_x(\bz) = \alpha_0 + \alpha_1 \bz$.
We write $\lambda_i(t) = \lambda(t|X_i(t),\bZ_i(t))$ and
$d\tilde{F}(t)=E[Y_{i} (t)\lambda _{i} (t)]dt$ .
We work under the main study/external reliability design.
Let us write $\bth = (\bgam^T, \beta, \omega)^T$ and $\sigma_w^2 = \Var(W(t)|\bZ(t)) = \sigma_x^2 + \sigma_u^2$.

The measurement error correction methods we consider involve the nuisance parameters $\alpha_0, \alpha_1, \sigma _{x}^{2}$, and $\sigma _{u}^{2}$. In practice, these parameters are unknown and must be estimated from suitable data. In particular, estimation of $\sigma_u^2$ requires replicate measurements of $W(t)$. In our simulation study, we consider first the simple setting without additional covariates
$\bZ(t)$, and then we extend it, for part of the considered scenarios, to the setting with additional covariates $\bZ(t)$. We assume that $\mu_x=\alpha_0$ and $\sigma_x^2$ and $\sigma_u^2$ are estimated from an external reliability study. The estimates are
computed by one-way random effects ANOVA, and we assume that the resulting conditional expectation and the conditional variance are transportable to the main study.

\subsection{The methods}

Let us write the relative risk function as $r(x,\bz,\bth) = \exp(\bgam^T \bz + \beta x + \omega (x - \tau)_+)$. If $X(t)$ was known, we would work with the standard Cox log partial likelihood, given by
\begin{align*}
l_{p} (\bth ) & = \sum _{i=1}^{n} \delta _{i}  [ \log r(X_i(t),\bZ_i(t),\bth)
-\log \sum _{j=1}^{n}Y_{j}(T_{i}) r(X_i(t),\bZ_i(t),\bth)] \\
& = \sum _{i=1}^{n} \delta _{i} [ (\bgam^T \bZ_i(t) + \beta X_i(t) + \omega (X_i(t) - \tau)_+)
- \log \sum _{j=1}^{n}Y_{j}(T_{i}) r(X_i(t),\bZ_i(t),\bth) ]
\end{align*}
Many methods for Cox regression analysis with covariate error involve replacing $r(x,\bz,\bth)$ with some substitute. The specific methods we examine are listed below. \\


\noindent\textbf{A. Naive Method}: The naive estimator is obtained by maximizing the log partial likelihood function with $W(t)$ plugged in for $X(t)$, ignoring the measurement error. \\

\noindent \textbf{B. Regression Calibration (RC) Methods} \\

\noindent \textbf{B1.  Simple RC Method (RC1)}: $X_{i}(t)$ is replaced throughout by
$\mu(W_i(t),\bZ_i(t))=E(X_{i}(t) |W_{i}(t), \bZ_i(t))$. We have $E( X_{i}(t)
|W_{i}(t), \bZ_i(t)) = (1-\lambda )\mu_{x}(\bZ_i(t)) + \lambda W_{i}(t)$ , where $\lambda = \Cov(X(t),W(t))/\Var(W(t))$
$ = \sigma _{x}^2 /\sigma _{w}^{2}$, the attenuation factor. \\

\noindent \textbf{B2.  Improved RC Method (RC2)}: $X_{i}(t)$ is replaced with $E( X_{i}(t) |W_{i}(t),
\bZ_i(t) )$ and $(X_{i}(t) -\tau )_{+} $ is replaced with $E( (X_{i}(t) -\tau)_{+}
|W_{i}(t), \bZ_i(t) )$. Define $\eta^2={\rm Var}(X_{i}(t)|W_{i}(t),\bZ_i(t))=\sigma_x^2(1-\lambda)$.
Then, by properties of the truncated normal distribution (see Johnson {\it{et al}}., 2014, Section 10.1), we have

\begin{align*}
E\left[\left. \left(X_{i}(t) -\tau \right)_{+} \right|W_{i}, \bZ_i(t) \right]
=\left(1-\Phi \left(\frac{-\mu \left(W_{i}(t), \bZ_i(t) \right)+\tau }{\eta } \right)\right)
\left(\mu \left(W_{i}(t), \bZ_i(t) \right)-\tau \right)\\
{+\; \phi \left(\frac{-\mu \left(W_{i}(t), \bZ_i(t) \right)+\tau }{\eta } \right)\eta},
\end{align*}
where $\phi$ and $\Phi$ denote the normal probability density function and the normal cumulative distribution function, respectively.
\\
\\
\noindent \textbf{C. Induced Relative Risk (RR) Methods} \\

\noindent \textbf{C1. Original RR Method (RR1)}: This is an extension of the method proposed by Prentice (1982) to threshold models.
Recall that we denote the hazard with respect to the true covariate $X(t)$ and the additional covariates $\bZ(t)$
by $\lambda (t|x,\bz)$. Denote the hazard with respect to the observed covariate $W(t)$ and $\bZ(t)$ by $\lambda^{*}(t|w,\bz)$.
As discussed in Prentice (1982),  $\lambda^{*}(t|w,\bz)=E[\lambda (t|X(t),\bz)|W(t)=w, \bZ(t)=\bz, \; T \geq t]$.
Under the assumption that the event is rare,
the conditioning on the event $\left\{T\ge t\right\}$ can be omitted,
leading to $\lambda^{*}(t|w,\bz)\approx E[\lambda (t|X(t))|W(t)=w,\bZ(t)=\bz]$. In the case of our model (\ref{CP}),
$\lambda (t|x,\bz)=\lambda _{0} (t)\exp (\beta x+\omega (x-\tau )_{+} + \bgam^T \bz)$, so that
$\lambda ^{*} (t|w,\bz)=\lambda _{0} (t)E[\exp (\beta X(t)+\omega (X(t)-\tau )_{+}) |W(t)=w, \bZ(t)=\bz]\exp(\bgam^T \bz)$.
Under our measurement error model, we have (proof in S.1 in the Supplemental Materials)
\begin{align*}
E[e^{\beta X(t)+\omega (X(t)-\tau )_{+} } |W(t)=w, \bZ(t)=\bz]
= e^{0.5\eta^{2} \beta^{2} +\beta \mu (w,\bz)} \times
 \Phi \left(\frac{\tau -\eta ^{2} \beta -\mu \left(w,\bz\right)}{\eta } \right) \\
 {+ e^{-\omega \tau +0.5\eta ^{2} \left(\beta +\omega \right)^{2} +\left(\beta +\omega \right)
\mu \left(w,\bz\right)} \times  \Phi \left(\frac{-\tau +\eta ^{2} \left(\beta +\omega \right)+\mu \left(w,\bz\right)}{\eta }
\right)},
\end{align*}
where $\Phi$ denotes the normal cumulative distribution function.
\\
\\
\noindent \textbf{C2. Modified RR Method (RR2)}: The RR1 method should theoretically yields estimates that are virtually
unbiased in the rare event scenario. However, in our simulations of RR1 under the rare event scenario, significant remaining bias was observed, as can be seen in the tables. We therefore
examined a bootstrap bias-correction procedure involving the following steps: \\

\noindent (i) Compute the RR1 estimate $\hat{\bth}$ based on the original data \\

\noindent (ii) Take $B$ bootstrap samples from the data using the weighted bootstrap algorithm as in Kosorok and Song (2007). We use the weighted bootstrap rather than the ordinary bootstrap in order to avoid the complexities that arise in survival analysis when there are ties.
\noindent The weighted bootstrap algorithm involves assigning a random weight to each observation, with the weights generated as follows: (a)
generate $n$ positive random variables  $\kappa _{1}^{} ,\kappa _{2}^{} ,...,\kappa _{n}^{} $ from the $\exp (1)$
distribution; (b) truncate these weights at 5, that is take $\kappa _{i}^{*} =\min (\kappa _{i} ,5)$; (c) divide each
weight with the sample average $\bar{\kappa }^{*}= n^{-1} \sum_{i=1}^n \kappa _{i}^{*}$ to obtain the standardized weight
$\kappa _{i}^{0}= \kappa _{i}^{*}/\bar{\kappa }^{*}$ (the standardized weights sum up to $n$).
Then, for a given function $f$, we replace expressions of
the form $\frac{1}{n} \sum _{i=1}^{n}f(X_{i},W_{i},\bZ_{i},T_{i},\delta _{i} ) $ with the weighted analogue
$\frac{1}{n} \sum _{i=1}^{n}\kappa _{i}^{0} f(X_{i},W_{i},\bZ_{i},T_{i},\delta _{i} ) $.
For each bootstrap sample $j$, $j=1,2,...,B$, compute the RR1 estimate $\hat{\bth}_j$
of $\bth$.
Let $\tilde{\bth}$ denote the mean of the estimates $\hat{\bth}_j$ over the $B$ bootstrap samples. \\

\noindent (iii) Estimate the bias of the RR1 estimate via ${\bf b}=\tilde{\bth}-\hat{\bth}$, and then compute the bias-corrected estimate
of $\bth$ as $\hat{\bth }_{corr} = \hat{\bth}-{\bf b} = \hat{\bth}- (\tilde{\bth} -\hat{\bth} ) = 2\hat{\bth}-\tilde{\bth}$. \\

\noindent \textbf{D. MPPLE Method}: The MPPLE method is based on the work of of Zucker (2005). The method applies to the case of time-independent covariates. It requires an assumption that any censoring of the event of interest
is independent of all other random variables in the model. Like Prentice's RR method, it works with the induced
hazard model $\lambda^{*}(t|w,\bz)=E[\lambda (t|X)|W=w, \bZ(t)=\bz, T \ge t]$, but it differs from Prentice's method
in that it avoids the rare disease assumption.
Let $f(x|w,\bz)$ denote the conditional density of $x$ given $w$ and $\bz$,
i.e., the normal density with mean $\mu(w,\bz)$ and variance $\eta^2$.
The induced hazard is then expressed  as $\lambda^*(t|w,\bz) = \lambda_0(t) \exp(\phi(\bth,w,\bz,\Lambda_0(t)))$ with
\begin{align*}
&\phi(\bth,w,\bz,c) \\
& \hspace*{24pt} =  \log \int \exp (-c\times r(x,\bz,\bth)) r(x,z,\bth) f(x|w,\bz) dx \\
& \hspace*{36pt} - \, \log \int \exp (-c\times r(x,\bz,\bth)) f(x|w,\bz) dx.
\end{align*}
The MPPLE estimate of $\bth$ is obtained by substituting this induced hazard into the
Cox partial likelihood and maximizing over $\bth$.  The various integrals required by this estimator were evaluated by 20-point Gauss-Hermite quadrature.
The induced hazard depends on the unknown cumulative hazard rate $\Lambda_0(t)$. Zucker (2005) proposed
a non-iterative forward recursion for estimating $\Lambda_0(t)$ for a given value of the regression
coefficients, and this estimate is then plugged into the partial likelihood expression. \\

\noindent \textbf{E. SIMEX Method}: SIMEX is a simulation-based method, obtained by adding additional measurement error to the data in a resampling-like stage, modeling the trend of the measurement error-induced bias as a function of the variance $\lambda$ of the added measurement error, and extrapolating this trend back to the case of no measurement error. The method is described in detail in Carroll {\it{et al}}.\ (2006, Chapter 5). In preliminary work, we examined three extrapolation methods :  rational linear extrapolation, simple quadratic extrapolation, and the third-degree polynomial extrapolant of the form  $G_{P3} (\lambda ,\Gamma )=\gamma _{1} +\gamma _{2} \lambda +\gamma _{3} \lambda ^{2} +\gamma_{4} \lambda ^{3}$. Examining plots with the fitted
extrapolation function superimposed on a scatterplot of the mean value of the estimate as a function of $\lambda$ (based on 1000 replications),
we found that the third-degree polynomial provided the best fit, and we used this extrapolation method in the implementation of the SIMEX estimator in our numerical studies. \\

\noindent \textbf{Remarks}:
The naive, RC1, RR1, the MPPLE and the SIMEX are existing methods in the literature, whereas the RC2 and the RR2 are new methods that we develop in this paper.
Generally speaking, the worst method is the naive method which ignores the measurement error, and therefore we expect it to perform poorly.
The RC1, RC2, and RR methods all rely on a rare disease approximation. Under the rare disease assumption, the representation of the induced relative
risk $\lambda^{*}(t|w,\bz)$ used in the RR methods is essentially exact, whereas the representation used in the RC1 and RC2 methods is approximate. We thus
expect RR1 and RR2 to perform better than RC1 and RC2. The MPPLE method avoids the rare disease approximation, and is thus expected to be the most accurate method.

The Cox model can be applied to settings where the covariates are time-independent and settings where the covariates are time-dependent.

In the time-dependent setting, if the rare disease approximation is reasonable, the RC1, RC2, RR1, and RR2 can be applied.
In discussions of the use of the Cox model with time-dependent covariates, a distinction is traditionally made between
exogenous time-dependent covariates, such as environmental conditions, and endogenous covariates, such as physiological or behavioral
characteristics of the individual under study (Kalbfleisch and Prentice, 2002, Section 6.3). The availability of a measurement of
an endogenous covariate at a given time entails that the individual was still alive at that time. Under the rare disease assumption
underlying the RC and RR methods, conditional expectations given survival up
to a given point are approximated by expectations that do not conditional on survival up to a given point. In this context, the distinction between
exogenous time-dependent covariates and endogenous time-dependent does not come into play. The MPPLE method avoids the rare-disease assumption, but is
applicable only to the case of time-independent covariates. To handle endogenous time-dependent covariates without a rare disease approximation,
one possible approach is the joint modeling approach as described by Rizopoulos (2012), and a few papers have considered joint models with changepoints
(Garre et al., 2007; Dantan et al., 2011; Jacqmin-Gadda, Commenges, and Dartigues, 2006; Ghosh, Ghosh, and Tiwari, 2011).
Another possible approach is to use a risk-set regression calibration strategy along the lines of Xie, Wang, and Prentice (2001);
developing this approach is a potential topic for future research.

\section{Asymptotic properties of the Naive, RC1, RC2, RR1, and MPPLE estimators}
\subsection{Naive, RC1 and RC2}
\noindent The naive, RC1, and RC2 estimators are all of a common form. Each involves replacing $X_{i}(t) $ in the term $\beta X_{i}(t) $ by a surrogate $g_{1} (W_{i}(t) ,\bZ_{i}(t) )$ and $(X_{i}(t) -\tau )_{+} $ in the term $\omega (X_{i}(t) -\tau )_{+} $ by a surrogate $g_{2} (W_{i}(t) ,\bZ_{i}(t) )$. The naive method takes $g_{1} (w,\bz)=w$ and $g_{2} (w,\bz)=(w-\tau )_{+} $, the RC1 method takes $g_{1} (w,\bz)=\mu (w,\bz)$ and $g_{2} (w,\bz)=(\mu (w,\bz)-\tau )_{+} $, and the RC2 method takes $g_{1} (w,\bz)=\mu (w,\bz)$ and $g_{2} (w,\bz)=E[ (X(t)-\tau )_{+} |W(t)=w,\bZ(t)=\bz]$. Let $\bg$ denote the function pair $(g_{1} ,g_{2} )$ and let $\bV_{i} (g(t))$ denote a vector of length $p+2$ in which the first $p$ components are the elements of $\bZ_{i}(t) $, the $(p+1)$-th component is $g_{1} (W_{i}(t) ,\bZ_{i}(t) )$, and the $(p+2)$-th component is $g_{2} (W_{i}(t) ,\bZ_{i}(t) )$.
\\
Define
\begin{align*}
& S^{(0)} (t,\bth ,g)=\frac{1}{n} \sum _{i=1}^{n}Y_{i} (t)\exp (\bth ^{T} \bV_{i} (g(t))) \\
& S^{(1)}t(t,\bth ,g)= \frac{1}{n} \sum _{i=1}^{n}Y_{i} (t)  \bV_{i} (g(t))\exp (\bth ^{T} \bV_{i}  (g(t)))
\end{align*}

\noindent The naive, RC1, and RC2 estimators are then obtained as the solution to $U^{(g)}(t^{*},\bth)=0$, where

$$
U^{(g)} (t^{*},\bth)=\sum _{i=1}^{n}\int _{0}^{t^{*} } \left[\bV_{i} (g(t))-\frac{S^{(1)} (t,\bth ,g)}{S^{(0)} (t,\bth ,g)} \right]  dN_{i} \left(t\right).
$$

\noindent Let $\hat{\bth}$ denote the resulting estimator. Additional notation is presented in Appendix A.1.
Using the methods of Andersen and Gill (1982), Struthers and Kalbfleisch (1986), Lin and Wei (1989), and Self and Prentice (1982), we obtain the following proposition:

\textbf{Proposition}: Under suitable technical conditions similar to those in Andersen and Gill (1982),
we have the following: \\
(i) \textit{Convergence to a Limit}:  Define $\bar{\bth}^{(g)}$ to be the solution of the equation
$q^{(g)}(t^{*},\bth)=0$. Then $\hat{\bth}\mathop{\to }\limits^{p} \bar{\bth}^{(g)}$. \\
(ii) \textit{Asymptotic Normality}: $n^{1/2} (\hat{\bth }-\bth ^{*} )$ converges in distribution
to a mean-zero multivariate normal distribution whose covariance matrix can be consistently estimated by
$$
\Omega (t,\, \hat{\bth },\, g)=
[n^{-1} I(t,\, \hat{\bth},\, g)]^{-1} \hat{A}(t,\, \hat{\bth },\, g) [n^{-1}I(t,\, \hat{\bth },\, g)]^{-1}
$$
%
%
where $\hat{A}(t,\bth ,g)= {n}^{-1} \sum _{i=1}^{n}H_{i} (t,\bth,g)^{\otimes 2}$  with
\begin{align*}
H_{i} (t^{*} ,\, \bth ,\, g) & = \int _{0}^{t^{*} }\left(\bV_{i} (g(t))-\frac{S^{(1)} (t,\, \bth ,\, g)}{S^{(0)} (t,\, \bth ,\, g)} \right) dN_{i} (t) \\
& \hspace*{24pt} - \,\int _{0}^{t^{*} }\frac{Y_{i}(t)\exp (\bth ^{T} \bV_{i} (g(t)))}{S^{(0)} (t,\, \bth ,\, g)} \left(\bV_{i} (g(t))-\frac{S^{(1)} (t,\, \bth ,\, g)}{S^{(0)} (t,\, \bth ,\, g)} \right) d\tilde{F}(t).
\end{align*}



For the RC methods, which involve the nuisance parameters $\bph=(\mu _{x} ,\sigma _{x}^{2} ,\sigma _{u}^{2} )$, these parameters are unknown and need to be estimated with the estimation error accounted for in the covariance matrix of the estimates. Suppose the estimation of $\bph$ is based on a sample of $m$ independent individuals, and denote by $\hat{\bph }$ the estimator of $\bph$. Then, the vector $\hat{\bph }$ is obtained as a solution of the estimation equations of the form $\sum _{i=1}^{m}\Psi _{i} (\bph ) =0$.
Denote by $\bph ^{*} $ the solution of  $\sum _{i=1}^{m}E(\Psi _{i} (\bph )) =0$.

Define
$$
\Phi = \Cov\left(\frac{1}{m} \sum _{i=1}^{m}{Y_{i} (t)}H_{i} (t,\bth ,\bph ,g),\sqrt{m} \, \dot{\Psi }_{\bph} (\bph ^{*} ) \right)
$$
where $\dot{\Psi }_{\bph }$ denote the first derivative of $\Psi$ respect to $\bph$. Also, denote by ${\ddot{\Psi}_{\bph }} $ the second derivative of $\Psi$ respect to $\bph$, and by $\dot{U}^{(g)}_{\bph } $ the first derivative of $U^{(g)}$ respect to $\bph $. $\Phi $ can be estimated asymptotically by $$\hat{\Phi }=\frac{1}{m} \sum _{i=1}^{m}{Y_{i}(t)}H_{i} (t,\hat{\bth},\hat{\bph},g) \frac{\partial }{\partial \bph } \Psi _{i} (\hat{\bph })^{T}.$$ As in Zucker and Spiegelman (2008), the corrected covariance matrix is
$$
\Omega _{corr} (t,\hat{\bth},\hat{\bph },g)=n^{-1} I(t,\, \hat{\bth },\, g)^{-1} A_{corr} (t,\, \hat{\bth},\hat{\bph },\, g)n^{-1} I(t,\, \hat{\bth},\, g)^{-1}
$$
where
$\hat{A}_{corr}(t,\hat{\bth},\hat{\bph },g)$ is equal to the sum of $\hat{A}(t,\hat{\bth},\hat{\bph },g)$ and the term
$$
\dot{U}^{(g)}_{\bph } (t, \hat{\bth},\hat{\bph })\Cov(\hat{\bph })\dot{U}^{(g)}_{\bph } (t,\hat{\bth},\hat{\bph })^{T} -\hat{\Phi }\ddot{\Psi }_{\bph } (\hat{\bph })^{-1} \dot{U}^{(g)}_{\bph } (t,\hat{\bth},\hat{\bph })^{T}.
$$

\subsection{RR1}
\noindent RR1 involves replacing $\exp (\beta X_{i} (t)+\omega (X_{i} (t)-\tau )_{+} )$ with
$E[ \exp (\beta X_{i} (t)+\omega (X_{i} (t)-\tau )_{+} )|$
\linebreak[4]
$W_{i}(t),\,\bZ_{i} (t)]$.
Denote $E( \exp (\beta X_{i} (t)+\omega (X_{i} (t)-\tau )_{+} )|W_{i} (t),\, \bZ_{i} (t))$ by $r(\bth ,\bV_{i} (t))$, where $\bV(t)$ denotes a vector of  length $p+1$ in which the first $p$ components are the elements of $\bZ_{i} (t)$, and the $(p+1)$-th component is $W_{i} (t)$. Then the partial log likelihood function is:
\begin{align*}
{l_{p}^{(RR1)} (t^{*} ,\bth )=\sum _{i=1}^{n}\int _{0}^{t^{*} }\log [Y_{i} (t)r(\bV_{i} (t),\bth  )\exp (\bgam ^{T} \bZ_{i} (t))]dN_{i} (t)  } \\ {-\sum _{i=1}^{n}\int _{0}^{t^{*} }\log [\sum _{j=1}^{n}Y_{j} (t)r(\bV_{j} (t),\bth  )\exp (\bgam ^{T} \bZ_{j} (t)) ] dN_{i}(t) } \\ {\, \, =\sum _{i=1}^{n}\int _{0}^{t^{*} }\log [r(\bV_{i} (t),\bth  )\exp (\bgam ^{T} \bZ_{i} (t))]dN_{i} (t)  } \\ {-\int _{0}^{t^{*} }\log [\sum _{j=1}^{n}Y_{j} (t)r(\bV_{j}(t),\bth  )\exp (\bgam ^{T} \bZ_{j}(t)) ] d\bar{N}(t)\, .}
\end{align*}
The RR1 estimator is the maximizer of $l_{p}^{RR1}t^{*},\bth)$, i.e., the solution
to the equation $U^{(RR1)}(t^{*},\bth) = \mathbf{0}$, where $U^{(RR1)}$ is the vector of derivatives
of $l_{p}^{(RR1)}$.
Let $r^{(1)} (\bth ,\bV_{i} (t))$ and $r^{(2)} (\bth ,\bV_{i} (t))$ denote the first and second derivative of $r(\bth  ,\bV_{i} (t))$ respect to $\bth$, respectively. Additional notation is presented in Appendix A.2.

\textbf{Proposition}: Under technical conditions similar to those in Andersen and Gill (1982),
we have the following: \\
\noindent
(i) \textit{Convergence to a Limit}: Define $\bar{\bth}^{(RR1)}$ to be the solution of the equation
$q^{(RR1)}\left(t^{*},\bth \right)=0$. Then $\hat{\bth }\mathop{\to }\limits^{p} \bar{\bth}^{(RR1)}$. \\
\noindent
(ii) \textit{Asymptotic Normality}: $n^{1/2} (\hat{\bth }-\bth ^{*})$ converges in distribution
to a mean-zero multivariate normal distribution whose covariance matrix can be consistently estimated by
\begin{align*}
\Omega(t^{*},\hat{\bth  })=n^{-1} I(t^{*},\hat{\bth })^{-1} \hat{A}(t,\hat{\bth })
n^{-1} I(t^{*},\hat{\bth })^{-1},
\end{align*}
where $\hat{A}(t,\bth)= n^{-1} \sum _{i=1}^{n}H_{i} (t,\bth)^{\otimes 2}$ with
$H_{i} (t^{*},\bth )$ is a vector of length $p+2$ in which the first $p$ components are
\begin{align*}
 H_{i} (t^{*},\bth)=\int _{0}^{t^{*} } \left(\bZ_{i} (t)-\frac{s^{(5)} (t,\bth  )}{s^{(0)} (t,\bth  )} \right) dN_{i} (t)-\int _{0}^{t^{*} }\frac{Y_{i} (t)\exp (\bgam ^{T} \bZ_{i} (t))}{s^{(0)} (t,\bth  )} \left(\bZ_{i} (t)-\frac{s^{(5)}(t,\bth  )}{s^{(0)} (t,\bth  )} \right)d\tilde{F}(t),\\
 \end{align*}

and the $(p+1)$-th and $(p+2)$-th components are
\begin{align*}
 {H_{i} (t^{*},\bth)=\int _{0}^{t^{*} }\left(\frac{r^{(1)}(\bth ,\bV_{i}(t) )}{r(\bth ,\bV_{i}(t))} -\frac{S^{(1)}(t,\bth)}{S^{(0)} (t,\bth )} \right) dN_{i} (t)} \\ {-\frac{1}{S^{(0)} (t,\bth)} \int _{0}^{t^{*} }\left(Y_{i} (t)r^{(1)} (\bth ,\bV_{i}(t) )-\frac{S^{(1)} (t,\bth )}{S^{(0)} (t,\bth)} Y_{i} (t)r(\bth ,\bV_{i}(t) )\right) d\tilde{F}(t) }.
 \end{align*}

\subsection{MPPLE}
\noindent Denote by $\hat{\Lambda }_{0} $ the estimator of $\Lambda _{0} $ , and let $\bg$ denote the function pair $(g_1,g_2)$ as in the naive method. The MPPLE estimator is obtained as the solution to $U_{r}^{(MP)} (t,\bth ,\hat{\Lambda }_{0} )=0$ , where
\[U_{r}^{(MP)} (t,\bth ,g,\hat{\Lambda }_{0} )=\frac{1}{n} \sum _{i=1}^{n} \delta _{i} \left(\xi_{r} (\bth ,\bV_{i}(g) ,T_{i})-\frac{\sum _{j=1}^{n}Y_{j} (T_{i}) \xi _{r} (\bth ,\bV_{j}(g) ,T_{i} )e^{\phi (\bth ,\bV_{j}(g) ,\Lambda _{0} (T_{i}))} }{\sum _{j=1}^{n}Y_{j} (T_{i} ) e^{\phi (\bth ,\bV_{j}(g) ,\Lambda _{0} (T_{i} ))} } \right) \]
where $\xi _{r} (\bth ,\bV_{i}(g) ,T_{i})=\frac{\partial }{\partial \bth _{r} } \phi (\bth ,v,\hat{\Lambda }_{0} (t,\bth ))$, $r=1\, ,\, 2\, ,...\, ,p+2$.

\noindent Define
\begin{align*}
{\Omega (t,\bth ,g,\Lambda )=\frac{1}{n} \sum _{i=1}^{n}\delta _{i} \left(\frac{\sum _{j=1}^{n}Y_{j} (T_{i} )\xi (\bth ,\bV_{j} (g),T_{i} ,\Lambda )^{\otimes 2} e^{\phi(\bth ,\bV_{j} (g),\Lambda _{0} (T_{i} ))}  }{\sum _{j=1}^{n}Y_{j} (T_{i} )e^{\phi (\bth ,\bV_{j} t(g),\Lambda _{0} (T_{i} ))}  } \right) } \\ {\, \, \, \, \, \, \, \, \, \, \, \, \, \, \, \, \, \, \, \, \, \, \, \, \, \, \, -\frac{1}{n} \sum _{i=1}^{n}\delta _{i} \left(\frac{\sum _{j=1}^{n}Y_{j} (T_{i} )\xi (\bth ,\bV_{j} (g),T_{i} ,\Lambda )^{\otimes 2} e^{\phi (\bth ,\bV_{j} (g),\Lambda _{0} (T_{i}))}  }{\sum _{j=1}^{n}Y_{j} (T_{i} )e^{\phi (\bth ,\bV_{j} (g),\Lambda _{0} (T_{i} ))}  } \right)^{\otimes 2} }
\end{align*}
with $\ba^{\otimes 2}=\ba \ba^{T}$.
Using the arguments of Zucker (2005), we obtain the following proposition: \\

\noindent \textbf{Proposition}\rm: Denote by $\bth _{0}$ the true value of $\bth$. Under suitable technical conditions similar to those in Andersen and Gill (1982), we have the following: \\
(i) \textit{Convergence to a Limit}\rm: $\hat{\bth }\mathop{\to }\limits^{p} \bth _{0}$. \\
(ii) \textit{Asymptotic Normality}\rm: $n^{{1 \mathord{/{\vphantom{1 2}}\kern-\nulldelimiterspace} 2} } (\hat{\bth }-\bth _{0})$ converges in distribution to a mean-zero multivariate normal distribution whose covariance matrix can be consistently estimated by $\hat{\Omega}^{-1} +\hat{\Omega}^{-1} \hat{H}\hat{\Omega}^{-1} $, where
$\hat{\Omega}=\Omega(t,\hat{\bth },g,\hat{\Lambda }_{0})$ and $\hat{H}$ is given in Zucker (2005, eq.\ (A.15)).

\noindent
\subsection{Asymptotic Bias }
As noted earlier, the asymptotic limits of the naive, RC1 and RC2 estimators are obtained as the solution $\bar{\bth}^{(g)}$ of $q^{(g)}(t^{*},\bth)=0$, where $q^{(g)}(t^{*},\bth)$ is the limit of $U^{(g)}(t^{*},\bth)$ as $n$ tends to infinity. Similarly, the asymptotic limit of the RR1 estimator is the solution $\bar{\bth}^{(g)}$ of $q^{(RR1)}(t^{*},\bth)=0$, where $q^{(RR1)} (t^{*},\bth)$ is the limit of $U^{(RR1)}(t^{*},\bth)$ as $n$ tends to infinity. The asymptotic bias is then $\bar{\bth}^{(g)}
-\bth_{0}$ for the relevant $\bar{\bth}^{(g)}$. Hughes (1993) previously presented similar asymptotic bias calculations for the naive estimator in the Cox model without a threshold effect.

\indent We computed the limiting values numerically for the naive, RC1, RC2, and RR1 estimators under the rare disease scenario where $n=50,000$ and cumulative incidence = 0.01. The Newton-Raphson method was used to find the point where the score function equals zero. Then, we compared the results with those obtained in the simulation studies, for the case when the measurement error parameters are known. Both the theoretical and empirical bias are based on a model with one covariate and true parameters of  $\beta =\log (1.5)=0.405$ and $\omega =\log (2)=0.693$. The starting values for the Newton Raphson calculation in all methods were (0,0). Table S.2 in the Supplemental Materials presents the results, where the asymptotic bias is labeled as {\textit{theoretical}}, and the simulation results are labeled as {\textit{empirical}}. The variable {\textit{pct}} denotes the percentage
of instances over the 1000 replications in which the estimation procedure converged. In addition, DELTA denotes the difference between the theoretical result and the simulation result. Generally, the RR1 method had the least bias, typically negligible, except at the lower extreme values of $\tau$, where the relative bias was $\pm0.05$ for $\rho_{xw}=0.8$ and becomes larger as $\rho_{xw}$ decreases. For the naive method, when $n=50,000$ and cumulative incidence = 0.01, the theoretical and simulation results agreed closely, as expected. For the RC1 and RC2 methods, the agreement between the theoretical and simulation results was better for $\tau<0$ with $n=200,000$ (keeping cumulative incidence of 0.01), where for $\tau>0$, the agreement was similar with $n=50,000$ and $n=200,000$. The results were close, except at the lower extreme values of $\tau$ in which case this difference was large. Regarding the RR1 method, the results were close, except at the lower extreme values of $\tau$ in which case this difference was large.
\section{Simulation Study}
In this section, we compare the finite sample properties of the various methods under several scenarios.
As mentioned previously, for the sake of simplicity we carry out the simulations in the setting of a single event
subject to right censoring. Tables 1-5 and Tables S.4-S.8 and S.10-S.13 in the Supplemental Materials present the results. As a benchmark, we also present the estimates under the case of no measurement error. Source code to reproduce the results is available as Supporting Information on the journal's web page (http://onlinelibrary.wiley.com/doi/xxx/suppinfo).



\subsection{Simulation Design}

We assumed a single time-independent covariate $W$, and fixed administrative censoring at time $t^{*} =10$. We set $\beta=\log(1.5)$ and $\omega=\log(2)$, representing
a scenario where the effect of the covariate on the hazard is initially moderate
and later becomes more pronounced.
The covariate $X$ was generated as standard normal and the event time was generated as exponential with parameter $\lambda =\lambda_{0} \exp (\beta X+\omega (X-\tau )_{+} )$.
\noindent The observed surrogate covariate value $W$ was generated using the classical measurement error model with $U\sim N(0,\sigma _{u}^{2})$. The values of $\sigma_{u}^{2}$ were set to yield $\rho_{xw}= 0.8, 0.6$, or 0.4 (with the resulting $\sigma_{u}^{2}$ values being 0.56, 1.77 and 5.25) in order to cover a range of $\rho_{xw}$ values commonly seen in real data. The changepoint $\tau$ was fixed at one of 5 points at various percentiles of the distribution of $X$:
$\Phi ^{-1} (0.1),\Phi ^{-1} (0.25),\Phi^{-1} (0.5),\Phi^{-1} (0.75),\Phi ^{-1} (0.9)$.
We examined the performance of the estimators under the common disease scenario where $n=3,000$ and the cumulative incidence was 0.5, and the rare disease scenario where $n=50,000$ and the cumulative incidence was 0.03, so that the value of the baseline hazard $\lambda_{0}$ was determined by the cumulative incidence and the value of the changepoint $\tau$, for each case.
\noindent We examined the setting of observing $W$ only for all the subjects. The simulations results are based on 1,000 replications, and in all cases, we report the relative bias of the median and the relative bias of the mean of the estimates.
In order to eliminate cases of divergence, we imposed the condition that
$|\hat{\bth}|\le 4.9$, and the results are based on replicates for which this condition was satisfied. Convergence problems arose more often when greater measurement error was considered, i.e. $\rho_{xw}=0.4$, and when the changepoint was at the lower or upper extreme of the covariate domain.
The convergence percent (percentage of replications in which the estimation procedure converged) is presented in Table S.6 in the Supplemental Materials for all scenarios considered.
The results of each method except the naive are presented in two versions: one assuming that the measurement error parameter is known, labeled (kn), and the second with the measurement error parameters estimated, labeled (unk). We assumed an external reliability study with a sample size of 500 and two measurements
of $W$ for each subject in order to estimate $\rho_{xw}$.

The starting values for the maximization were chosen as follows. For the naive method we arbitrarily took starting
values of $\beta=\omega=0$. Since the Cox partial likelihood is concave, the choice of starting
values for this method is not so critical. For the other methods, we used as starting values the estimate
yielded by the method one degree lower in complexity (with the hierarchy being naive, RC1, RC2, RR1, RR2,
MPPLE).
The estimation of $\bth$ in the RR1 method was done by maximizing the log likelihood $l_{p}^{(RR1)} (t^{*},\bth)$ that was described in Section 3.2. We used the
Matlab routine \emph{'fmincon'} with $\bth$ constrained to the range $[-5,5]$.
The MPPLE method was not examined in the rare disease case because in this scenario it is approximately equivalent to the the RR1 method (and therefore to the RR2 method), since
$\exp(-\Lambda_0(t)r(x,\bz,\bth))$ $\approx 1$ for all $t$ and $x$, so that
$\exp(\phi(\bth,w,\bz,\Lambda_0(t))) \approx E[r(X,\bz,\bth)|W=w, \bZ=\bz]$.
The computational burden of RR1 is low and can be described in units of minutes (about 5-30 minutes, depends on the sample size), whereas this of the RR2 is much longer and can be described in units of hours (about 5 to 24 hours, depends on the sample size).

The computational burden of RR1 is relatively light and the runtime is on the scale of minutes (about 5-30 minutes, depending on the sample size), whereas the burden of the RR2 method is much heavier and the runtime is on the scale of hours (about 5 to 24 hours, depending on the sample size). Thus, we recommend using the RR1 method in initial analyses and the RR2 method
in the final definitive analysis.

\subsection{RC2 vs. RC1}
\normalsize
In this paper we presented an improved RC method, RC2, over the basic RC, RC1.
As described in the the next subsection, we found that this method indeed performed better than RC1 in the setting of Cox model. The RC2 is not specific to the Cox model, but can be used in other settings. At a reviewer's suggestion, we examined the relative performance of RC2 and RC1 in a simpler setting of a simple linear regression model $Y=X \beta$. We used a simulation study which is similar in its design to that of our main simulation study in this paper: We set $\beta=\log(1.5)$ and $\omega=\log(2)$. The covariate $X$ was generated as standard normal, and the observed surrogate covariate value $W$ was generated using the classical measurement error model with $U\sim N(0,\sigma _{u}^{2})$. The values of $\sigma_{u}^{2} $ were set so as to yield $\rho_{xw}= 0.8, 0.6$, or 0.4, corresponding to $\sigma_{u}^{2}$ values of 0.56, 1.77 and 5.25. The changepoint $\tau$ was fixed at one of 5 points at various percentiles of the distribution of $X$:
$\Phi ^{-1} (0.1),\Phi ^{-1} (0.25),\Phi^{-1} (0.5),\Phi^{-1} (0.75),\Phi ^{-1} (0.9)$. The sample size was $n=1,500$. We examined the setting of observing $W$ only for all the subjects. The vector parameters is then $\bth=(\beta,\omega)$. The simulation study contains the calculations of the RC1 and the RC2 under the version that assuming the measurement error parameters estimated, when we assumed an external reliability study with a sample size of 500 and two measurements
of $W$ for each subject in order to estimate $\rho_{xw}$. The simulations results are based on 1,000 replications, and in all cases, we report the average and the median of the estimate.
In order to eliminate cases of divergence, we imposed the condition that
$|\hat{\bth}|\le 4.9$, and the results are based on replicates for which this condition was satisfied. Convergence problems arose more often when greater measurement error was considered, i.e. $\rho_{xw}=0.4$.
The starting values in the estimation process were $\beta_{0}=0$ and $\omega_{0}=0$, arbitrarily.
Table 8 presents the results, including the convergence percent (percentage of replications in which the estimation procedure converged) in the problematic scenarios where the percentage is lower than 1. We can see that the RC2 performed better than RC1 for all values of the changepoint and the error variance.

\subsection{Summary of Simulation Results} 
\subsubsection{Changepoint vs. Two Variables}
\noindent
Table 1 presents a comparison of the means, between the results with the naive, RC1, and RC2 methods in the current setting with the results of naive analysis and the results with
regression calibration in the setting of two functionally unrelated covariates $X_1$ and $X_2$ measured with error, with parameters chosen so
that the mean and covariance structure match those of the current setting
(``2 variables" setting). For estimation of $\beta$, we see that the RC1 estimates for
the changepoint setting are often worse than the RC estimates in the ``2 variables" setting. For estimation of $\omega$,
the RC1 estimates for the changepoint setting are worse than the RC estimates in the ``2 variables" setting, while with RC2, which
is a specialized RC method for the changepoint setting, the results are close to those for the the RC estimates in the ``2 variables" setting.
\subsubsection{Main Results for the Changepoint Model}
\begin{landscape}
\begin{figure}[h!]
\bc
\includegraphics[width=1.45in, height=1.3in]{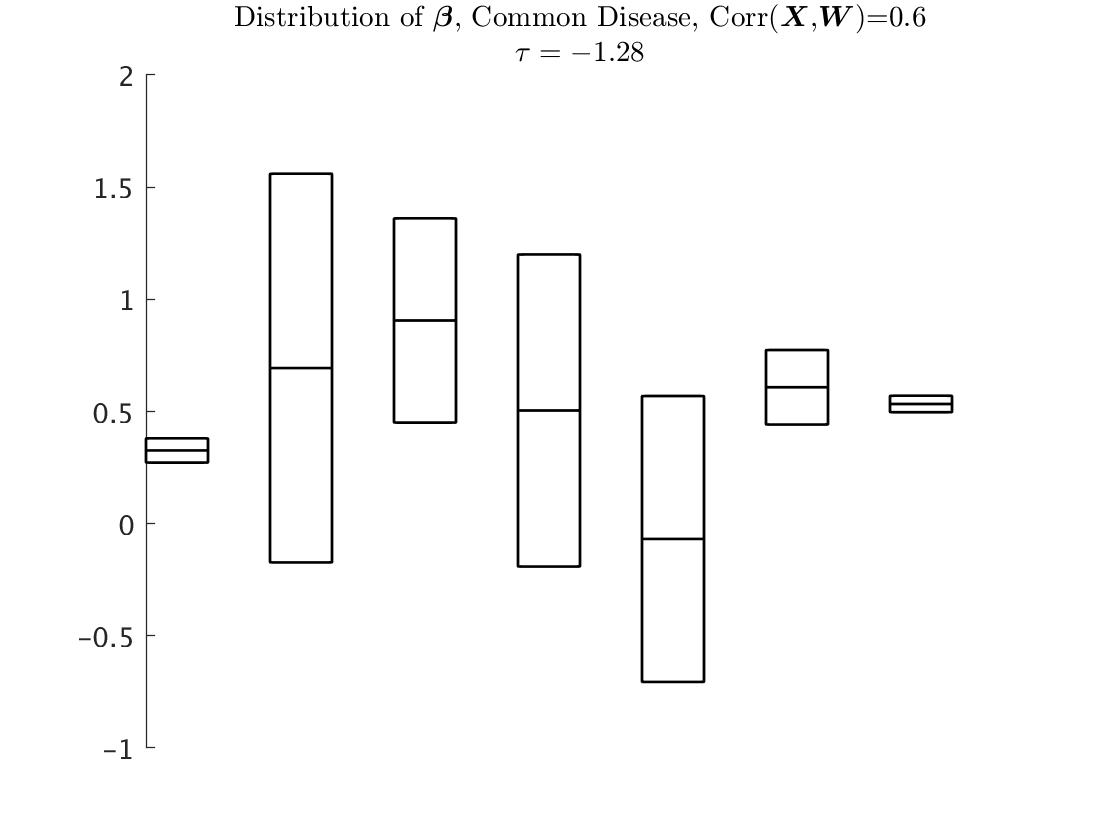} \hskip0.01truein
\includegraphics[width=1.45in, height=1.3in]{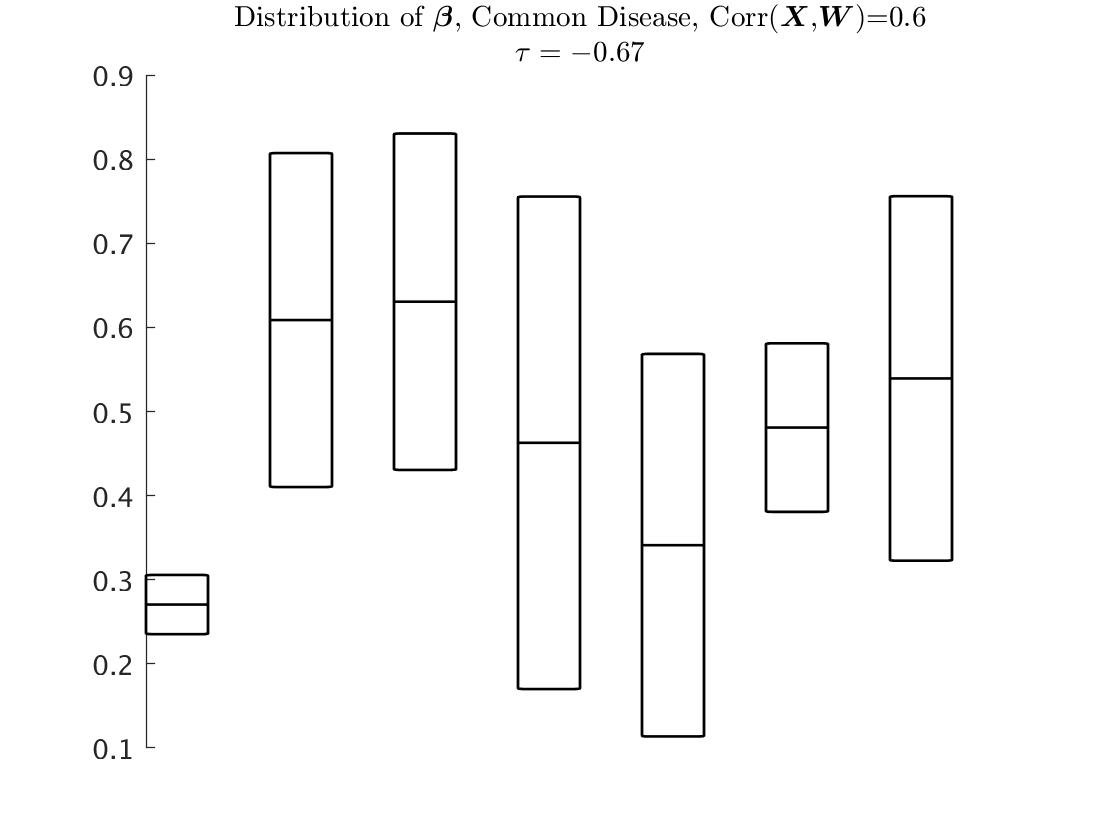} \hskip0.01truein
\includegraphics[width=1.45in, height=1.3in]{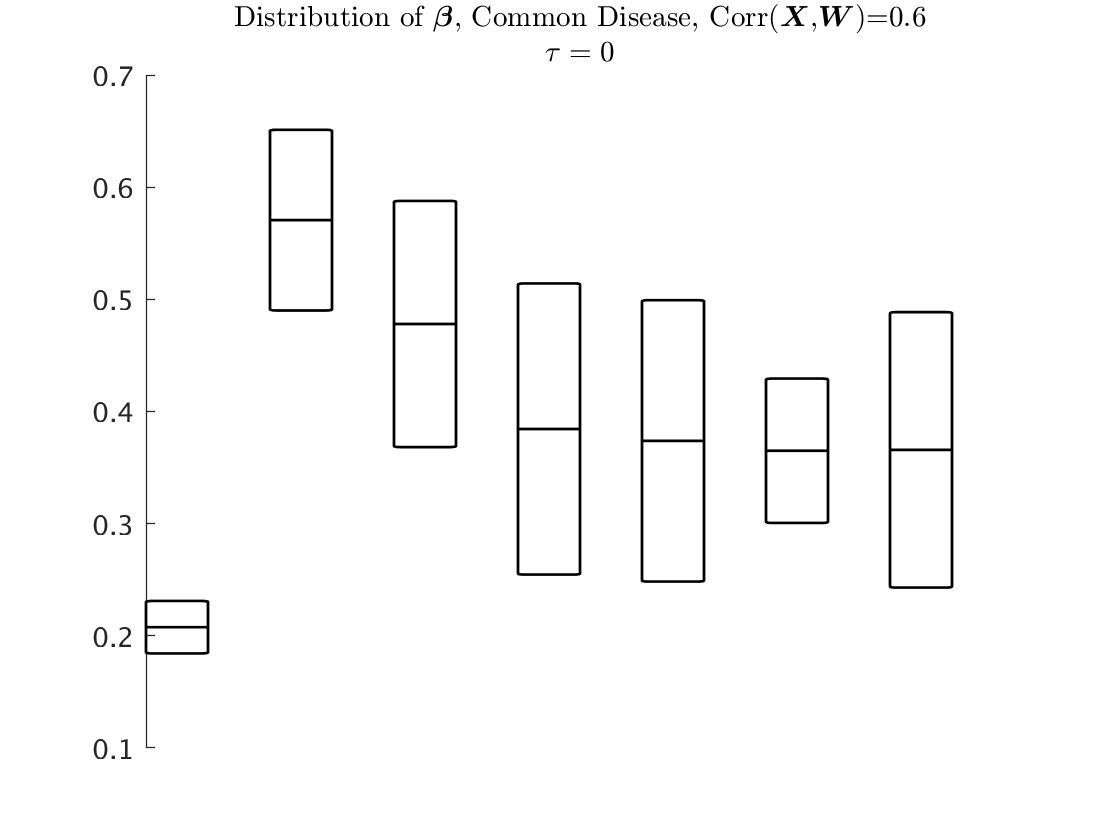} \hskip0.01truein
\includegraphics[width=1.45in, height=1.3in]{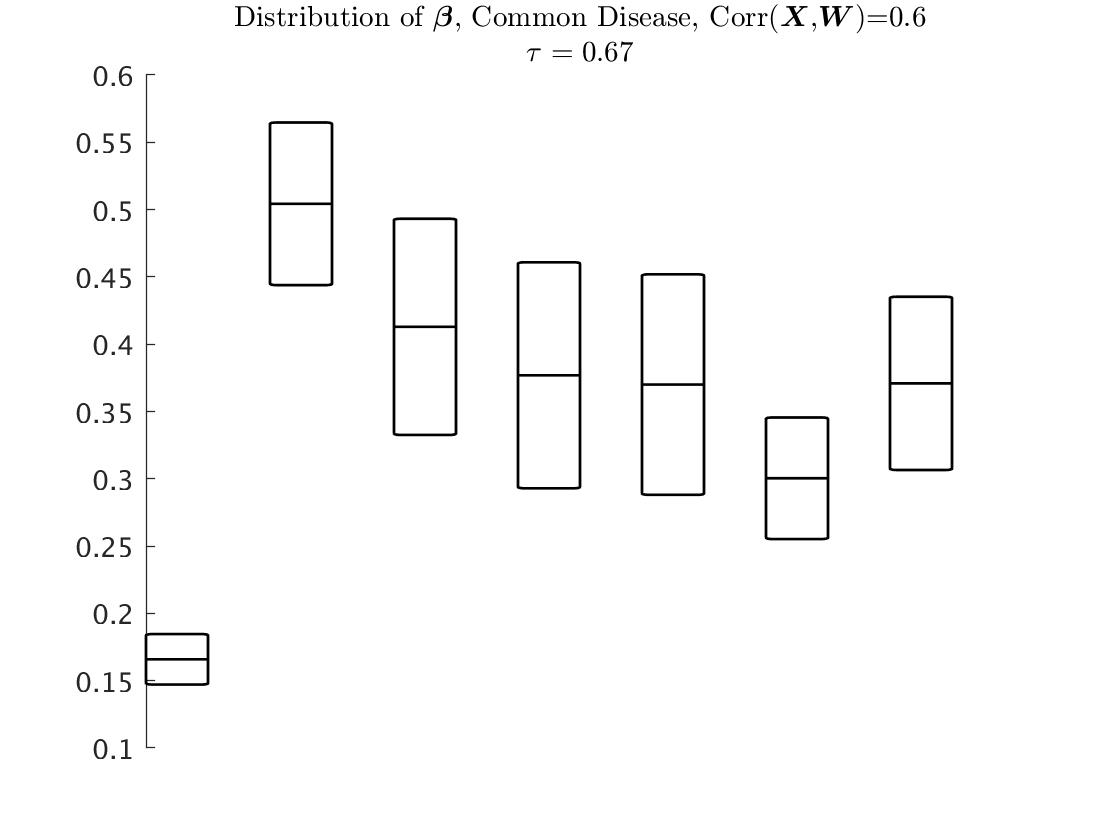} \hskip0.01truein
\includegraphics[width=1.45in, height=1.3in]{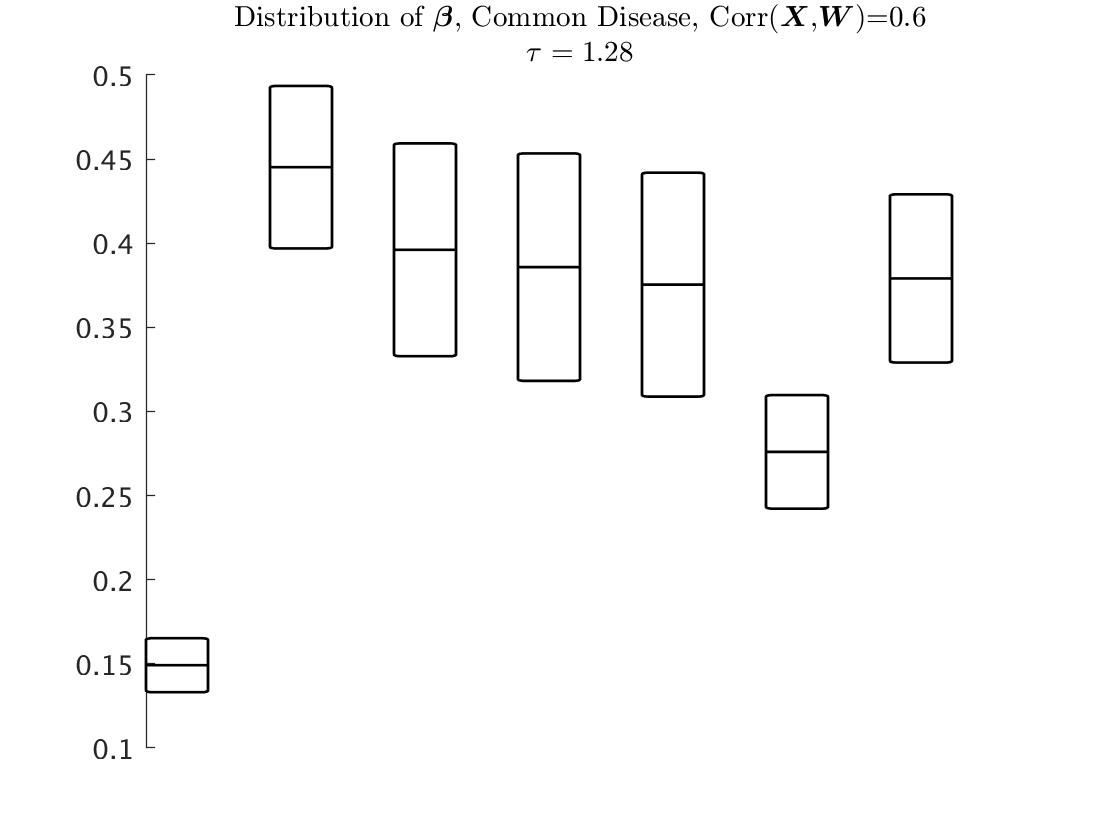} \hskip0.01truein
\vspace{0.5cm}
\\
\includegraphics[width=1.45in, height=1.3in]{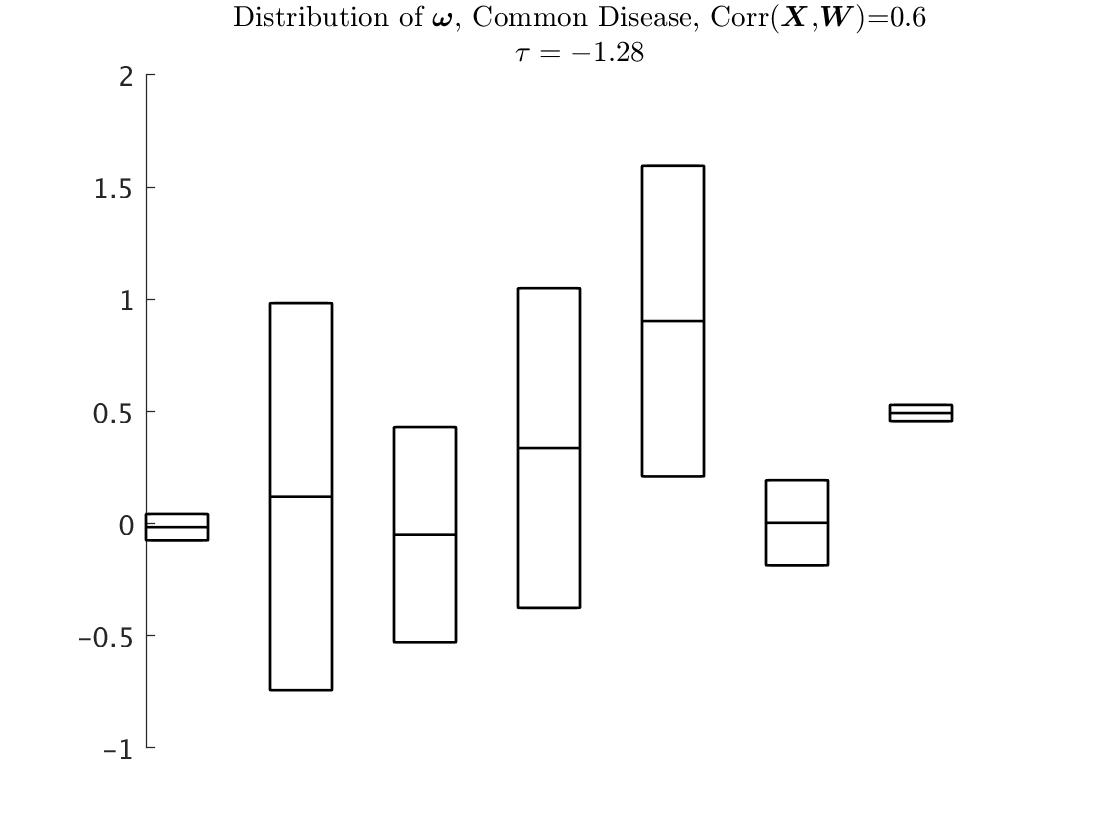} \hskip0.01truein
\includegraphics[width=1.45in, height=1.3in]{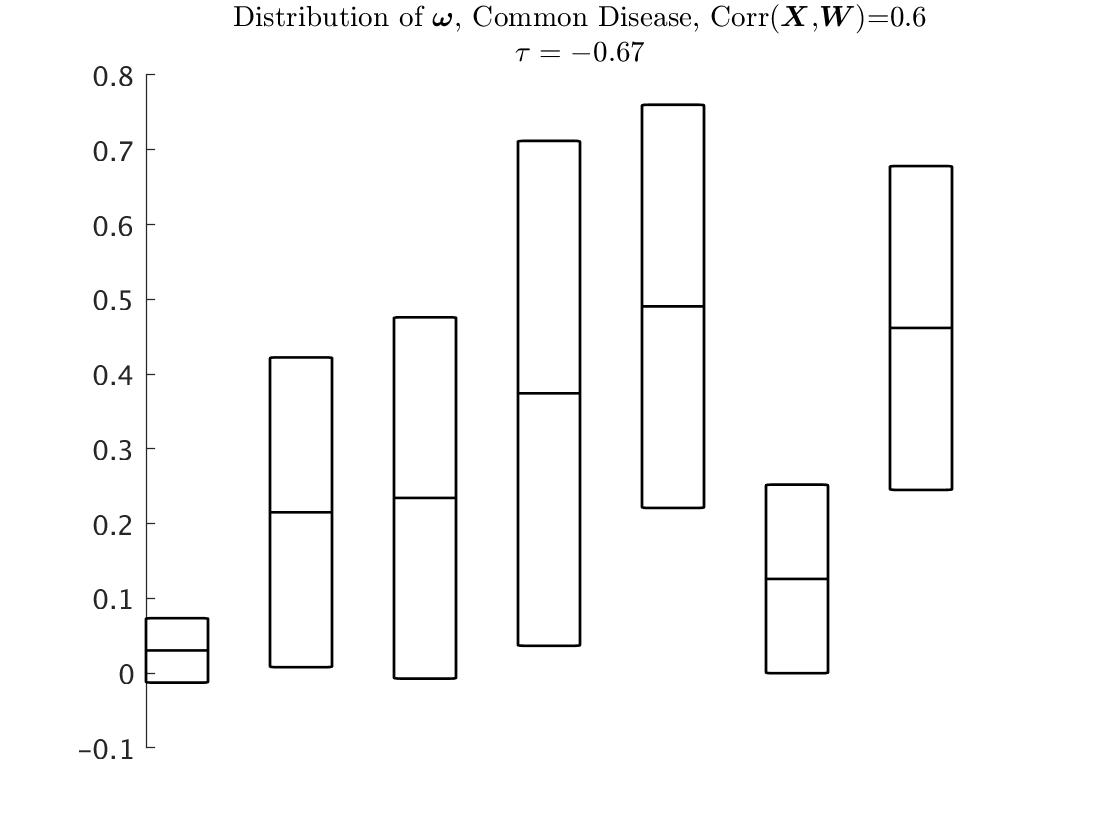} \hskip0.01truein
\includegraphics[width=1.45in, height=1.3in]{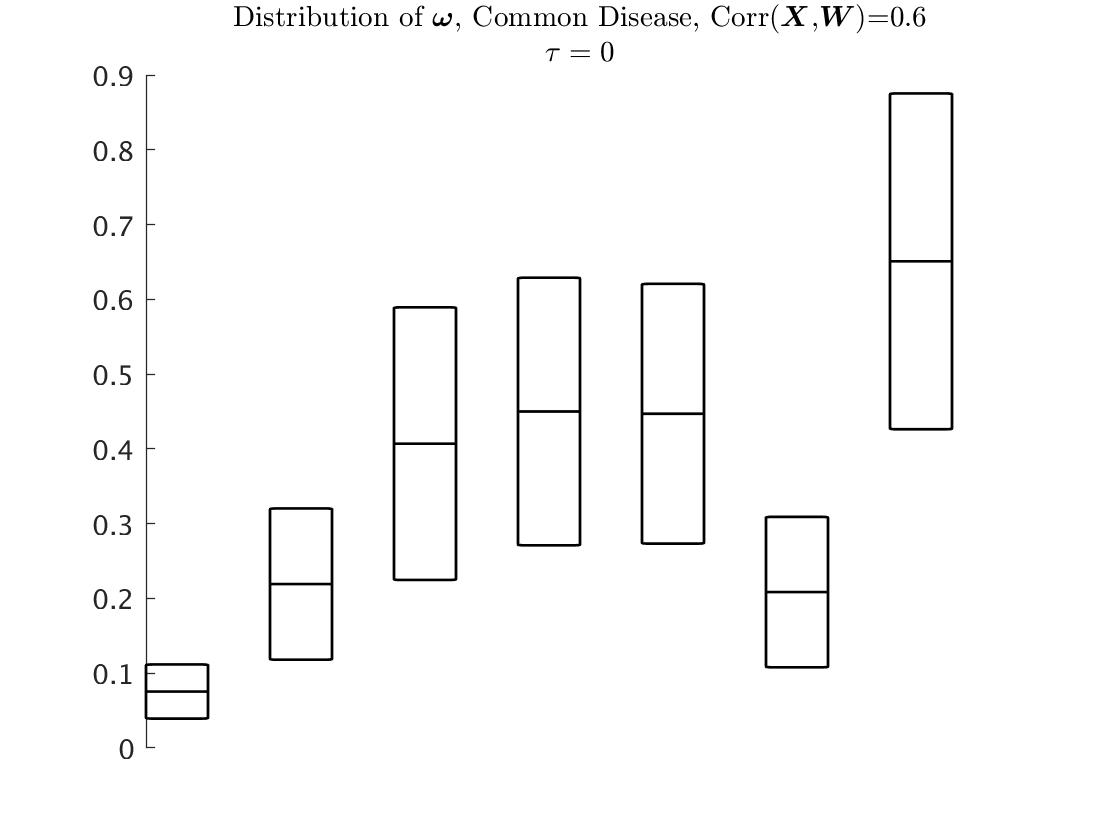} \hskip0.01truein
\includegraphics[width=1.45in, height=1.3in]{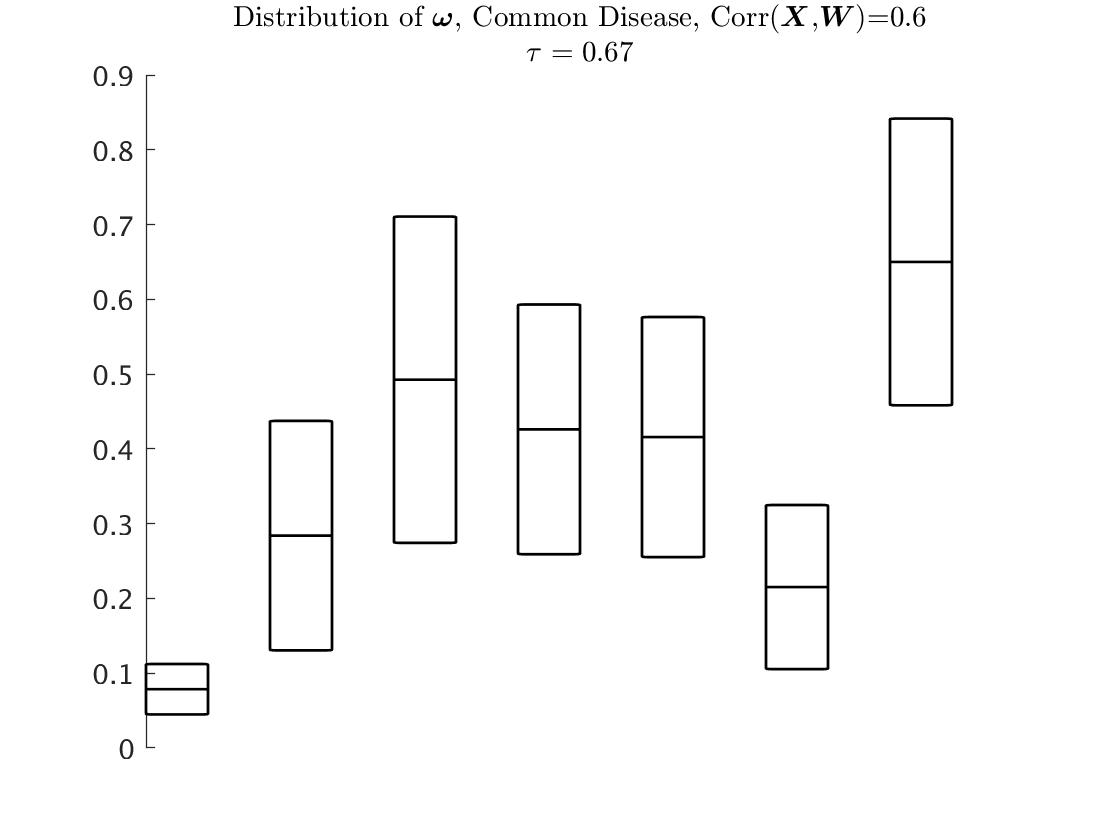} \hskip0.01truein
\includegraphics[width=1.45in, height=1.3in]{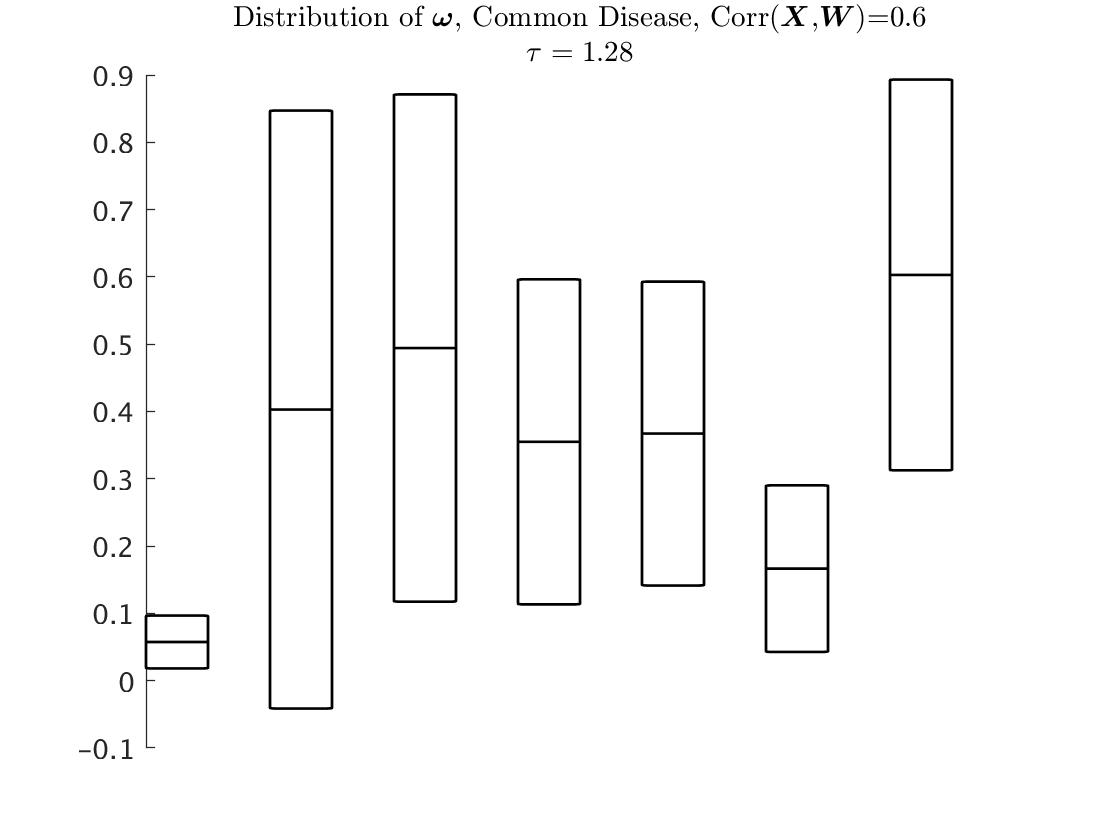} \hskip0.01truein
\vspace{0.5cm}
\ec
\caption{\footnotesize
Boxplots of each estimator over 1000 replications. The range of each box is between the median of the estimator minus 0.5 of its interquartile range, to the median of the estimator plus 0.5 of its interquartile range. The horizontal line inside each box is the median of the estimator. For a given value of the changepoint $\tau$, each plot describes from left to right the naive, RC1, RC2, RR1, RR2, SIMEX, and MPPLE methods. For RC1, RC2, RR1, RR2 and MPPLE methods, the plots are based on the estimates under unknown nuisance parameters which were estimated by an external reliability sample of size 500 with 2 replications/person. The boxplots describe the results under the common disease with $n=3,000$ and cumulative incidence of 0.5. The plots on the top present the behaviour of $\hat{\beta}$, and the plots on the bottom present the behaviour of $\hat{\omega}$. The value of $\rho_{xw}$ is 0.6.}

\label{fig:simubox}
\end{figure}
\end{landscape}

\textbf{ (i) General Results}
\\
Figure \ref{fig:simubox}
provides a general comparison of all estimators examined using boxplots. The range of each box is between the median of the estimator minus 0.5 of its interquartile range, to the median of the estimator plus 0.5 of its interquartile range. The horizontal line inside each box is the median of the estimator. We use this format of boxplot rather than the standard one because a reviewer recommended displaying the simulation results using boxplots but unfortunately we did not store the first and third quartiles but only the interquartile range.
We present results separately for $\beta$ and $\omega$ for the common disease scenario, and separately for each value of the changepoint considered. The figure presents results under $\rho_{xw}=0.6$. The corresponding results for  $\rho_{xw}=0.8$ and $\rho_{xw}=0.4$, and the results for the rare disease scenario under $\rho_{xw}=0.8, 0.6, 0.4$ are presented in S.3 in the Supplemental Materials. Clearly, the best performing method in the common disease case was the MPPLE, and the best performing method in the rare disease case was RR2.
The detailed results of the simulations are presented in Tables S.4-S.5 in the Supplemental Materials.
A number of common trends were seen:
(a) The estimators of $\beta$ performed well for $\tau$ values in the middle to upper end of the covariate domain. The estimators of $\omega$ performed well when $\tau$ was in the middle of the covariate domain ($\tau =0$), and substantially less well when the changepoint was at the upper or lower extreme of the covariate domain.
(b) As expected, the estimators performed progressively less well as measurement error increased.
(c) RC2 performed better than RC1 for all values of the changepoint and error variance, under both known and unknown nuisance parameters, and under common and rare disease.
(d) The RC2 estimator was considerably better than the naive estimator.

\normalsize
\textbf{ (ii) Detailed Results}
\\
(a) The advantage of RC2 relative to the naive estimator was particularly prominent in regard to the estimation of $\omega $, where the bias of the naive estimate ranged from -102.44\% to -64.52\% in the common disease case and from -100.05\% to -51.38\% in the rare disease case (the bias of the naive estimate of $\beta$ ranged from -83.33\% to 37.15\% in the common disease case and from -80.57\% to -7.86\% in the rare disease case).
(b) In the common disease case, the SIMEX method performed better for $\tau > 0$ than for $\tau < 0$, but even for the $\tau $$>$ 0 the bias was substantial, particularly for the estimating $\omega $. Because of this poor performance and the heavy computational burden of the SIMEX method,
we did not examine SIMEX in the rare disease case. This poor performance may be due to the fact that the relative risk function is not differentiable.
(c) Comparing the results for the common disease case with those for the rare disease case, we found that the naive estimate of $\beta$ and the naive, RC1 and RC2 estimates of $\omega$ performed better in the rare disease case than in the common disease case for all values of $\tau $ under both known and estimated measurement error parameters, whereas the RC1 and RC2 estimates of $\beta$ performed better in the common disease case than in the rare disease case for all values of $\tau$ under both known and estimated measurement error parameters.
(d) For the RR methods, the results were better in the common disease case for $\tau<0$ but were better in the rare disease case when $\tau>0$.
(e) With known nuisance parameters, RC2 yielded better results than RR1 for estimation of $\beta$ in the common disease case with $\tau > 0$, but worse results in the rare disease case with $\tau > 0$.
When the measurement error parameters were estimated, RR1 had less bias than RC2 for all values of $\tau$.
(f) In regard to the estimation of $\omega$ with $\tau < 0$, under both known and estimated measurement error parameters, RR1 had less bias than RC2. For $\tau \ge 0$, RC2 had less bias than RR1. Interestingly, for $\tau>0$ in the rare disease case, the mean RC2 estimate was generally greater than the true $\omega $ value of 0.69 (overestimation), whereas the mean RR1 estimate was less than the true value (underestimation). This phenomenon was prominent with substantial measurement error.

\textbf{ (iii) Coverage Rates}
\\
We also examined the empirical coverage rates of nominal Wald-type 95\% confidence intervals. Tables 2-3 and Table S.7 in the Supplemental Materials present the results. For the RR2 method, we used the estimated asymptotic variance in order to calculate the confidence interval. This is an appropriate approximation since the empirical variances of the RR1 and the RR2 that were obtained in the simulation study were generally close. The empirical coverage rate was calculated based on 1,000 simulation replications. The coverage was good with the RR and the MPPLE methods.


\subsection{Robustness} 
One of the assumptions made (Section 2.1) is the assumption that $X(t)$ and $U(t)$ are both normally distributed. This assumption is required for the methods RC2, RR1, RR2 and the MPPLE methods. We
examined the robustness of the first three methods to violation of this assumption, Table 4 and Table S.8 in the Supplemental Materials present the results. We first reran the simulations under a heavy-tails scenario where $X(t)$ and $U(t)$ were generated under the $t$ distribution with degrees of freedom (df) of 6 or 15. We matched the mean and the variance to those in the simulations under the normal
distribution. The results under the \emph{t}-distribution with df=15 were close to those under the normal distribution, whereas the results under the \emph{t}-distribution with df=6 were far from those under the normal distribution. Afterward, we reran the simulations under a skewed distribution scenario with $X(t)$ and $U(t)$ both generated according to the log gamma distribution with parameters $\alpha=\beta=1$. As before, we matched the mean and the variance to that used in the simulations under the normal distribution. The results were noticeably worse that those obtained under the normal distribution. Thus, the methods
are robust to mild heavy-tailedness, but not to severe heavy-tailedness or skewness.
The RR method (Section 3.2) can be adapted to the non-normal case by using numerical integration in place of the formula we presented for the normal case to evaluate
the conditional expectation $E[\lambda (t|X(t),\bz)|W(t)=w, \bZ(t)=\bz]$ under the relevant distribution. The MPPLE method can be adapted
similarly.
\\

\subsection{Additional Error-Free Covariates}
\begin{landscape}
\begin{figure}[h!]
\bc
\includegraphics[width=1.45in, height=1.3in]{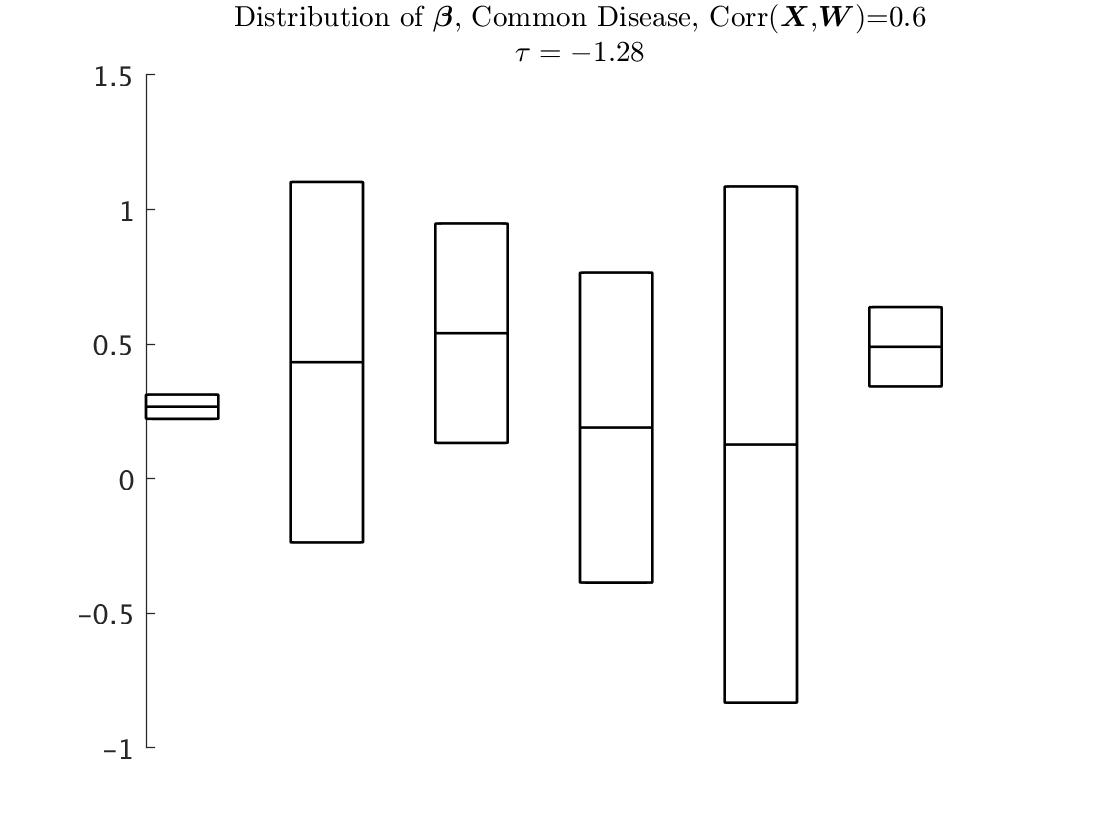} \hskip0.01truein
\includegraphics[width=1.45in, height=1.3in]{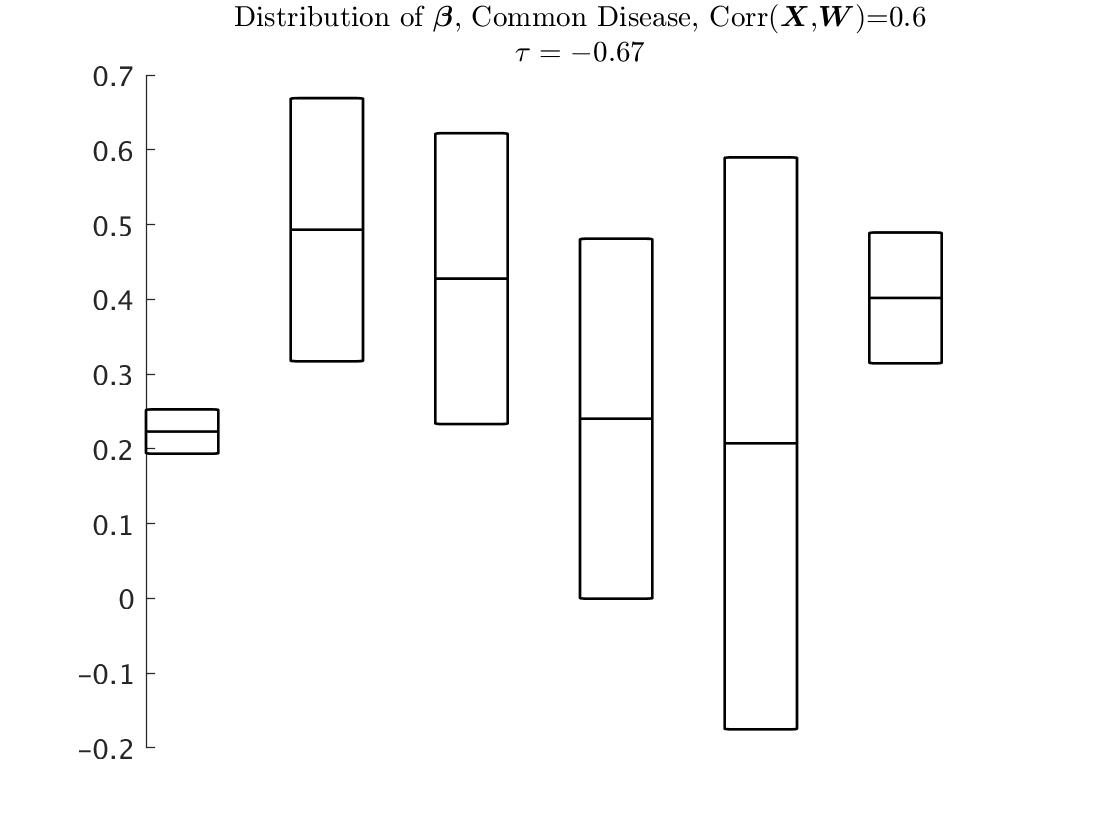} \hskip0.01truein
\includegraphics[width=1.45in, height=1.3in]{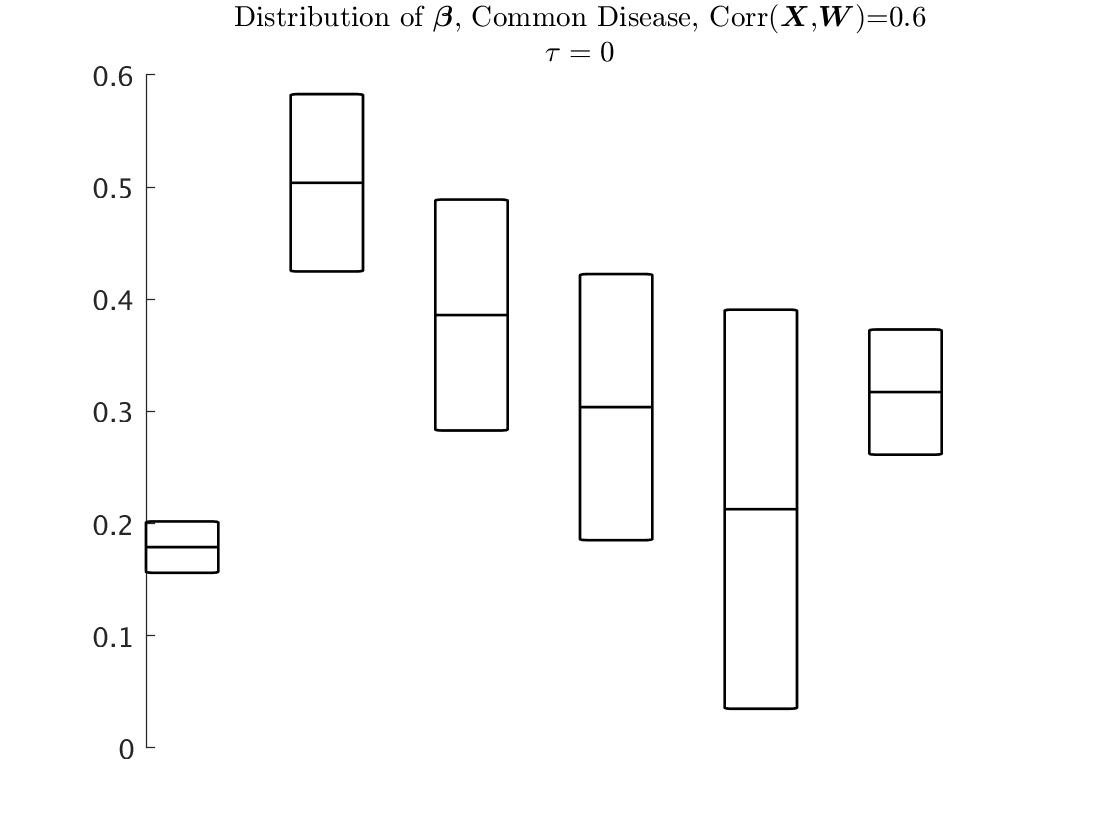} \hskip0.01truein
\includegraphics[width=1.45in, height=1.3in]{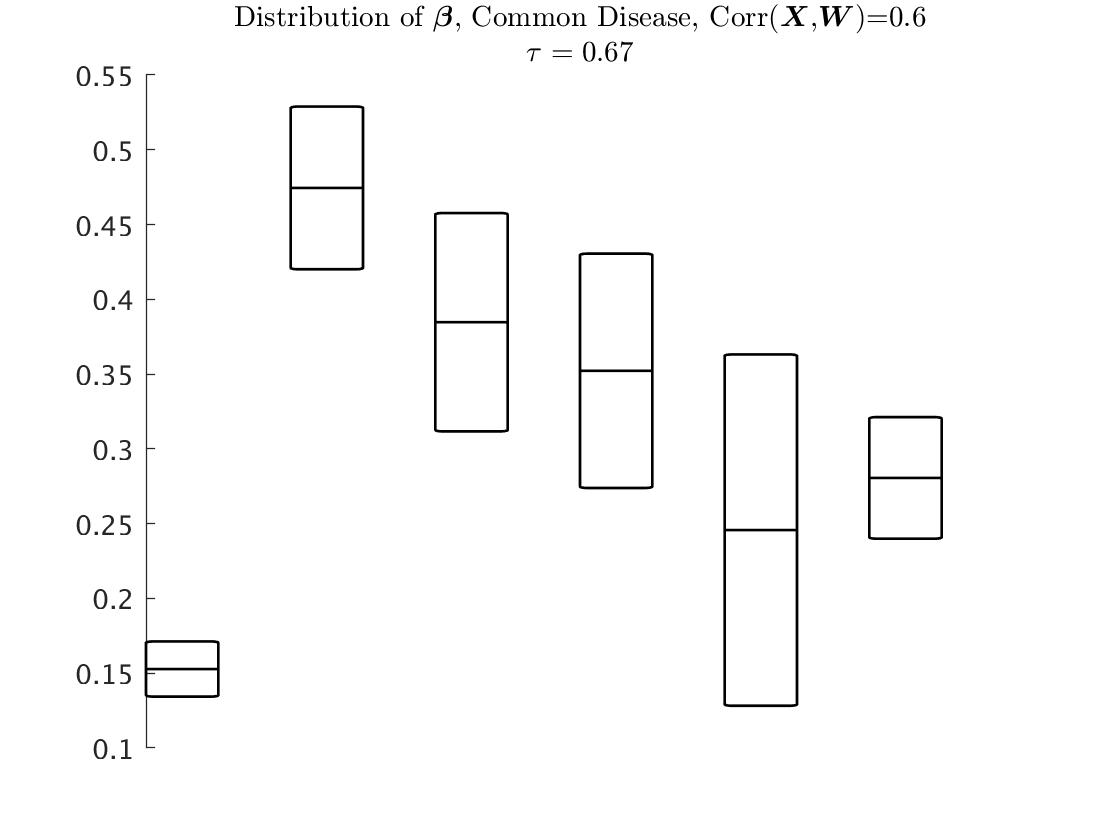} \hskip0.01truein
\includegraphics[width=1.45in, height=1.3in]{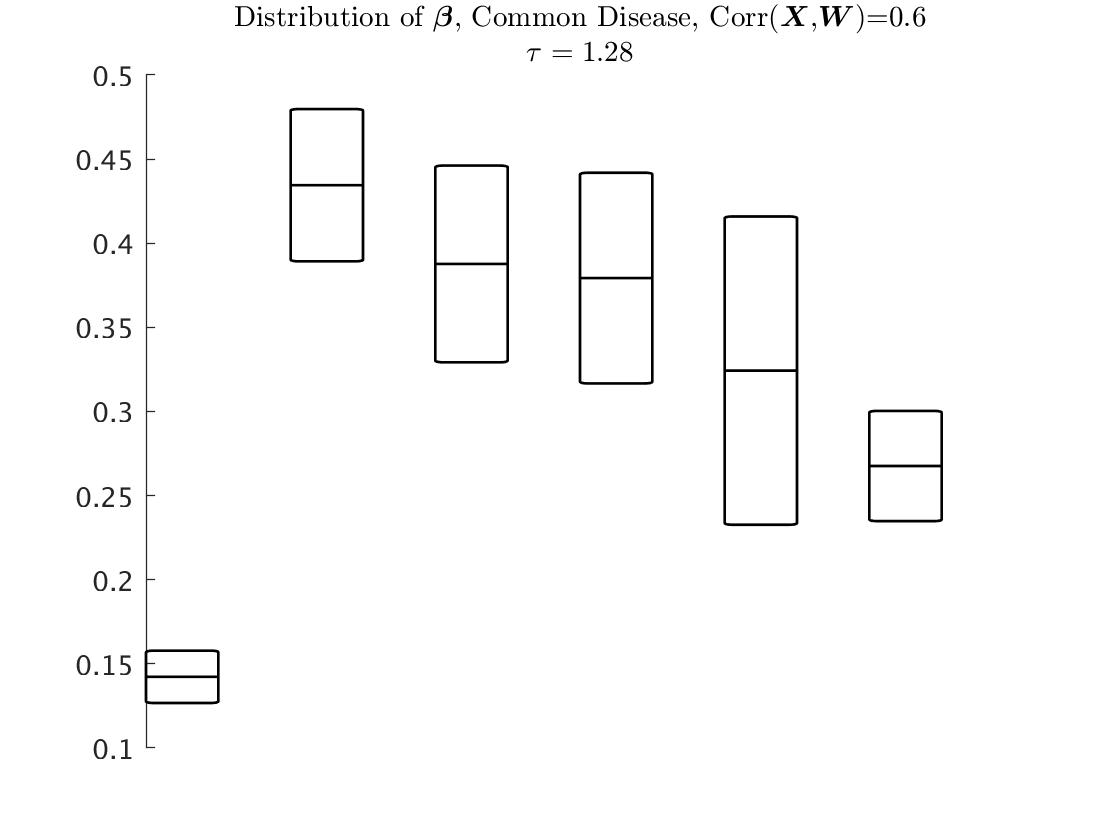} \hskip0.01truein
\vspace{0.5cm}
\\
\includegraphics[width=1.45in, height=1.3in]{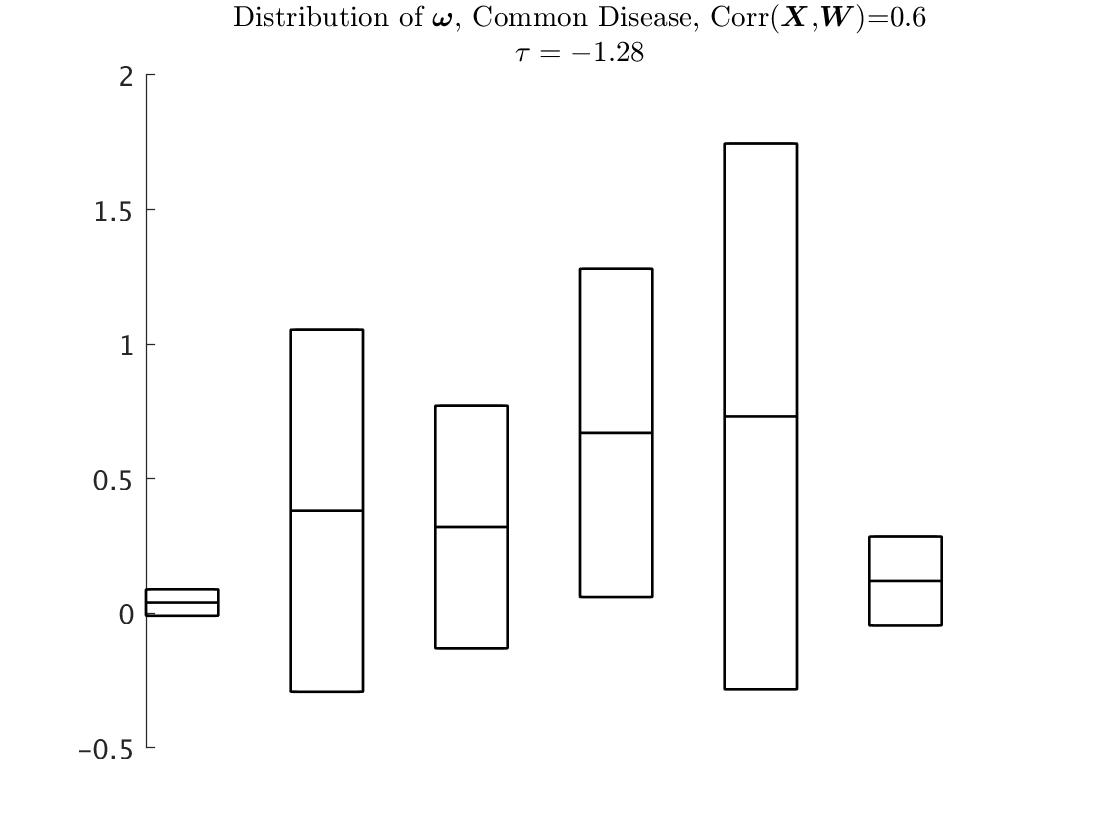} \hskip0.01truein
\includegraphics[width=1.45in, height=1.3in]{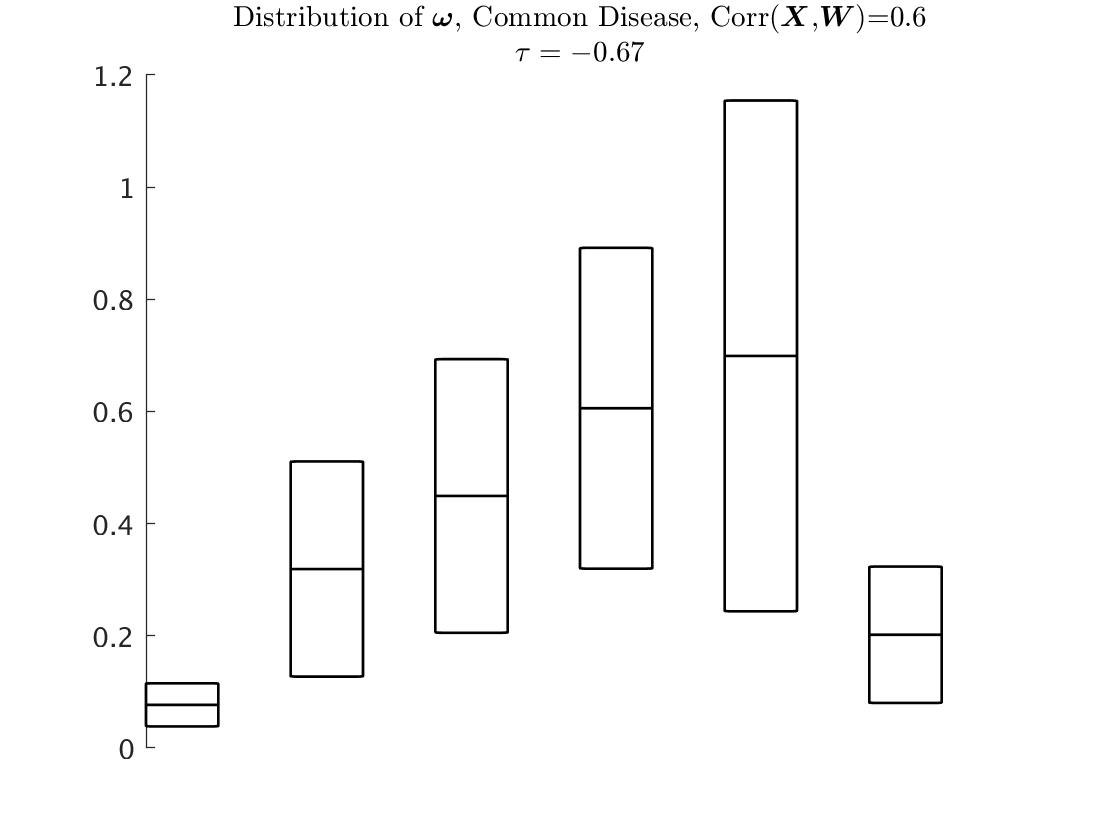} \hskip0.01truein
\includegraphics[width=1.45in, height=1.3in]{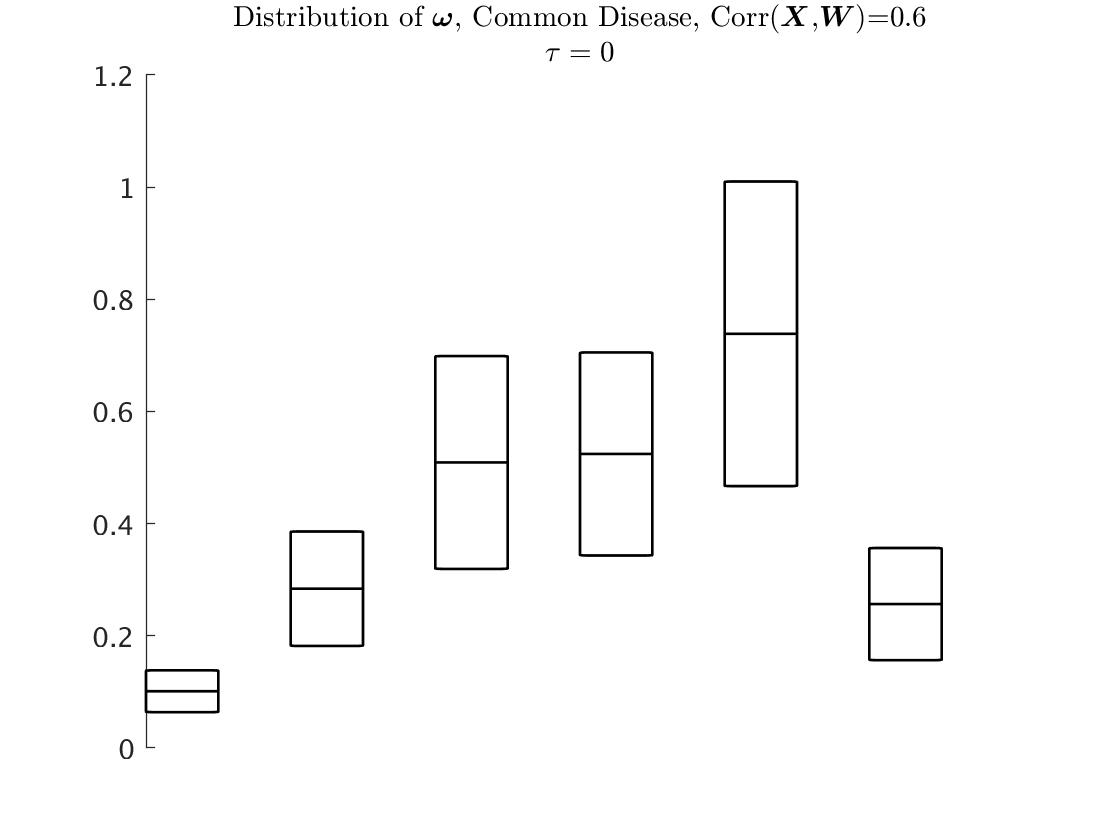} \hskip0.01truein
\includegraphics[width=1.45in, height=1.3in]{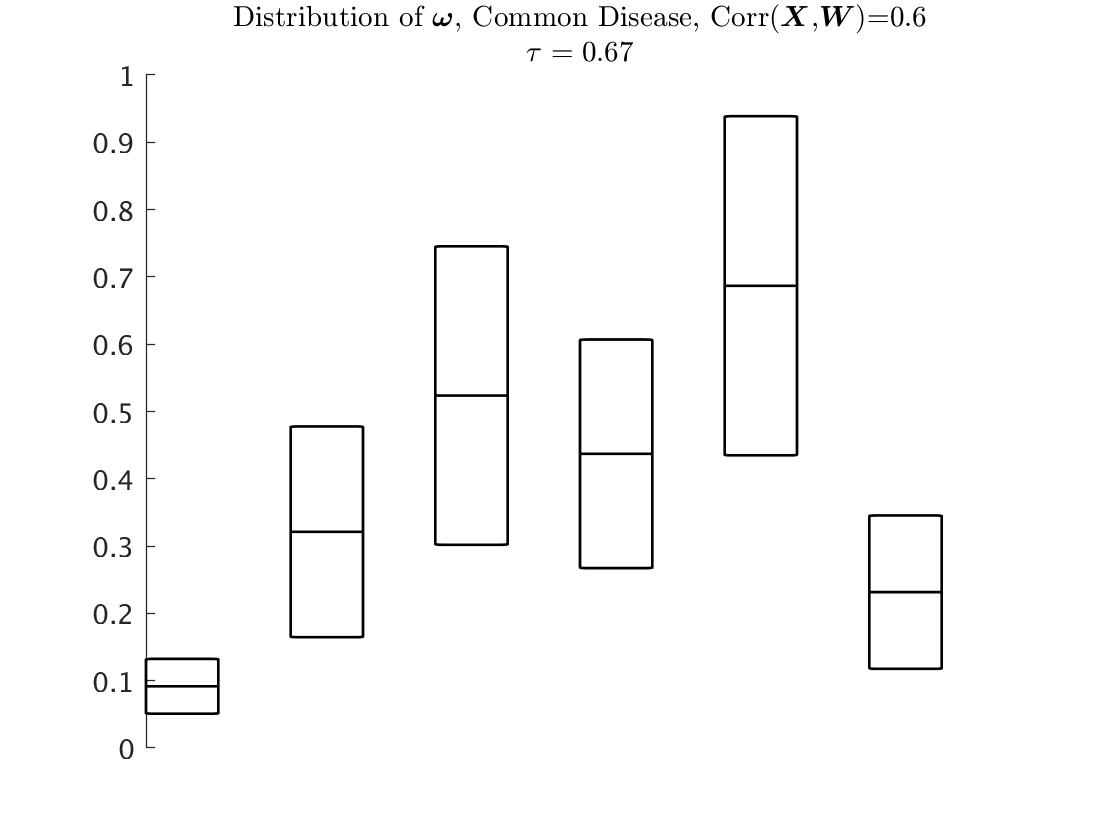} \hskip0.01truein
\includegraphics[width=1.45in, height=1.3in]{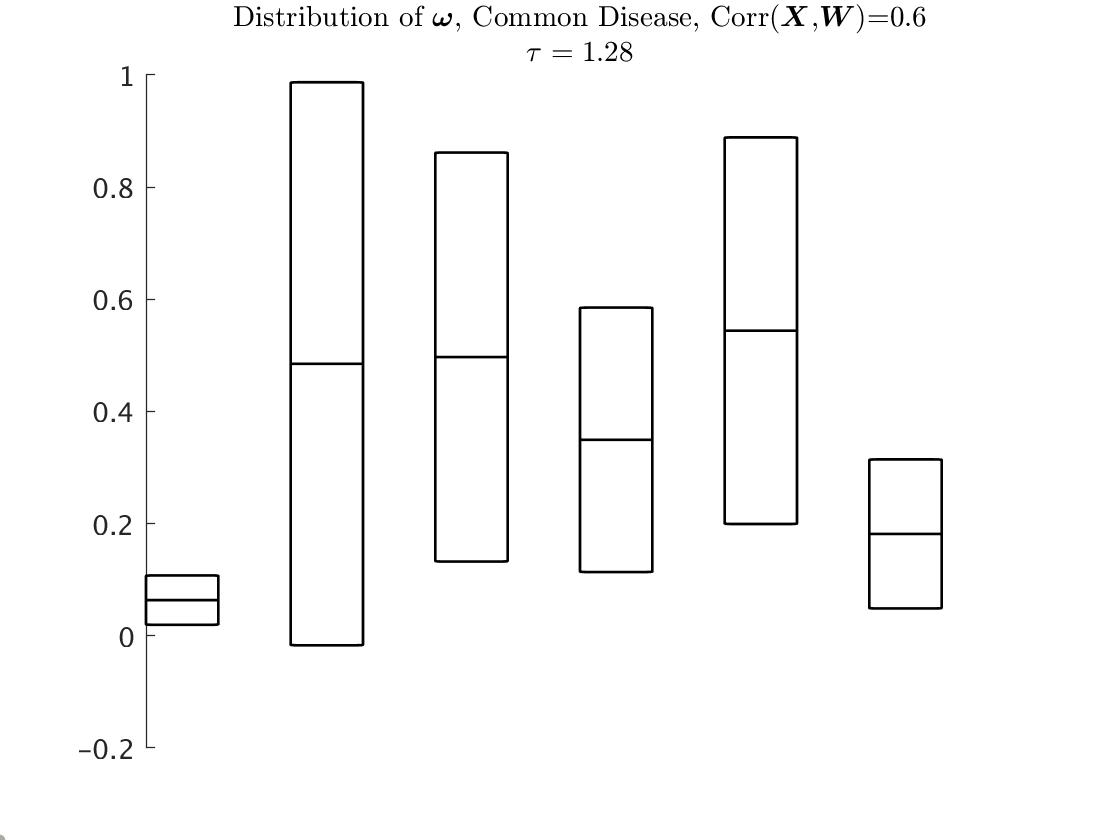} \hskip0.01truein
\vspace{0.5cm}
\\
\includegraphics[width=1.45in, height=1.3in]{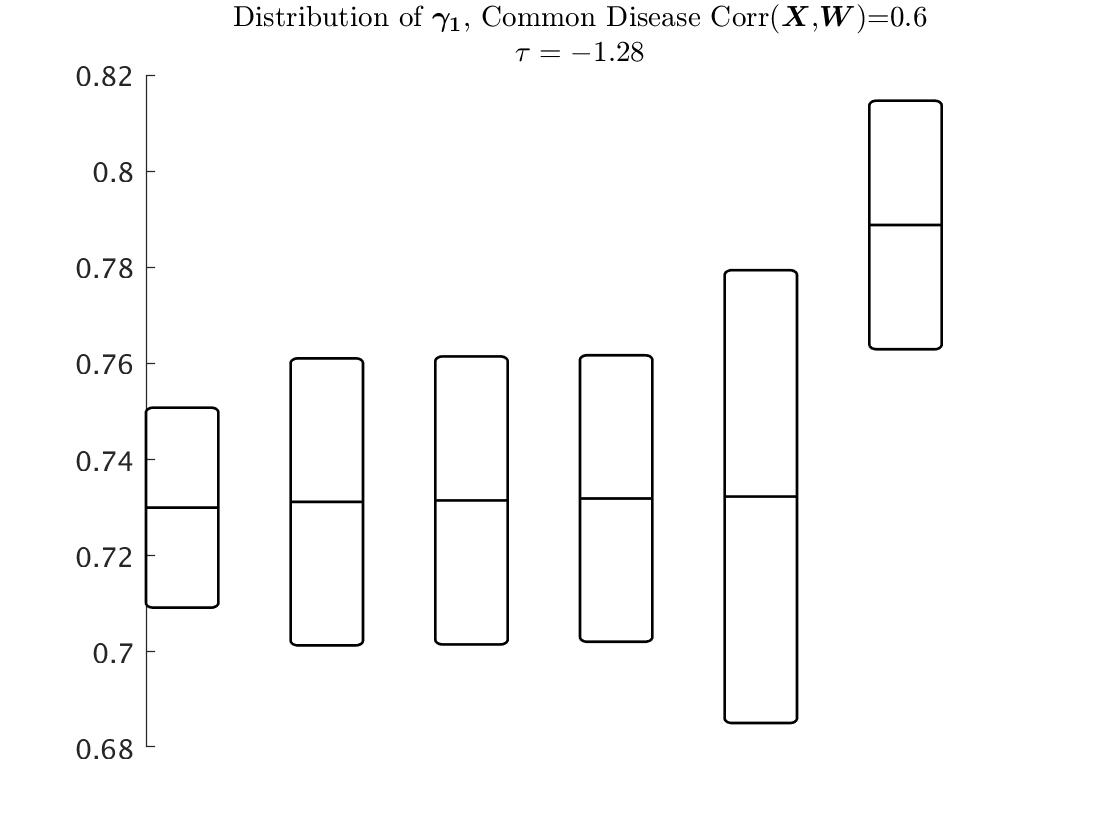} \hskip0.01truein
\includegraphics[width=1.45in, height=1.3in]{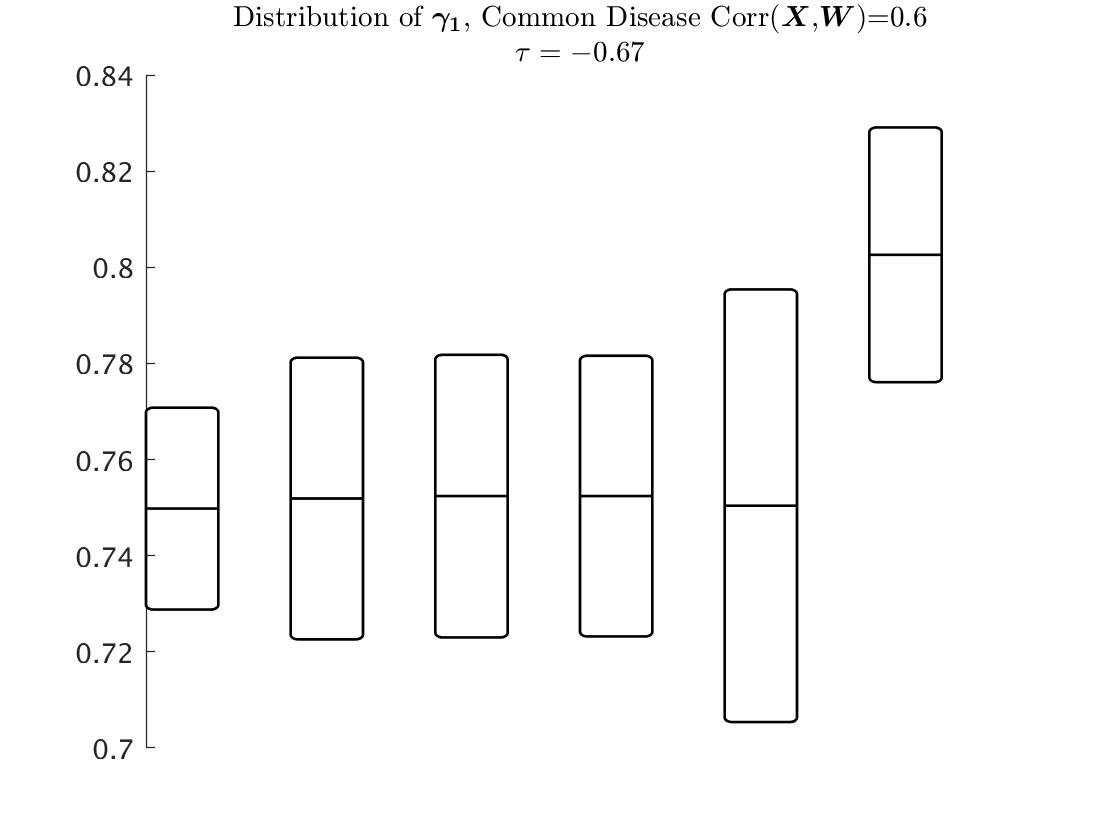} \hskip0.01truein
\includegraphics[width=1.45in, height=1.3in]{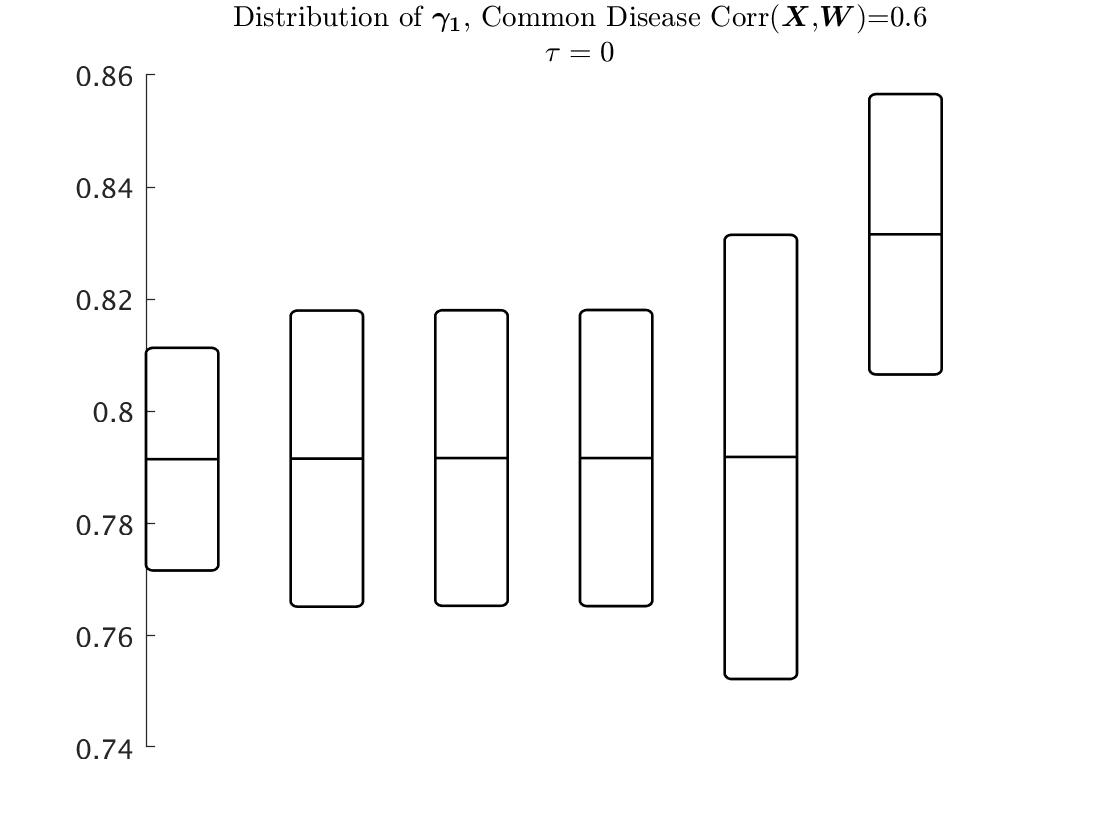} \hskip0.01truein
\includegraphics[width=1.45in, height=1.3in]{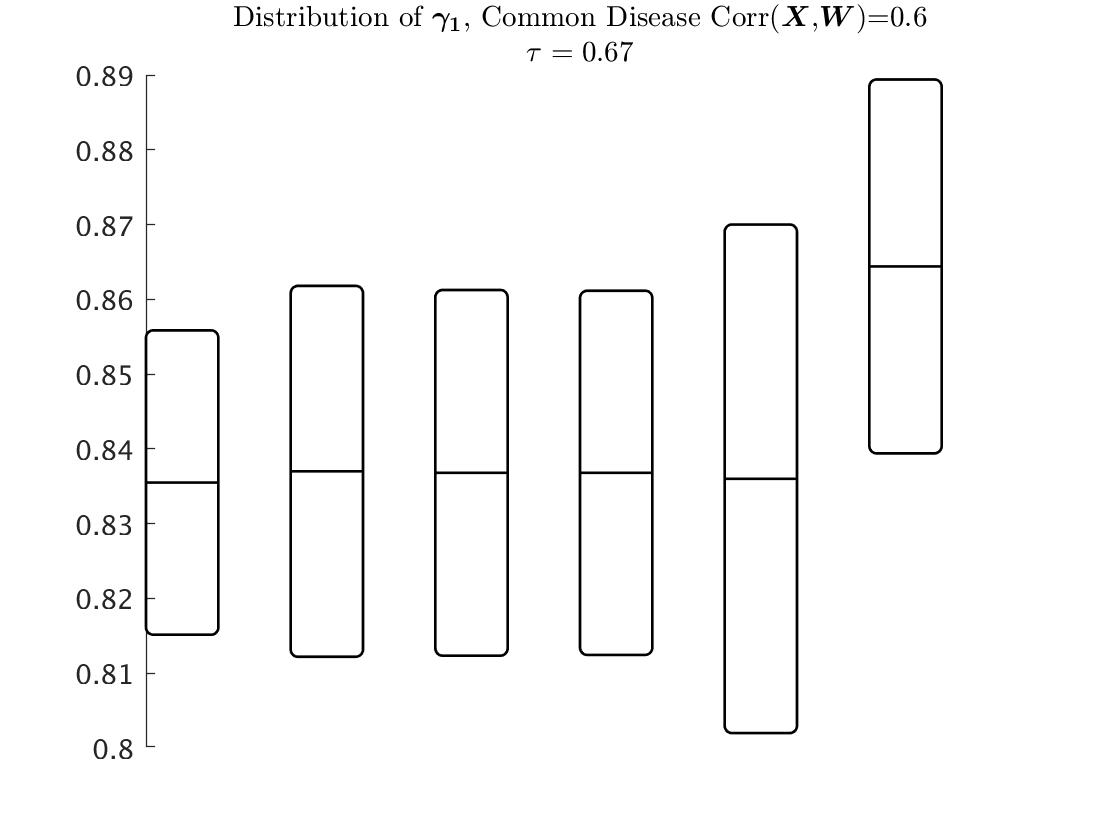} \hskip0.01truein
\includegraphics[width=1.45in, height=1.3in]{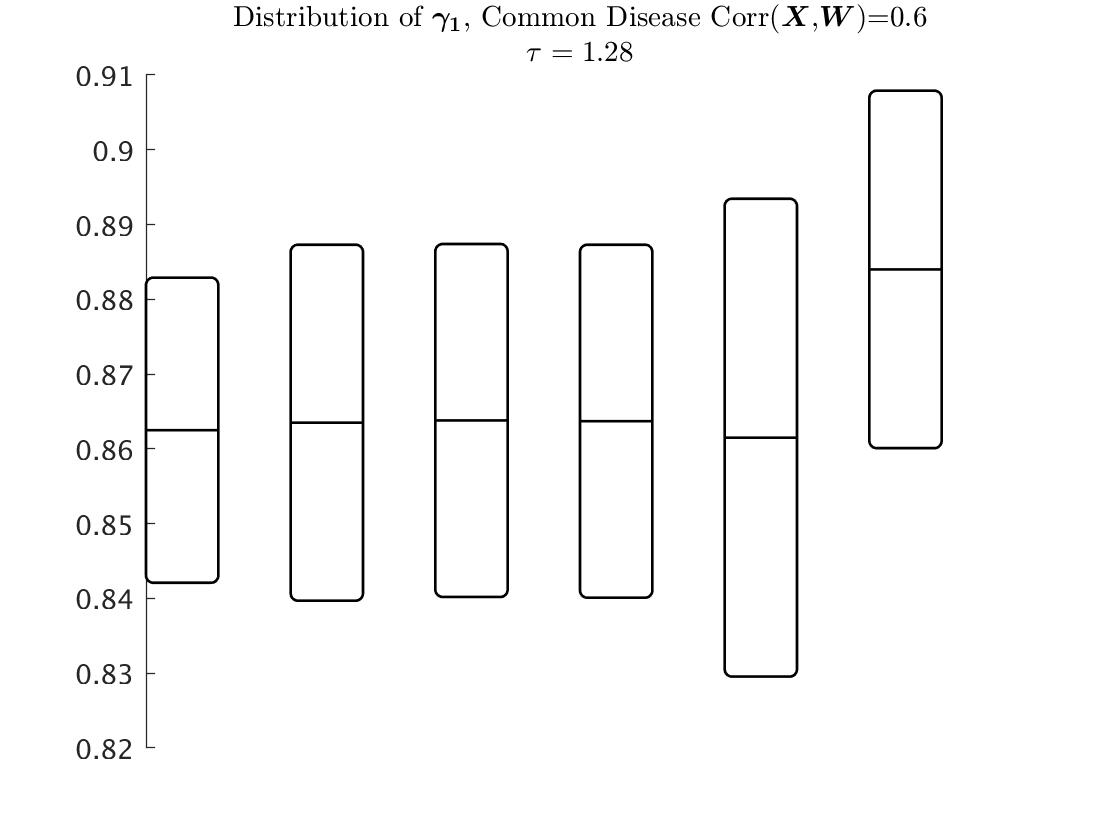} \hskip0.01truein
\vspace{0.5cm}
\ec
\caption{\footnotesize
Boxplots of each estimator over 1000 replications. The range of each box is between the median of the estimator minus 0.5 of its interquartile range, to the median of the estimator plus 0.5 of its interquartile range. The horizontal line inside each box is the median of the estimator. For a given value of the changepoint $\tau$, each plot describes from left to right the naive, RC1, RC2, RR1, RR2, and the SIMEX methods. For RC1, RC2, RR1, and the RR2 methods, the plots are based on the estimates under unknown nuisance parameters which were estimated by an external reliability sample of size 500 with 2 replications/person. The boxplots describe the results under the common disease with $n=3,000$ and cumulative incidence of 0.5. The plots from top to bottom present the behaviour of $\hat{\beta}$, $\hat{\omega}$ and $\hat{\gamma_1}$, respectively, where the behaviour of $\hat{\gamma_2}$ is the same as the the behaviour of $\hat{\gamma_1}$. The value of $\rho_{xw}$ is 0.6.}

\label{fig:simuboxZ}
\end{figure}
\end{landscape}

The previous simulation assumed a single covariate $X$ which is measured with error. But in typical applications there are additional error-free covariates $Z$ as well. We therefore extended the previous simulation to include two additional time-independent error-free $\bZ=[Z_1,Z_2]$. The simulation design was the same as before, but with the following additional settings: We set the coefficients of $\bZ$ to be $\gamma_1=\log(2.5)$ and $\gamma_2=\log(3)$, the covariates vector $[X,Z_1,Z_2]$ was generated as multivariate normal with zero mean vector and the identify covariance matrix. The event time was generated as exponential with parameter $\lambda =\lambda_{0} \exp (\beta X+\omega (X-\tau )_{+} )+\gamma_1 Z_1+\gamma_2 Z_2$. Here we examined the performance of the estimators except the MPPLE, under the common disease scenario only. Tables S.10-S.11 in the Supplemental Materials present the simulation results for all the methods examined.
Figure \ref{fig:simuboxZ}
provides a general comparison between all estimators examined using boxplots. The building of the boxplot is the same as we did before, that is, the range of each box is between the median of the estimator minus 0.5 of its interquartile range, to the median of the estimator plus 0.5 of its interquartile range. The horizontal line inside each box is the median of the estimator.
Results are presented separately for $\beta$, $\omega$ and $\gamma_1$, and separately for each value of the changepoint considered, for $\rho_{xw}=0.6$. The behaviour of $\gamma_2$ is the same as the  behaviour of $\gamma_1$ for all the considered scenarios. The comparisons for $\rho_{xw}=0.8$ and $\rho_{xw}=0.4$ are presented in S.9 in the Supplemental Materials.

%
%
%

The best performing method for $\beta$ was RC2, and for $\omega$ the RR2. The trend were seen for $\beta$ and $\omega$ are the same as those in the setting of one covariate $X$ only without additional covariates $\bZ$ (Section 4.3.2., (i)). The best performing method for $\gamma_1$ and $\gamma_2$ was the SIMEX. The general results for the estimators of $\gamma_1$ and $\gamma_2$ are: (a) They performed well as $\tau$ increased (b) As expected, they performed progressively less well as measurement error increased (c) They had close values over the all methods for each scenario considered.

\indent We also examined the robustness as in the setting of one covariate $X$ only (Section 4.4): Table 5 and Table S.13 in the Supplemental Materials present the results. The results under the \emph{t}-distribution with df=15 were close to those under the normal distribution, whereas the results under the \emph{t}-distribution with df=6 were far from those under the normal distribution. The results under the log gamma distribution with parameters $\alpha=\beta=1$ were noticeably worse that those obtained under the normal distribution. The estimates of $\gamma_1$ and $\gamma_2$ of \emph{t}-distribution with df=6 and df=15 were close, whereas those of the log gamma distribution were a little far from them. The estimates of $\omega$ under \emph{t}-distribution with df=6 were lower than those of \emph{t}-distribution with df=15. The estimates of $\beta$ under \emph{t}-distribution with df=6 were higher than those of \emph{t}-distribution with df=15.
\normalsize
\section{Illustrative Examples}
\subsection{NHS Example}
As noted in the introduction, our work was motivated by some possible instances of threshold effects observed in the Nurses' Health Study (NHS), including threshold effects observed in the NHS's investigation of the long term health effects of air pollution. We considered an analysis of the effect of exposure to particulate matter of diameter 10 $\mu g$/$m^3$ or less (PM$_{10}$) in relation to fatal myocardial infarction (MI).
Thus, the event of main interest is death due to MI, with other causes of death operating as competing events.

Here, 93,013 female nurses were followed from June 1992 to June 2006, with 1,073 fatal MI events observed. PM$_{10}$ exposure was assessed for each individual by linking the individual's residential address to her predicted PM$_{10}$ exposure using a spatio-temporal model derived from data from Environmental Protection Agency (EPA) area monitors (Yanosky {\it{et al}}., 2008; Paciorek and Liu, 2009). The time scale in the analysis was age in months, so that the data are subject to left truncation.
We fit a stratified Cox model, with stratum defined by age in months. For each stratum we used all of the cases and $10\%$ random sample of the nurses who did not experience the event (we took this random sample in order to reduce the heavy computational burden). Thus, we worked with
a data set comprising 11,793 female nurses. Aside from the PM$_{10}$, the Cox model adjusted for calendar year, indicator variables for season, and indicator variables for US state of residence.
All covariates were time-varying.
In this study, the measurement error is largely of Berkson form since
the surrogate value was determined by the nearest EPA monitoring site, so that all the participants for whom a given site was the closest monitoring site were assigned
the same surrogate value. The methods that work with the conditional distribution of $X$ given $W$, including the RC1, RC2, RR1, RR2, and MPPLE are applicable,
but the SIMEX method, which assumes the classical measurement error model and works with the conditional distribution of $W$ given $X$ is not applicable.
In addition, the MPPLE method
was not included because it not appropriate for the setting of time-dependent covariates.
For the determination of the changepoint value, we used the a priori information regarding the PM$_{10}$, that the annual standard recommended by the World Health Organization for PM$_{10}$ particles is a concentration of 20 micrograms per cubic meter (World Health Association (2006)), and therefore we set the changepoint at 25 $mg/m^3$. In the following analysis, we consider PM$_{10}$ minus its mean as the covariate of interest rather than PM$_{10}$ itself, and we refer it as the standardized PM$_{10}$. This difference yields more stable results of the estimates than those based on PM$_{10}$ itself.



\begin{figure}[h]
\bc
\includegraphics[width=1.43in, height=1.3in]{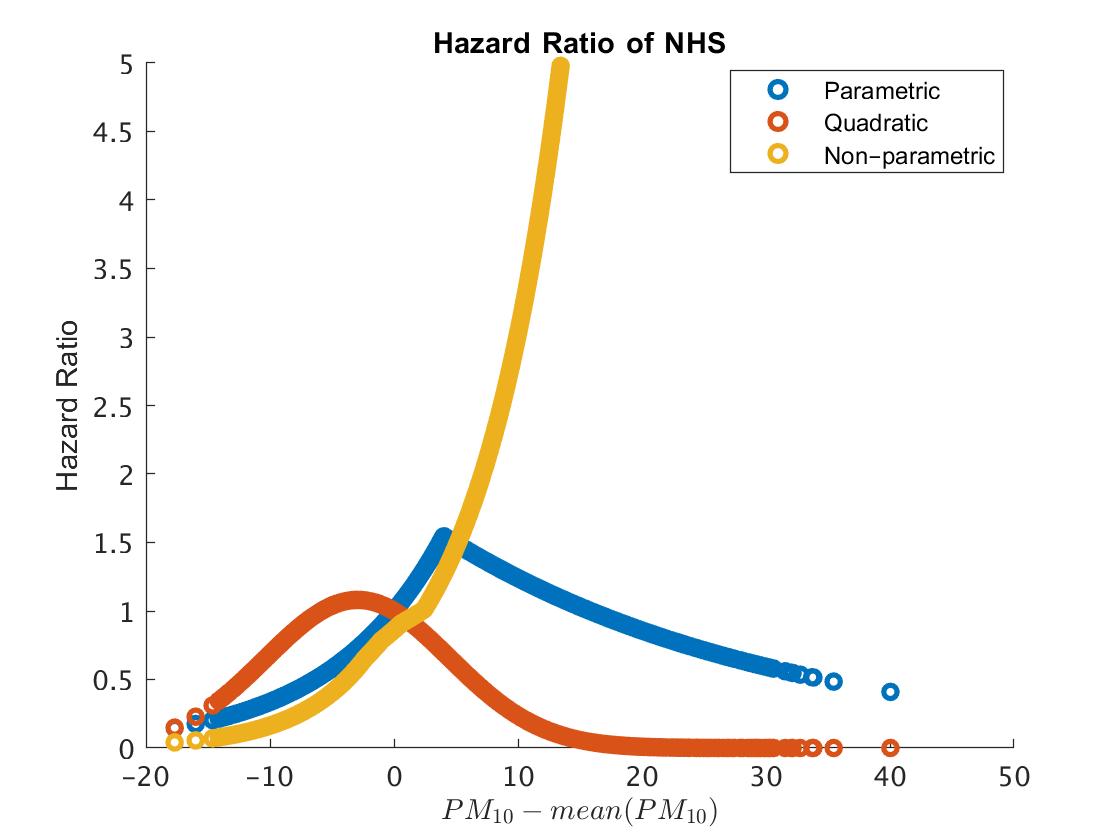} \hskip0.01truein
\ec
\caption{\footnotesize
Hazard Ratio (HR) for the NHS data with event of fatal myocardial infarction, based on (1) the fitted parametric model, under the RR2 method (blue curve) (2) quadratic function (red curve) (3) non-parametric approach, which is a flexible parametric approach involving
an extended version of the RR2 method with 5 changepoints (yellow curve). The standardized changepoint for the first two plots is 4 $mg/m^3$.}
\label{fig:nhr_hr}
\end{figure}

Denote by $\tilde{X}$ the difference of $X$ minus its mean, and denote by $\tilde{W}$ the difference of $W$ minus its mean.
To estimate the conditional expectation $E(\tilde{X}|\tilde{W})$ and conditional variance, $\Var(\tilde{X}|\tilde{W})$, needed for the correction methods, we used an external validation study of 98 person-months in 4 cities of personal PM$_{10}$ measurements, which included personal environmental monitors and a surrogate exposure based on the spatio-temporal model of Yanosky (Kioumourtzoglou {\it{et al}}., 2014). We fit a mixed linear model of personal difference PM$_{10}$ minus its mean on the surrogate exposure, and we obtained  $E(\tilde{X}|\tilde{W})=1.181+0.635\tilde{W} $ and $\Var(\tilde{X}|\tilde{W})=58.169$.
For RR2, the estimate that we used for the standard deviation for a given parameter estimate was
the standard deviation estimate obtained from the asymptotic theory for RR1, since the empirical standard deviations of RR1 and RR2 in our simulation study were close.
Table 6 summarizes the results of the analysis. We report the results for the standardized PM$_{10}$ and for the standardized (PM$_{10} -\tau)_+$ only,
although we included background variables in the Cox analysis as stated previously. We give the estimate, the
standard deviation in brackets, the p-value and the $95\%$ confidence interval of the relevant coefficient.
The estimate of $\omega$ was found to be statistically significant under all of the methods considered except the RR2, that is, there is an evidence that PM$_{10}$ of 25 $mg/m^3$ is a changepoint.
\\
\indent For characterizing the predictive accuracy of the various methods, we use an adapted version of the concordance index ($C$-index) presented by Harrell et al. (1996), with the predictive marker taken to be the induced relative risk. Details are provided in the S.14 in the Supplemental Materials.
For the RR2 method, we used the formula for the induced risk derived for the RR1 and RR2 methods, with the estimates of $\beta$ and $\omega$
taken to be those obtained with the RR2 method.
%
%
The results are presented in Table 6. The $C$-index for all the methods is around 0.78, indicating good prediction performance with all the methods considered.
Figure \ref{fig:nhr_hr} presents the results graphically in terms of standardized PM$_{10}$, by plotting the hazard ratio HR($\tilde{x}$)$=\exp(\beta \tilde{x}+\omega (\tilde{x}-\tilde{\tau})_{+})$ for the standardized changepoint $\tilde{\tau}=4 mg/m^3$ with $\beta$ and $\omega$ estimated using the under the RR2 method, the method that performed best in our simulation study among those available for the case of time-dependent covariates (blue curve).
At the request of a reviewer, we also plot the HR($\tilde{x}$) obtained with two other approaches: (1) expressing
HR($\tilde{x}$) as a quadratic function of $\tilde{x}$ (red curve), and (2) a flexible parametric approach involving
an extended version of the RR2 method with 5 changepoints, which we refer it as a non-parametric approach (yellow curve).  	
The first two approaches are relative close, whereas the third approach seems to fit the data poorly.

We can see that the HR increases before the standardized changepoint of 4 $mg/m^3$, and above the changepoint it stabilizes. Therefore for policy implementation of health care, the recommended PM$_{10}$ should not increase the 25 $mg/m^3$.

\newpage
\subsection{FHS Example}
Here we consider the FHS data set (Gordon and Kannel (1968)), which involved 664 men aged 35-44 with no history of high blood pressure or cardiovascular disease at the beginning of the study. The participants were followed for up to 48 years. The aim of the analysis was to examine the effect of an individual long-term underlying the systolic blood pressure (SBP) level at the beginning of the study on the risk of deathfrom cardiovascular disease.
Thus, the event of main interest is death due to cardiovascular disease, with other causes of death operating as competing events.	
In this example, we used only the baseline SBP, so that the
covariate, $X$, is time-independent.
The covariate $X$ used was a transformed version of SBP defined by TSBP$=\log((\mathrm{SBP}-75)/25)$, which has been found to be approximately normally distributed. Due to random fluctuations in blood pressure, the SBP measured on a given
occasion is an error-prone version of individual long-term underlying SBP. The surrogate $W$ for $X$ was the average of the TSBP values from the first two exams, which were 2 years apart. A total of 208 deaths occurred over the follow-up period.
The analysis here was conducted as in Zucker (2005) with $X$ taken to be distributed $N(\mu,\sigma^2_{X})$ and $W$ assumed to be given by $W=X+U$, where $U$ is distributed $N(0, \sigma^2_U)$. From the two initial TSBP values, the estimates of $\mu_X$, $\sigma^2_X$ and $\sigma^2_U$ were 0.71, 0.045 and 0.013, respectively.
The a priori knowledge about SBP is that both low and high SBP increase the risk for cardiovascular disease, where low SBP is level below 90 $mmHg$ (hypotension), normal SBP is level of 90 to $<$120 $mmHg$, elevated SBP is level of 120 to $<$130, 
and for high SBP there are two stages: "stage 1" includes the range of $[130,140]$, and "stage 2" includes SBP which is higher than 140 (National Heart Lung and Blood Institute (2008), Whelton {\it{et al}}.\ (2017)). We expect to see a small change from one category to the next, but it should become apparent when the new level is somewhere in the middle of the next category. Here we considered moving from low to normal SBP using the changepoint of 105 $mmHg$. We applied all the methods discussed in this paper to the TSBP changepoint value of 0.182, which corresponds to 105 $mmHg$.

\begin{figure}[h]
\bc
\includegraphics[width=1.43in, height=1.3in]{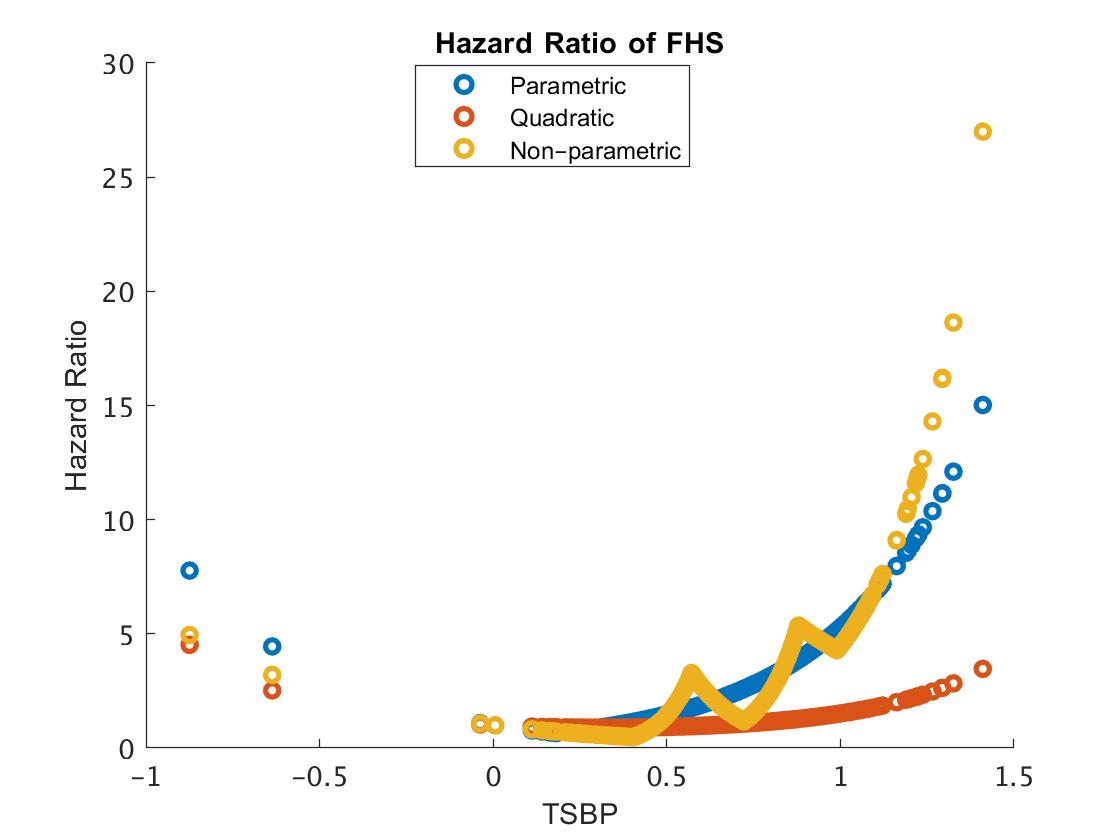} \hskip0.01truein
\ec

\caption{\footnotesize
Hazard Ratio (HR) for the FHS data with event of cardiovascular disease death, based on (1) the fitted parametric model, under the MPPLE method (blue curve) (2) quadratic function (red curve) (3) non-parametric approach, which is a flexible parametric approach involving
an extended version of the RR2 method with 5 changepoints (yellow curve). The changepoint $\tau$ for the first two plots is 0.182.}

\label{fig:fhs_hr}
\end{figure}

The variance of the SIMEX estimates of the regression coefficients was estimated by weighted bootstrap. In this example, the RR1 and RR2 methods may not be appropriate since the event was not rare (31.33\% of the participants experienced the event), but we still report the value of the estimate for comparison with the MPPLE.
Table 7 summarizes the results of the analysis. We report the results for $W$ and for $(W -\tau)_+$. We give the estimate, the standard deviation in brackets, the p-value and the $95\%$ confidence interval of the relevant coefficient. The estimate of $\omega$ was found to be statistically significant under all of the methods considered, that is, there is an evidence that the effect of SBP changes at 105 $mmHg$.
\\
For characterizing the predictive accuracy of the various methods, we use the concordance index ($C$-index)
as presented by Harrell at al. (1996), again taking the predictive marker to be the induced relative risk.
For the SIMEX method, the induced relative risk used was given by the formula for the induced relative risk for the naive estimate, substituting for
$\beta$ and $\omega$ the estimates obtained with the SIMEX method.
The results are presented in Table 7. The $C$-index for all the methods except the SIMEX and the MPPLE is around 0.62 (where for the SIMEX the $C$-index is 0.58), which indicates that the model fits the data not poorly, and also indicated that these methods behave close. The $C$-Index for the MPPLE is 0.85, which indicates on a good fitting of the model, and this is the best method in this data.
\\
Figure \ref{fig:fhs_hr} presents the results graphically in terms of TSBP, by plotting the hazard ratio HR for TSBP = 0.182 with $\beta$ and $\omega$ estimated using the MPPLE method. In this example we choose to show the results of the MPPLE since this is the best method over the considered methods according to our simulation study.
As in the previus example, we also plot the HR obtained with the quadratic (red curve) and non-parametric (yellow curve) approaches. The three approaches agree: We can see that the HR is high in very low values of TSBP, decreases as TSBP increases, and start to increase at TSBP = 0 which corresponds to SBP of 100 $mmHg$. Because the parameters estimates of $\beta$ and $\omega$ correspond to the transformed variable TSBP and not SBP, they are difficult to direct interpret. Figure \ref{fig:fhs_hr} is the best way to understand the findings. It is evident from this Figure that there are adverse effects of both low and high SBP. For policy implementation of health care, these results should be replicated in widely in order to suggest that the blood pressure that the patients should maintain is around 105 $mmHg$.

\newpage
\section{Potential extensions of the model}

\indent We describe four potential extensions of our model. The first involves allowing more than one changepoint in the main covariate. Specifically, the model is of the form:
\begin{align*}
\lambda (t|x(t),\bz(t))=\lambda _{0} (t)\exp(\gamma ^{T} \bz(t)+\beta x(t)+\omega _{1}(x(t)-\tau _{1})_{+} +...+\omega _{k} (x(t)-\tau _{k} )_{+}) ,
\end{align*}
where $\tau _{1} ,...,\tau _{k} $ are the potential changepoints in the covariate domain.
The second extension involves allowing changepoints in the background covariates in addition to a changepoint in the main covariate. That is, the model is of the form:
\begin{align*}
\lambda (t|x(t),\bz(t))=\lambda _{0} (t)\exp (\sum _{j=1}^{p}\gamma _{j} z_{j}(t)+\sum _{j=1}^{p}\sum _{k_{j} =1}^{K_{j} }\omega _{k_{j} } (z_{j} (t)-\tau _{k_{j} })_{+}+\beta x(t)+\omega (x(t)-\tau)_{+}  ).
\end{align*}
In principle, all the methods discussed in this paper can be extended to handle the above two models. The theory can be extended straightforwardly. The resulting increase in the number of parameters, however, may lead to a substantial increase in the run time or the rate of convergence failure. Also, the variances of the parameter estimates will increase,
and it may take a large sample size to estimate all the parameters with reasonable precision.

\indent
The third extension concerns the measurement error structure.
In our model, in the case with time-dependent covariates we assumed that the measurement errors $U_i(t_j)$ are i.i.d.\ across $j$ with
distribution $N(0,\sigma_u^2)$.
Often this is a plausible assumption. In some situations, however, there may be serial correlation among the $U_i(t_j)$'s. Under the internal validation design, we can extend
our methods to accommodate such correlation. Suppose we assume that $\bX \sim N(\bmu,\bSig_X)$ and $\bU \sim N(0,{\bf D})$, where $\bSig_X$ and ${\bD}$ are
covariance matrices that can be estimated from the internal validation data, possibly using parametric models. Define $\bSig_W = \bSig_X + {\bf D}$.
The distribution of $\bX$ given $\bW$ is then $N(\bmu + \bG(\bW - \bmu), \bH)$, with $\bG =  \bSig_X \bSig_W^{-1} $ and $\bH = \bSig_X - \bSig_X \bSig_W^{-1} \bSig_X$.
We can then compute the conditional expectations $E[X(t_j)|\bW]$, $E[(X(t_j)|-\tau)_+|\bW]$, and $E[\exp(\beta X(t_j) + \omega (X(t_j)|-\tau)_+|\bW]$ and thereby
apply the RC1, RC2, and RR methods. A similar development can be carried out for the internal replicate measures design.

\indent
The fourth extension concerns the nature of the main covariate
$X(t)$. In our model, we assumed that $X(t)$ is continuous.
A reviewer inquired about the case where $X(t)$ is an ordinal categorical variable.
If numerical codes are assigned to the different levels of the covariate,
the model (\ref{CP}) can be used, and then the issue is how to deal with the
misclassification. The MPPLE approach, which works with the conditional distribution
of the true covariate given the observed covariate, can be extended in a straightforward
way to a categorical variable with misclassification. Other possible ways of handling
the misclassification include the corrected score approach of Zucker and Spiegelman
(2008), which works with the conditional distribution of the observed covariate given the true covariate or the misclassification SIMEX of K\"{u}chenhoff et al. (2006).
Some work has been done on optimal data-driven choice of the numerical scores used
for the categories (see Willems et al., 2017, for a discussion in the Cox model setting),
and this approach can in principle be combined with one of the aforementioned approaches
to handling misclassification.

\section{Summary}
\normalsize
We have developed point and interval estimators for the regression coefficients in a Cox survival model with a changepoint, in a
setting where the covariate values are subject to measurement error. This type of analysis is of interest in many
epidemiological studies. We considered the case where the changepoint is known and where the covariate of main interest is a scalar. All the methods developed in this paper can be extended to the multi-dimensional case. In addition to the naive method
ignoring the measurement error, we examined the following methods: two versions of regression calibration (RC1 and the new RC2 ), SIMEX,
the induced
relative risk approach of Prentice (1982) (in two versions: Prentice's original proposal (RR1) and a new version using a bootstrap
bias correction (RR2)), and the MPPLE method of Zucker (2005).

We derived the asymptotic properties of the estimators and carried
out a simulation study under rare and common disease settings to evaluate their bias and confidence
interval coverage. The simulation study mainly considered the simple case of one main covariate only, but also included the extended case of one main covariate with additional error-free covariates. For the former setting, we examined the both common disease and the rare diseases cases, where for the later setting we examined the common disease case only.
Under one main covariate only: In general, all the correction methods performed better than the naive analysis with no correction, but
the methods that performed best were the RR2 method for the rare disease setting and the MPPLE method for the common disease setting. In the common disease case, the average relative bias over all simulation replications of the estimator of $\omega$ from its true value of 0.69 for $\rho_{xw}=0.8$ ranged
(over the various choices of the measurement error parameters) from
$-0.48$ to 0.02 with the RR2 and MPPLE methods, compared with $-0.93$ to $-0.16$ with the naive, SIMEX, and RC methods, respectively. In the rare disease case, the average relative bias over all simulation replications ranged from $-0.06$
to 0.07 with RR2 method and from $-0.85$ to 0.12 with the naive and RC methods.
\\
\indent The key factors determining the performance of the methods was the correlation between the true variable and its surrogate and
the location of the changepoint. As expected, the estimators and the coverage probabilities performed better with less measurement error and a centrally located changepoint.
It is interesting to note that, in contrast with the standard Cox model without a threshold, in the current setting with a threshold the
RC and SIMEX methods generally performed poorly, even with a modest degree of measurement error ($\rho_{xw}=0.8$). Thus, when the covariate is
measured with error, measurement error correction is substantially harder in the model with a threshold
than in the model without a threshold, even when the threshold is known.
In a follow-up paper we will present methodology for the case where the threshold is unknown and has to be estimated.

Analyses such as that included in Section 5.1 are used to estimate attributable disease burden and set public policy on maximum exposure limits
(U.S.\ EPA, 2009; Fann {\it{et al}}., 2011). Accurate characterization of the exposure-response relationship is critical for accurate policy-making. In this paper, we have developed methods to improve the methodology for dose-response characterization in the present of exposure measurement error, addressing a key limitation in previously available methods.



\vspace*{1pc}

\noindent {\bf{Conflict of Interest}}

\noindent {\it{The authors have declared no conflict of interest.}}


\section*{Appendix}

\subsection*{A.1. \enspace Notation for the naive, RC1 and RC2 estimates}

The following is additional notation relevant for the asymptotic theory of the naive, RC1 and RC2 estimators (with $\ba^{\otimes 2} $ for a vector $\ba$ defined as $\ba \ba^{T} $):
\[S^{(2)} (t,\, \bth ,\, g)=\frac{1}{n} \sum _{i=1}^{n}Y_{i} (t)\,  \bV_{i}(g(t))^{\otimes 2} \exp (\bth ^{T} \bV_{i} (g(t)))\]
\[\lambda _{i}(t)=\lambda _{0} (t)\exp \{\beta X_{i}(t) +\omega (X_{i}(t) -\tau )_{+} +\bgam ^{T} \bZ_{i}(t) \}\]
\[\tilde{\lambda }_{i} (t)dt=Y_{i} (t)\lambda _{i} (t)dt\]
\[\tilde{S}^{(0)} (t,\bth)=\frac{1}{n} \sum _{i=1}^{n}Y_{i} (t)\lambda _{i}(t)dt \]
\[\tilde{S}^{(1)} (t,\bth,g)=\frac{1}{n} \sum _{i=1}^{n}\bV_{i} (g(t))Y_{i} (t)\lambda _{i} (t)dt\]
\[\tilde{S}^{(2)}(t,\, \bth ,\, g)=\frac{1}{n} \sum _{i=1}^{n}\bV_{i} (g(t))^{\otimes 2} Y_{i} (t)\,  \lambda _{i} (t)dt\]
 \[\tilde{s}^{(0)} (t,\, \bth )=E(\tilde{S}^{(0)} (t,\, \bth )) ,
 \\ \tilde{s}^{(j)} (t,\, \bth ,\, g)=E(\tilde{S}^{(j)} (t,\, \bth ,\, g))\, \, ,\, \, j=1\, ,\, 2\]
\[I(t,\bth ,g)=\sum _{i=1}^{n}\delta _{i} \left(\frac{S^{(2)} (T_{i} ,\bth ,g)}{S^{(0)} (T_{i} ,\bth ,g)} -\left( \frac{S^{(1)} (T_{i} ,\bth ,g)}{S^{(0)}(T_{i} ,\bth,g)}\right)^{\otimes 2} \right)\]
\[\Sigma (t^{*},\bth ,g)=\int _{0}^{t^{*} }\left(\frac{s^{(2)} (t,\bth,g)}{s^{(0)} (t,\bth ,g)} -\left( \frac{s^{(1)}(t,\bth,g)}{s^{(0)} (t,\bth,g)} \right)^{\otimes 2}\right) \tilde{s}^{(0)} (t,\bth )dt\]
\[q^{(g)}(t^{*},\bth)\ ={\int _{0}^{t^{*} }\tilde{s}^{(1)} (t,\bth ,g)dt-\int _{0}^{t^{*} }\frac{s^{(1)} (t^{*},\bth ,g)}{s^{(0)} (t,\bth,g)} \tilde{s}^{(0)} (t,\bth) dt }.\]

\subsection*{A.2. \enspace Notation for the RR estimator}
The following is additional notation relevant for the asymptotic theory for the RR estimator.
\[S^{(0)}(t,\bth  \, )=\frac{1}{n} \sum _{i=1}^{n}Y_{i} (t)r(\bth  \, ,\bV_{i} (t)) \exp (\bgam ^{T} \bZ_{i} (t))\]
\[S^{(1)}(t,\bth  \, )=\frac{1}{n} \sum _{i=1}^{n}Y_{i} (t)r^{(1)} (\bth  ,\bV_{i} (t)) \exp(\bgam ^{T} \bZ_{i}(t))\]
\[S^{(2)} (t,\bth  \, )=\frac{1}{n} \sum _{i=1}^{n}Y_{i} (t)r^{(2)} (\bth  ,\bV_{i} (t)) \exp (\bgam ^{T} \bZ_{i} (t))\]
\[S^{(3)}(t,\bth  \, )=\frac{1}{n} \sum _{i=1}^{n}{r^{(1)}(\bth  ,\bV_{i} (t)) \mathord{/{\vphantom{r^{(1)} (\bth  ,\bV_{i} (t)) r(\bth  ,\bV_{i} (t))}}\kern-\nulldelimiterspace} r(\bth  ,\bV_{i} (t))} \tilde{\lambda }_{i} (t) \]
\[S^{(4)}(t,\bth  \, )=\frac{1}{n} \sum _{i=1}^{n}({r^{(2)} (\bth  ,\bV_{i} (t)) \mathord{/{\vphantom{r^{(2)} (\bth  ,\bV_{i} (t)) r(\bth  ,\bV_{i} (t))}}\kern-\nulldelimiterspace} r(\bth  ,\bV_{i} (t))} -({r^{(1)} (\bth  ,\bV_{i} (t)) \mathord{/{\vphantom{r^{(1)} (\bth  ,\bV_{i} (t)) r(\bth  ,\bV_{i} (t))}}\kern-\nulldelimiterspace} r(\bth  ,\bV_{i} (t))} ))^{\otimes 2} \tilde{\lambda }_{i}(t) \]
\[S^{(5)} (t,\bth  \, )=\frac{1}{n} \sum _{i=1}^{n}Y_{i} (t)r(\bth  \, ,\bV_{i} (t)) \exp (\bgam ^{T} \bZ_{i} (t))\bZ_{i} (t)\]
\[S^{(6)} (t,\bth  \, )=\frac{1}{n} \sum _{i=1}^{n}Y_{i} (t)r(\bth  \, ,\bV_{i} (t)) \exp (\bgam ^{T} \bZ_{i} (t))\bZ_{i} (t)^{\otimes 2} \]
\[S^{(7)} (t,\bth  \, )=\frac{1}{n} \sum _{i=1}^{n}Y_{i} (t)r(\bth \, ,\bV_{i} (t)) \exp (\bgam ^{T} \bZ_{i} (t))\bZ_{i} (t)\]
\[S^{(8)} (t,\bth  \, )=\frac{1}{n} \sum _{i=1}^{n}Y_{i} (t)r^{(1)} (\bth  \, ,\bV_{i} (t)) \exp (\bgam ^{T} \bZ_{i} (t))\bZ_{i} (t)\]
\[S^{(9)} (t,\bth  \, )=\frac{1}{n} \sum _{i=1}^{n}\bZ_{i} (t) \]
\[\tilde{S}^{(0)} (t,\bth  \, )=\frac{1}{n} \sum _{i=1}^{n}Y_{i} (t)\lambda _{i} dt \]
\[s^{(j)} (t,\bth  \, )=E(S^{(j)} (t,\bth  \, ))\, ,\, \, j=0,...,9  , \tilde{s}^{(0)} (t,\bth  \, )=E(S^{(0)} (t,\bth  \, ))\, \,  .\]

Denote  $-\frac{\partial ^{2} l_{p} (t,\bth )}{\partial \bth ^{2} } $ by $I(t,\bth )$. This is a matrix of  size $(p+2)\times (p+2)$, where
\begin{align*}
{I_{kl} (t^{*},\bth  )=-\int _{0}^{t^{*} }\ \left(\frac{S^{(6)} (t,\bth  )}{S^{(0)}(t,\bth  )} -\frac{S^{(5)} (t,\bth  )^{\otimes 2} }{S^{(0)} (t,\bth  )^{\otimes 2} } \tilde{S}^{(0)}(t,\bth  )\ \right)dt \, ,\, \, \, k,l=1,...,p}
\end{align*}
\begin{align*}
 {I_{kl} (t^{*},\bth  )\, =\int _{0}^{t^{*} }\ \left(-\frac{S^{(8)} (t,\bth  )}{S^{(0)} (t,\bth  )} +\frac{S^{(7)} (t,\bth  )}{S^{(0)} (t,\bth  )} \frac{S^{(1)} (t,\bth  )}{S^{(0)} (t,\bth )} \tilde{S}^{(0)} (t,\bth  )\ \right)dt \, \, ,\, \, \, k=1,...,p\, ,\, l=p+1,p+2}
 \end{align*}
\begin{align*}
{I_{kl} (t^{*},\bth  )=-\int _{0}^{t^{*} }\ \left(S^{(4)} (t,\bth  )-\frac{S^{(2)} (t,\bth  )}{S^{(0)} (t,\bth  )} -\frac{S^{(1)}(t,\bth  )^{\otimes 2} }{S^{(0)} (t,\bth  )^{\otimes 2} } \tilde{S}^{(0)}(t,\bth )\ \right)dt \, \, ,\, \, \, k\, ,\, l=p+1,p+2\, \, .}
\end{align*}

In addition define $\Sigma (t^{*},\bth)$ to be a matrix of  size $(p+2)\times (p+2)$, where
\begin{align*}
{\Sigma _{kl} (t^{*},\bth  )=-\int _{0}^{t^{*} }\ \left(\frac{s^{(6)} (t,\bth  )}{s^{(0)} (t,\bth  )} -\frac{s^{(5)} (t,\bth )^{\otimes 2} }{s^{(0)} (t,\bth )^{\otimes 2} } \tilde{s}^{(0)} (t)\ \right)dt \, ,\, \, \, k,l=1,...,p}
 \end{align*}
\begin{align*}
{\Sigma _{kl} (t^{*},\bth )\, =\int _{0}^{t^{*} }\ \left(-\frac{s^{(8)} (t,\bth  )}{s^{(0)} (t,\bth )} +\frac{s^{(7)} (t,\bth  )}{s^{(0)} (t,\bth  )} \frac{s^{(1)} (t,\bth )}{s^{(0)} (t,\bth  )} \tilde{s}^{(0)} (t,\bth  )\ \right)dt \, \, ,\, \, \, k=1,...,p\, ,\, l=p+1,p+2}
 \end{align*}
\begin{align*}
{\Sigma _{kl} (t^{*},\bth )=-\int _{0}^{t^{*} }\ \left(s^{(4)} (t,\bth  )-\frac{s^{(2)} (t,\bth  )}{s^{(0)} (t,\bth  )} -\frac{s^{(1)} (t,\bth  )^{\otimes 2} }{s^{(0)} (t,\bth  )^{\otimes 2} } \tilde{s}^{(0)} (t,\bth )\ \right)dt \, \, ,\, \, \, k\, ,\, l=p+1,p+2\, \, .}
\end{align*}

Define $q^{(RR1)}(t^{*} ,\bth  )$ to be a vector of length $p+2$ in which
\noindent
\begin{align*}
{q_{k}^{(RR1)} (t^{*} ,\bth )\, =\int _{0}^{t^{*} }\ \left(s^{(9)}(t,\bth  )-\frac{s^{(5)} (t,\bth  )}{s^{(0)} (t,\bth  )} \tilde{s}^{(0)} (t,\bth  )\ \right)dt \, ,\, \, \, k=1,...,p}
\end{align*}
\begin{align*}
 {q_{k}^{(RR1)} (t^{*} ,\bth  )\, =\int _{0}^{t^{*} }\ \left(s^{(3)} (t,\bth  )-\frac{s^{(1)} (t,\bth )}{s^{(0)} (t,\bth  )} \tilde{s}^{(0)} (t,\bth  )\ \right)dt \, ,\, \, k=p+1,p+2\, .}
 \end{align*}


\newpage

\begin{center}
\fontsize{8.5}{0.5}\selectfont
\begin{tabular}{  m{1.15cm}  m{0.85cm} m{1.45cm} m{1.45cm} m{1.45cm} m{1.45cm} m{1.45cm}}
\mc{7}{c}{\bf{TABLE 1.} Comparing changepoint with two error-prone covariates without a changepoint} \\
\\
\\
\\
\\
\\
\\
\\
\\
\rowcolor{lightgray} \mc{7}{c} {\textbf{\emph{Two Variables and Changepoint Estimates$^a$ of \boldmath{$\beta$}, (\boldmath{$\beta$}, \boldmath{$\omega$}) = (0.405, 0.693)}}} \\
\\
\\
\\
\\
\\
 \toprule

\rowcolor{lightgray} & & Naive  &Naive  & RC1$^b$& RC1$^b$& RC2$^b$\\
 \\
 \\
 \\

\rowcolor{lightgray} $\tau$ & $\rho _{xw}$& 2 variables$^c$ & changepoint$^d$& 2 variables & changepoint & changepoint\\
\\
\toprule
$\Phi^{-1}(0.25)$ & 0.8  & 0.337& 0.435&	0.430& 0.589&0.582\\
\\
                  & 0.6  & 0.215& 0.272&	0.439& 0.674&0.674\\
                  \\
                  & 0.4  & 0.104& 0.123&	0.411& 0.781&0.739\\

 \\
 \\
$\Phi^{-1}(0.5)$ & 0.8   & 0.293& 0.334&	0.456& 0.521&0.457\\
\\
                  & 0.6  & 0.180& 0.209&	0.498& 0.578&0.489\\
\\
                  & 0.4  & 0.086& 0.098&	0.540& 0.612&0.516\\
\\
 \\

$\Phi^{-1}(0.75)$ & 0.8  & 0.268& 0.282&	0.448& 0.471&0.412\\
\\
                  & 0.6  & 0.156& 0.167&	0.483& 0.505&0.420\\
\\
                  & 0.4  & 0.073& 0.077&	0.513& 0.522&0.430\\
\\
%

\\
\\
\\
\\
\\
\\
\\
\\
\rowcolor{lightgray} \mc{7}{c}{\textbf{\emph{Two Variables and Changepoint Estimates$^a$ of \boldmath{$\omega$}, (\boldmath{$\beta$}, \boldmath{$\omega$}) = (0.405, 0.693)}}} \\
\\
\\
\\
\\
\\
 \toprule
\rowcolor{lightgray} & & Naive  &Naive  & RC1$^b$& RC1$^b$& RC2$^b$\\
 \\
 \\
 \\
\rowcolor{lightgray} $\tau$ & $\rho _{xw}$& 2 variables$^c$ & changepoint$^d$& 2 variables & changepoint & changepoint\\
\\
\toprule
$\Phi^{-1}(0.25)$ & 0.8  & 0.279& 0.162&	0.502&0.333& 0.364\\
\\
                  & 0.6  & 0.102& 0.031&	0.396& 0.157&0.181\\
\\
                  & 0.4  & 0.029& 0.017&	0.360&0.001&0.054 \\

 \\
 \\
$\Phi^{-1}(0.5)$ & 0.8   & 0.311& 0.241&	0.486&0.376& 0.499\\
\\
                  & 0.6  & 0.123& 0.076&	0.342& 0.209&0.384\\
\\
                  & 0.4  & 0.036& 0.003&	0.224& 0.107&0.297\\
\\
 \\

$\Phi^{-1}(0.75)$ & 0.8  & 0.297& 0.246&	0.535& 0.422&0.554\\
\\
                  & 0.6  & 0.113& 0.080&	0.438& 0.284&0.468\\
\\
                  & 0.4  & 0.032& 0.019&	0.404& 0.221&0.396\\
\\
  \end{tabular}
  \end{center}
\footnotesize{$^a$ The values
in the cells are means over 1000 replications for the common disease scenario. $^b$ The estimates of the RC are calculated under known nuisance parameters with $\sigma_w^2=1$ and with $\sigma_u^2$ that was determined according to the value of $\rho_{xw}$. $^c$ The setting is a model with two error-prone covariates without a changepoint. $^d$ The setting is a model with a single error-prone covariate with a changepoint.}

\begin{landscape}
\begin{center}
\small
\fontsize{8.5}{0.5}\selectfont
\begin{tabular}{llcccccccccc}
\mc{10}{c}{\bf{TABLE 2.} Nominal coverage probability of \boldmath{$\beta$}} \\
\\
\\
\\
\\
\\
\\
\\
\\
\\
\\
\\
\\
\\
\\
\\
\\
\rowcolor{lightgray} \mc{10}{c}{\bf \underline{Case A}\rm: Common Disease$^a$}\\
\\
\\
\\
\\
\\
\\
\mc{10}{c}{$\rho _{xw} =0.6$ } \\
\\
\\
\\
\\
\\
\\
$\tau$ Value      & Naive          & RC1(kn)$^b$     &RC1(ukn)$^c$   & RC2(kn)$^b$ &RC2(ukn)$^c$& RR1(kn)$^b$    & RR1(ukn)$^c$ &MPPLE(kn)$^b$   &MPPLE(ukn)$^c$ \\
\\
\\
\\
\\
\\
$\Phi^{-1}(0.1)$    &\textbf{0.793}  &\textbf{0.972}  &\textbf{0.997}          & \textbf{0.871} & \textbf{0.977}   & 0.957  & 0.956  &\textbf{0.998} &\textbf{0.994} \\
$\Phi^{-1}(0.25)$   &\textbf{0.276}  &\textbf{0.882}  &0.951  & \textbf{0.871} & 0.952           & \textbf{0.973}  & \textbf{0.972}  &0.955           &0.955  \\
$\Phi^{-1}(0.5)$    &\textbf{0.002}  &\textbf{0.616}  &\textbf{0.773}  & \textbf{0.929} & 0.962   & 0.949          & 0.964  &0.947           &0.944  \\
$\Phi^{-1}(0.75)$   &\textbf{0.000}  &\textbf{0.593}  &\textbf{0.831}  & 0.948          & \textbf{0.983}   & 0.938  & 0.956          &0.946           &0.937  \\
$\Phi^{-1}(0.9)$    &\textbf{0.000}  &\textbf{0.866}  &0.957  & 0.950          & \textbf{0.976}   & 0.940  & 0.956          & 0.942
               &0.937  \\
\\
\\
\\
\\
\\
\\
\\
\\
\\
\\
\\
\\
\\
\\
\\
\\
\\
\\
\rowcolor{lightgray} \mc{10}{c}{\bf \underline{Case B}\rm: Rare Disease$^e$}\\
\\
\\
\\
\\
\\
\\
\mc{10}{c}{$\rho _{xw} =0.6$ } \\
\\
\\
\\
\\
\\
\\
$\Phi^{-1}(0.1)$    &\textbf{0.923}  &0.964  &\textbf{0.992} &\textbf{0.908} &\textbf{0.988}  &\textbf{0.884}  &\textbf{0.877} \\
$\Phi^{-1}(0.25)$   &\textbf{0.741}  &\textbf{0.873}  &0.942 &\textbf{0.819} &\textbf{0.870}  &0.951  &0.951  \\
$\Phi^{-1}(0.5)$    &\textbf{0.098}  &\textbf{0.211}  &\textbf{0.359} &\textbf{0.714} &\textbf{0.770}  &0.953  &\textbf{0.970}  \\
$\Phi^{-1}(0.75)$   &\textbf{0.000}  &\textbf{0.013}  &\textbf{0.278} &\textbf{0.831} &\textbf{0.839}  &0.949  &\textbf{0.983}  \\
$\Phi^{-1}(0.9)$    &\textbf{0.000}  &\textbf{0.139}  &\textbf{0.660} &\textbf{0.917} &0.940  &0.940  &\textbf{0.975}  \\
\\
\end{tabular}
\end{center}
\footnotesize{$^a$ $n=3,000$ with cumulative incidence of 0.5. $^b$ (kn) indicates estimates under known nuisance parameters with $\sigma_w^2=1$ and with $\sigma_u^2$ that was determined according to the value of $\rho_{xw}$. $^c$ (ukn) indicates estimates under unknown nuisance parameters which were estimated by an external reliability sample of size 500 with 2 replications/person. $^d$ values in bold format are outside of the band $0.95\pm 1.96\sqrt{\frac{0.95\times 0.05}{1000} }=[0.936, 0.964]$. $^e$ $n=50,000$ with cumulative incidence of 0.03.}
\end{landscape}

\begin{landscape}
\begin{center}
\small
\fontsize{8.5}{0.5}\selectfont
\begin{tabular}{llccccccccc}
\mc{10}{c}{\bf{TABLE 3.} Nominal coverage probability of \boldmath{$\omega$}} \\
\\
\\
\\
\\
\\
\\
\\
\\
\\
\\
\\
\\
\\
\\
\\
\\
\rowcolor{lightgray} \mc{10}{c}{\bf \underline{Case A}\rm: Common Disease$^a$}\\
\\
\\
\\
\\
\\
\\
\mc{10}{c}{$\rho _{xw} =0.6$ } \\
\\
\\
\\
\\
\\
\\
$\tau$ Value      & Naive          & RC1(kn)$^b$     &RC1(ukn)$^c$   & RC2(kn)$^b$ &RC2(ukn)$^c$& RR1(kn)$^b$    & RR1(ukn)$^c$ &MPPLE(kn)$^b$   &MPPLE(ukn)$^c$ \\
\\
\\
\\
\\
\\
$\Phi^{-1}(0.1)$    &\textbf{0.000}  &\textbf{0.975}  &\textbf{0.990} &\textbf{0.808} & \textbf{0.861}   &\textbf{0.975}   &\textbf{0.977} &\textbf{0.998} &\textbf{0.995}\\
$\Phi^{-1}(0.25)$   &\textbf{0.000}  &\textbf{0.624}  &\textbf{0.684} &\textbf{0.742} & \textbf{0.770}   &\textbf{0.997}   &\textbf{0.985} &\textbf{0.965} &\textbf{0.971}\\
$\Phi^{-1}(0.5)$    &\textbf{0.000}  &\textbf{0.099}  &\textbf{0.154} &\textbf{0.801} & \textbf{0.783}   &\textbf{0.860}   &\textbf{0.899} &0.955 &0.954\\
$\Phi^{-1}(0.75)$   &\textbf{0.000}  &\textbf{0.476}  &\textbf{0.514} &\textbf{0.888} & \textbf{0.863}   &\textbf{0.782}   &\textbf{0.823} &0.954 &0.955\\
$\Phi^{-1}(0.9)$    &\textbf{0.000}  &\textbf{0.895}  &0.937 &\textbf{0.913} & \textbf{0.916}   &\textbf{0.860}   &\textbf{0.860} & 0.950     &0.946\\
\\
\\
\\
\\
\\
\\
\\
\\
\\
\\
\\
\\
\\
\\
\\
\\
\\
\\
\\
\\
\rowcolor{lightgray} \mc{10}{c}{\bf \underline{Case B}\rm: Rare Disease$^e$}\\
\\
\\
\\
\\
\\
\\
\mc{10}{c}{$\rho _{xw} =0.6$ } \\
\\
\\
\\
\\
\\
\\
$\Phi^{-1}(0.1)$    &\textbf{0.000} &0.963 &\textbf{0.993} & \textbf{0.905} &0.942 &\textbf{0.891}&\textbf{0.891}\\
$\Phi^{-1}(0.25)$   &\textbf{0.000} &\textbf{0.855} &\textbf{0.859} & \textbf{0.837} &\textbf{0.833} &0.960&0.949\\
$\Phi^{-1}(0.5)$    &\textbf{0.000} &\textbf{0.275} &\textbf{0.347} & \textbf{0.878} &\textbf{0.893} &0.950         &\textbf{0.966}\\
$\Phi^{-1}(0.75)$   &\textbf{0.000} &\textbf{0.584} &\textbf{0.605} & 0.941 &\textbf{0.980} &0.934&\textbf{0.972}\\
$\Phi^{-1}(0.9)$    &\textbf{0.000} &\textbf{0.946} &\textbf{0.974} & \textbf{0.877} &\textbf{0.988} &0.947&\textbf{0.970}\\
\\
\\
\\
\end{tabular}
\end{center}
\footnotesize{$^a$ $n=3,000$ with cumulative incidence of 0.5. $^b$ (kn) indicates estimates under known nuisance parameters with $\sigma_w^2=1$ and with $\sigma_u^2$ that was determined according to the value of $\rho_{xw}$. $^c$ (ukn) indicates estimates under unknown nuisance parameters which were estimated by an external reliability sample of size 500 with 2 replications/person. $^d$ values in bold format are outside of the band $0.95\pm 1.96\sqrt{\frac{0.95\times 0.05}{1000} }=[0.936, 0.964]$. $^e$ $n=50,000$ with cumulative incidence of 0.03.}
\end{landscape}

\begin{landscape}
{\bf Table 4. Robustness - median bias }
\begin{center}
\fontsize{8.5}{0.5}\selectfont
\begin{tabular}{m{0.98cm}  m{0.98cm} m{0.68cm}| m{0.95cm}  m{0.95cm} m{0.95cm}| m{0.95cm} m{0.95cm} m{0.95cm}| m{0.95cm} m{0.95cm}  m{0.95cm} m{0.95cm}}
\mc{12}{c}{\textbf{Finite Sample Bias$^a$ in \boldmath{$\beta$}, (\boldmath{$\beta$}, \boldmath{$\omega$})=(0.405, 0.693)}} \\
\\
\\
\\
&&&&Gamma&(1,1) &&$t(6)$ &&& $t(15)$\\
\hline
\\
\\
$\tau$ &Disease& $\rho _{xw}$ & RC2$^b$ &  RR1$^b$ & RR2$^b$ & RC2$^b$ &  RR1$^b$ & RR2$^b$ & RC2$^b$ &  RR1$^b$ & RR2$^b$\\
\\
\\
$\Phi^{-1}(0.1)$ &Common$^c$ & 0.6  & -1.434&	-1.786&	-1.834& 0.536&	-0.520&	-1.230&	0.792&	-0.088&	-1.105 \\
 \\
 $\Phi^{-1}(0.25)$ &Common$^c$ &  0.6  & -0.669&	-0.948&	-0.958 & 0.574&	0.087&	-0.039&	0.483&	0.037&	-0.119\\
 \\
 $\Phi^{-1}(0.5)$ &Common$^c$ &  0.6  & -0.376&	-0.506&	-0.509 & 0.321&	0.062&	0.027&	0.205&	-0.034&	-0.053\\
 \\
 $\Phi^{-1}(0.75)$ &Common$^c$ & 0.6  & -0.257&	-0.300&	-0.312 & 0.156&	0.039&	0.024&	0.056&	-0.038&	-0.047 \\
 \\
 $\Phi^{-1}(0.9)$ &Common$^c$ &  0.6  & -0.189	&-0.209&	-0.225& 0.062&	0.024&	0.016&	0.011&	-0.018&	-0.032 \\
 \\
  \\
  \\
  \\
  \\
  \\
  \mc{12}{c}{\textbf{Finite Sample Bias$^d$ in \boldmath{$\omega$}, (\boldmath{$\beta$}, \boldmath{$\omega$})=(0.405, 0.693)}} \\
  \\
\\
\\
&&&&Gamma&(1,1) &&$t(6)$ &&& $t(15)$\\
\hline
\\
\\
$\tau$ &Disease& $\rho _{xw}$ & RC2$^b$ &  RR1$^b$ & RR2$^b$ & RC2$^b$ &  RR1$^b$ & RR2$^b$ & RC2$^b$ &  RR1$^b$ & RR2$^b$\\
\\
$\Phi^{-1}(0.1)$ &Common$^c$ & 0.6  & 0.626&	0.762&	0.781&	-0.717&	-0.067&	0.343&	-0.792&	-0.272&	0.293 \\
 \\

 $\Phi^{-1}(0.25)$ &Common$^c$ & 0.6  & 0.287&	0.318&	0.308&	-0.789&	-0.473&	-0.413&	-0.631&	-0.365&	-0.302 \\
 \\
$\Phi^{-1}(0.5)$ &Common$^c$ & 0.6  & 0.236&	0.064&	0.053&	-0.677&	-0.523&	-0.526&	-0.455&	-0.372&	-0.369 \\
 \\
 $\Phi^{-1}(0.75)$ &Common$^c$ & 0.6  & 0.164&	-0.141&	-0.142&	-0.649&	-0.610&	-0.614&	-0.366&	-0.422&	-0.425 \\
 \\
 $\Phi^{-1}(0.9)$ &Common$^c$ & 0.6  & 0.018&	-0.304&	-0.284&	-0.690&	-0.695&	-0.711&	-0.364&	-0.503&	-0.502 \\
 \\
 \\
\end{tabular}
\end{center}
\footnotesize{$^a$ The values
in the cells are relative bias of the median of $\beta$, i.e., (median-0.405)/0.405. $^b$ The estimates were obtained under unknown nuisance parameters which were estimated by an external reliability sample of size 500 with 2 replications/person. $^c$ $n=3,000$ with cumulative incidence of 0.5. $^d$ The values
in the cells are relative bias of the median of $\omega$, i.e., (median-0.693)/693.}
\end{landscape}

\begin{landscape}
{\bf Table 5. Additional Covariates Robustness - median bias}
\begin{center}
\fontsize{8.5}{0.5}\selectfont
\begin{tabular}{m{0.98cm}  m{0.98cm} m{0.68cm}|  m{0.95cm} m{0.95cm} m{0.95cm}| m{0.95cm} m{0.95cm} m{0.95cm} | m{0.95cm} m{0.95cm} m{0.95cm} }
\mc{12}{c}{\textbf{Finite Sample Bias$^a$ in \boldmath{$\beta$}, (\boldmath{$\beta$}, \boldmath{$\omega$}, \boldmath{$\gamma_1$}, \boldmath{$\gamma_2$})=(0.405, 0.693, 0.916, 1.099)}} \\
\\
\\
\\
&&&&Gamma&(1,1) &&$t(6)$ &&& $t(15)$\\
\hline
\\
\\
$\tau$ &Disease& $\rho _{xw}$ & RC2$^b$ &  RR1$^b$ & RR2$^b$ & RC2$^b$ &  RR1$^b$ & RR2$^b$ & RC2$^b$ &  RR1$^b$ & RR2$^b$\\
\\
\\
$\Phi^{-1}(0.1)$ &Common$^c$ &  0.6  & -1.291&	-1.694&	-1.982& 0.227&	-0.485&	-0.231&	0.186&	-0.531&	-0.534 \\
 \\
 $\Phi^{-1}(0.25)$ &Common$^c$ & 0.6  & -0.620&	-0.888&	-1.123 & 0.205&	-0.180&	-0.065&	0.065&	-0.342&	-0.308\\
 \\
 $\Phi^{-1}(0.5)$ &Common$^c$ & 0.6  & -0.321&	-0.428&	-0.719 & 0.118&	-0.091&	-0.159&	-0.012&	-0.218&	-0.402\\
 \\
 $\Phi^{-1}(0.75)$ &Common$^c$ & 0.6  & -0.181&	-0.224&	-0.454 & 0.063&	-0.033&	-0.194&	-0.031&	-0.112&	-0.324 \\
 \\
 $\Phi^{-1}(0.9)$ &Common$^c$ & 0.6  & -0.111&	-0.126&	-0.277& 0.025&	0.010&	-0.091&	-0.024&	-0.058	&-0.180 \\
 \\
  \\
  \\
  \\
  \\
  \\
  \mc{12}{c}{\textbf{Finite Sample Bias$^d$ in \boldmath{$\omega$}, (\boldmath{$\beta$}, \boldmath{$\omega$},\boldmath{$\gamma_1$},\boldmath{$\gamma_2$})=(0.405, 0.693, 0.916, 1.099)}} \\
  \\
\\
\\
&&&&Gamma&(1,1) &&$t(6)$ &&& $t(15)$\\
\hline
\\
\\
$\tau$ &Disease& $\rho _{xw}$ & RC2$^b$ &  RR1$^b$ & RR2$^b$ & RC2$^b$ &  RR1$^b$ & RR2$^b$& RC2$^b$ &  RR1$^b$ & RR2$^b$\\
\\
$\Phi^{-1}(0.1)$ &Common$^c$ & 0.6  & 0.559&	0.740&	0.986&	-0.546&	-0.147&	-0.256&-0.452&	-0.054&	-0.034 \\
 \\

 $\Phi^{-1}(0.25)$ &Common$^c$ & 0.6  & 0.270& 	0.297& 	0.564&	-0.559&	-0.321&	-0.328&	-0.383&	-0.178&	-0.122 \\
 \\
$\Phi^{-1}(0.5)$ &Common$^c$ & 0.6  & 0.166&	-0.020&	0.409&	-0.531&	-0.431&	-0.276&	-0.311&	-0.266&	-0.021 \\
 \\
 $\Phi^{-1}(0.75)$ &Common$^c$ & 0.6  & 0.052&	-0.236&	0.143&	-0.546&	-0.536&	-0.294&	-0.301&	-0.371&	-0.049 \\
 \\
 $\Phi^{-1}(0.9)$ &Common$^c$ & 0.6  & -0.129&	-0.390&	-0.117&	-0.633&	-0.687&	-0.436&	-0.310&	-0.467&	-0.172 \\
 \\
 \\
 \\
  \\
  \\

\end{tabular}
\end{center}
\footnotesize{$^a$ The values
in the cells are relative bias of the median of $\beta$, i.e., (median-0.405)/0.405. $^b$ The estimates were obtained under unknown nuisance parameters which were estimated by an external reliability sample of size 500 with 2 replications/person. $^c$ $n=3,000$ with cumulative incidence of 0.5. $^d$ The values
in the cells are relative bias of the median of $\omega$, i.e., (median-0.693)/0.693. }
\end{landscape}

\begin{landscape}
\begin{center}
\fontsize{8.5}{0.5}\selectfont
\begin{tabular}{m{0.98cm}  m{0.98cm} m{0.68cm}| m{0.95cm}  m{0.95cm} m{0.95cm} | m{0.95cm} m{0.95cm} m{0.95cm}  | m{0.95cm} m{0.95cm}  m{0.95cm} m{0.95cm}}

  \mc{12}{c}{\textbf{Finite Sample Bias$^e$ in \boldmath{$\gamma_1$}, (\boldmath{$\beta$}, \boldmath{$\omega$},\boldmath{$\gamma_1$},\boldmath{$\gamma_2$})=(0.405, 0.693, 0.916, 1.099)}} \\
  \\
\\
\\
&&&&Gamma&(1,1) &&$t(6)$ &&& $t(15)$\\
\hline
\\
\\
$\tau$ &Disease& $\rho _{xw}$ & RC2$^b$ &  RR1$^b$ & RR2$^b$ & RC2$^b$ &  RR1$^b$ & RR2$^b$ & RC2$^b$ &  RR1$^b$ & RR2$^b$\\
\\
$\Phi^{-1}(0.1)$ &Common$^c$ & 0.6  & -0.150&	-0.155&	-0.157&	-0.179&	-0.184&	-0.186&	-0.187&	-0.186&	-0.188 \\
 \\

 $\Phi^{-1}(0.25)$ &Common & 0.6  & -0.126&	-0.127&	-0.126&	-0.159&	-0.159&	-0.161&	-0.166&	-0.165&	-0.168 \\
 \\
$\Phi^{-1}(0.5)$ &Common & 0.6  & -0.083	&-0.083	&-0.084&	-0.123&	-0.123&	-0.124&	-0.127&	-0.127&	-0.129 \\
 \\
 $\Phi^{-1}(0.75)$ &Common & 0.6  & -0.051&	-0.051&	-0.054&	-0.086&	-0.086&	-0.086&	-0.083&	-0.083&	-0.086 \\
 \\
 $\Phi^{-1}(0.9)$ &Common & 0.6  & -0.037&	-0.041&	-0.042&	-0.059&	-0.060&	-0.062&	-0.055&	-0.055&	-0.057 \\
 \\
 \\
 \\
 \\
 \\
  \\
  \\
  \\
  \\
  \\
  \mc{12}{c}{\textbf{Finite Sample Bias$^f$ in \boldmath{$\gamma_2$}, (\boldmath{$\beta$}, \boldmath{$\omega$},\boldmath{$\gamma_1$},\boldmath{$\gamma_2$})=(0.405, 0.693, 0.916, 1.099)}} \\
  \\
\\
\\
&&&&Gamma&(1,1) &&$t(6)$ &&& $t(15)$\\
\hline
\\
\\
$\tau$ &Disease& $\rho _{xw}$ & RC2$^b$ &  RR1$^b$ & RR2$^b$ & RC2$^b$ &  RR1$^b$ & RR2$^b$ & RC2$^b$ &  RR1$^b$ & RR2$^b$\\
\\
$\Phi^{-1}(0.1)$ &Common$^c$ & 0.6  & -0.150&	-0.150&	-0.151&	-0.179&	-0.179&	-0.180&	-0.185&	-0.186&	-0.186 \\
 \\

 $\Phi^{-1}(0.25)$ &Common & 0.6  & -0.127&	-0.127&	-0.127&	-0.159&	-0.159&	-0.161&	-0.164&	-0.166&	-0.168\\
 \\
$\Phi^{-1}(0.5)$ &Common & 0.6  & -0.083&	-0.083&	-0.085&	-0.123&	-0.123&	-0.125&	-0.125&	-0.125&	-0.124 \\
 \\
 $\Phi^{-1}(0.75)$ &Common& 0.6  & -0.051&	-0.051&	-0.053&	-0.085&	-0.085&	-0.085&	-0.081&	-0.081&	-0.081 \\
 \\
 $\Phi^{-1}(0.9)$ &Common & 0.6  & -0.037&	-0.038&	-0.039&	-0.058&	-0.058&	-0.059&	-0.055&	-0.055&	-0.055 \\
 \\
 \\
\end{tabular}
\end{center}
\footnotesize{$^e$ The values
in the cells are relative bias of the median of $\gamma_1$, i.e., (median-0.916)/0.916. $^b$ The estimates were obtained under unknown nuisance parameters which were estimated by an external reliability sample of size 500 with 2 replications/person. $^c$ $n=3,000$ with cumulative incidence of 0.5. $^f$ The values
in the cells are relative bias of the median of $\gamma_2$, i.e., (median-1.099)/1.099. }
\end{landscape}

\begin{landscape}
\begin{center}
\small
\fontsize{7.6}{1.15}\selectfont
\begin{tabular}{llcccccccccc}

\mc{10}{c}{\bf{TABLE 6.} Results for the NHS Study of Chronic PM\boldmath{$_{10}$} (\boldmath{$mg$/$m^3$}) Exposure in Relation to Fatal MI $^a$} \\
\\
\\
\\
\\
\\
\\
\\
\mc{10}{c}{Assuming Known Threshold $\tau$$^b$, Event is Fatal MI} \\
\\
\\
\\
\\
\\
\\
\\
\\
\boldmath{$\tau$} &  \textbf{coefficient} & \textbf{Naive} & \textbf{RC1} & \textbf{RC2} & \textbf{RR1}& \textbf{RR2} \\
\\
\\
\\
\\
\\
\\
\\
\rowcolor{lightgray} \textbf{4} &\boldmath{$\beta$}  &0.034 (0.008)&0.052(0.027) &0.087 (0.047)  &0.120 (0.052) &0.109 (0.052) \\[1pt] \\
\\
                    &&8.02e-06        &0.058   &0.063         &0.023         &0.038     &      \\[1pt]   \\
\\
                &&[0.019,0.049]    &[-0.002,0.106]        &[-0.005,0.179] &[0.017,0.222]&[0.006,0.212]  & \\[1pt]\\
\\
\\
\\

\rowcolor{lightgray}   &\boldmath{$\omega$} &-0.039 (0.013) &-0.059 (0.024) &-0.123 (0.060)       &-0.157 (0.077)&-0.146 (0.077) \\[1pt] \\
\\
                    &&0.003    &0.015       & 0.040   &0.042   &0.060       &   \\[1pt]      \\
\\
                    &&[-0.064,-0.013]    &[-0.106, -0.011] &[-0.240,-0.006]&[-0.309,-0.006]&[-0.298,0.006]& \\[1pt]\\
\\
\\
\\

\rowcolor{lightgray}   &\textbf{C-index} &0.781 &0.783       &0.777& 0.777   &0.778 \\[1pt] \\
\\

\end{tabular}
\end{center}
\footnotesize{$^a$ The results are in terms of standardized PM$_{10}$ (that is, PM$_{10}$ minus its mean). Each cell contains (in that order): estimate (standard deviation), p-value, $95\%$ confidence interval of the relevant coefficient). $^b$ The threshold is in terms of standardized PM$_{10}$ with the value 4. This value is corresponded to PM$_{10}$ = 25 $mg$/$m^3$.}\\
\end{landscape}

\begin{landscape}
\begin{center}
\small
\fontsize{7.6}{1.18}\selectfont
\begin{tabular}{llccccccccccc}

\mc{10}{c}{\bf{Table 7.} Results for the FHS Study of the Effect of Systolic Blood Pressure on Cardiovascular Disease Death$^a$} \\
\\
\\
\\
\\
\\
\\
\\

\mc{10}{c}{Assuming Known Threshold $\tau$$^b$, Event is Cardiovascular Disease Death} \\
\\
\\
\\
\\
\\
\\
\\
\\
\boldmath{$\tau$} &\textbf{coefficient} & \textbf{Naive} & \textbf{RC1} & \textbf{RC2} & \textbf{RR1} & \textbf{RR2} & \textbf{SIMEX} & \textbf{MPPLE} \\
\\
\\
\\
\\
\\
\\
\\
\\

\rowcolor{lightgray} \textbf{0.182$^b$} &\boldmath{$\beta$} &-1.121 (1.200)  &-2.405 (1.666) &-2.360 (1.731)       &-2.356 (1.637)    &-3.214 (1.637) &-4.173 (1.756) &-2.345 (1.631) \\[1pt]  \\
\\
                    &&0.350   &0.149         &0.173    &0.150              &0.050         &0.018 &0.150 \\[1pt] \\
\\
                    &&[-3.474 ,1.232]&[-5.670, 0.859]&[-5.752, 1.033] &[-5.565, 0.852]  &[-6.423, -0.006]&[-7.616,-0.731] &[-5.541,0.851] \\[1pt]  \\
\\
\\
\\
\\
\rowcolor{lightgray}   &\boldmath{$\omega$} &3.055 (1.328) &4.900 (1.842)    &4.858 (1.899)    &4.855 (1.816)  &5.742 (1.816) &4.415 (1.842) &4.900  (1.800) \\[1pt] \\
\\
                    &&0.021            &0.008          &     0.011      &0.008         &0.002       &0.017        &0.007 \\[1pt] \\
                    &&[0.453 ,5.657]& [1.289,8.510] &[1.137,  8.579] &[1.296,8.413]  &[2.183,  9.300]&[0.804    8.026]&[1.371,8.429] \\[1pt] \\
\\
\\
\\

\rowcolor{lightgray}   &\textbf{C-index} &0.619 &0.619       &0.619& 0.619   &0.621&0.578&0.854 \\[1pt] \\
\\
\\
\end{tabular}
\end{center}
\footnotesize{$^a$ Each cell contains (in that order): estimate (standard deviation), p-value, $95\%$ confidence interval of the relevant coefficient). $^b$ This
value corresponds to SBP = 105 $mmHg$.}\\
\end{landscape}

\begin{center}
\fontsize{8.0}{0.07}\selectfont
\begin{tabular}{  m{1.15cm}  m{0.85cm} m{1.45cm} m{1.45cm} m{1.45cm} m{1.45cm}}
\mc{6}{c}{\bf{TABLE 8.} Comparing RC2 with RC1 in a Simple Linear Regression} \\
\\
\\
\\
\\
\rowcolor{lightgray} \mc{6}{c} {\textbf{\emph{Mean Estimates$^a$ of \boldmath{$\beta$}, (\boldmath{$\beta$}, \boldmath{$\omega$}) = (0.405, 0.693)}}} \\
\\
\\
\\
 \toprule

\rowcolor{lightgray} & & $\beta$  &$\beta$  & $\omega$& $\omega$\\
 \\
 \\

\rowcolor{lightgray} $\tau$ & $\rho _{xw}$& RC1$^b$ & RC2$^b$& RC1$^b$ & RC2$^b$\\
\\
\toprule
$\Phi^{-1}(0.1)$ & 0.8  & 0.363& 0.409&	0.708& 0.693\\
\\
                  & 0.6  & 0.340& 0.419&	0.716& 0.693\\
                  \\
                  & 0.4  & 0.408& 0.501&	0.718& 0.692\\

 \\
 \\
$\Phi^{-1}(0.25)$ & 0.8   & 0.338& 0.408&	0.737& 0.693\\
\\
                  & 0.6  & 0.250& 0.417&	0.790& 0.693\\
\\
                  & 0.4  & 0.227& 0.487&	0.828& 0.695\\
\\
 \\

$\Phi^{-1}(0.5)$  & 0.8  & 0.383& 0.407&	0.743& 0.695\\
\\
                  & 0.6  & 0.321& 0.414&	0.884& 0.697\\
\\
                  & 0.4  & 0.177& 0.468&	1.277& 0.716\\
\\
 \\
$\Phi^{-1}(0.75)$ & 0.8   & 0.447& 0.406&	0.672& 0.698\\
\\
                  & 0.6  & 0.494& 0.412&	0.728& 0.710\\
\\
                  & 0.4  & 0.549& 0.449&	0.804& 0.764\\
\\
 \\

$\Phi^{-1}(0.9)$  & 0.8  & 0.439& 0.406&	0.690& 0.705\\
\\
                  & 0.6  & 0.465& 0.409&	0.835& 0.745\\
\\
                  & 0.4  & 0.491& 0.427&	0.130& 0.722\\
\\

\\
\\
\\
\\
\rowcolor{lightgray} \mc{6}{c} {\textbf{\emph{Median Estimates$^c$ of \boldmath{$\beta$}, (\boldmath{$\beta$}, \boldmath{$\omega$}) = (0.405, 0.693)}}} \\
\\
\\
\\
 \toprule

\rowcolor{lightgray} & & $\beta$  &$\beta$  & $\omega$& $\omega$\\
 \\
 \\

\rowcolor{lightgray} $\tau$ & $\rho _{xw}$& RC1$^b$ & RC2$^b$& RC1$^b$ & RC2$^b$\\
\\
\toprule
$\Phi^{-1}(0.1)$ & 0.8  & 0.360& 0.406&	0.708& 0.693\\
\\
                  & 0.6  & 0.323& 0.403&	0.716& 0.692\\
                  \\
                  & 0.4  & 0.298& 0.396&	0.719& 0.687\\

 \\
 \\
$\Phi^{-1}(0.25)$ & 0.8   & 0.336& 0.405&	0.736& 0.693\\
\\
                  & 0.6  & 0.237& 0.404&	0.788& 0.690\\
\\
                  & 0.4  & 0.120& 0.399&	0.822& 0.680\\
\\
 \\

$\Phi^{-1}(0.5)$ & 0.8  & 0.383& 0.405&	0.740& 0.692\\
\\
                  & 0.6  & 0.313& 0.405&	0.863& 0.688\\
\\
                  & 0.4  & 0.148& 0.400&	1.134& 0.670\\
\\
 \\

$\Phi^{-1}(0.75)$ & 0.8  & 0.446& 0.405&	0.670& 0.694\\
\\
                  & 0.6  & 0.485& 0.405&	0.682& 0.685\\
\\
                  & 0.4  & 0.505& 0.401&	0.667& 0.639\\
\\
\\
\\
 \\

$\Phi^{-1}(0.9)$ & 0.8  & 0.437& 0.405&	0.676& 0.697\\
\\
                  & 0.6  & 0.457& 0.403&	0.708& 0.678\\
\\
                  & 0.4  & 0.455& 0.393&	0& 0.557\\
\\

\\
\\
\\
\rowcolor{lightgray} \mc{4}{c} {\textbf{\emph{Percent of convergence over 1000 replications$^d$}}} \\
\\
\\
\\
 \toprule

 \\
 \\
 \\

\rowcolor{lightgray} $\tau$ & $\rho _{xw}$& RC1$^b$ & RC2$^b$\\
\\
\toprule
$\Phi^{-1}(0.1)$ & 0.4  & 0.999& 0.999\\
 \\
 \\
$\Phi^{-1}(0.25)$ & 0.4   & 0.999& 0.999\\
\\

$\Phi^{-1}(0.5)$ & 0.4  & 0.993& 0.999&\\
\\
\\

$\Phi^{-1}(0.75)$ & 0.4  & 0.947& 0.993\\
\\
\\
$\Phi^{-1}(0.9)$ & 0.8  &       & 0.999     \\
\\
                 & 0.6  &0.992       &      \\
                 \\
                 & 0.4 & 0.900 &0.949       \\
\\

  \end{tabular}
  \end{center}
\footnotesize{$^a$ The values
in the cells are means over 1000 replications for a sample of $n=1500$. $^b$ The estimates of RC1 and RC2 are calculated under  unknown nuisance parameters which were estimated by an external reliability sample of size 500 with 2 replications/person. $^c$ The values
in the cells are medians over 1000 replications for a sample of $n=1500$. $^d$ The table presents the problematic cases only where the percent of convergence was not 100\%. }

\end{document}